\documentclass{emulateapj}
\usepackage{amsmath}
\usepackage{paralist}
\usepackage{color}

\shorttitle{Host Galaxies of PTF SNe Ic and Ic-bl \& Implications for Jet-Production}

\begin{document}

\title{Host Galaxies of Type Ic and Broad-lined Type Ic Supernovae from the Palomar Transient Factory: Implication for Jet Production}


\author{Maryam Modjaz\altaffilmark{1,2},  
Federica B. Bianco\altaffilmark{1,3}, 
Magdalena Siwek\altaffilmark{1,4}, 
Shan Huang\altaffilmark{1}, 
Daniel A. Perley\altaffilmark{5}, 
David Fierroz\altaffilmark{1}, 
Yu-Qian Liu\altaffilmark{1},  
Iair Arcavi\altaffilmark{6,7,8}, 
Avishay Gal-Yam\altaffilmark{9},
Nadia Blagorodnova\altaffilmark{10},
Bradley S. Cenko\altaffilmark{11,12}, 
Alexei V. Filippenko\altaffilmark{13},  
Mansi Kasliwal\altaffilmark{10}, 
Shri Kulkarni\altaffilmark{10}, 
Steve Schulze\altaffilmark{9,14,15}, 
Kirsty Taggart\altaffilmark{5}, 
Weikang Zheng\altaffilmark{13}
}

\altaffiltext{1}{Center for Cosmology and Particle Physics, New York University, 726 Broadway, New York, NY 10003, USA; {\it email:} mmodjaz@nyu.edu}
\altaffiltext{2}{Center for Computational Astrophysics, Flatiron Institute,162 5th Avenue, 10010, New York, NY, USA}
\altaffiltext{3}{Center for Urban Science and Progress, New York University, 1 MetroTech Center, Brooklyn, NY 11201, USA}
\altaffiltext{4}{Harvard-Smithsonian Center for Astrophysics, 60 Garden Street, Cambridge, MA 02138, USA}
\altaffiltext{5}{Astrophysics Research Institute, Liverpool John Moores University, IC2, Liverpool Science Park,146 Brownlow Hill, Liverpool L3 5RF, UK}
\altaffiltext{6}{Department of Physics, University of California, Santa Barbara, CA 93106-9530, USA}
\altaffiltext{7}{Las Cumbres Observatory Global Telescope, 6740 Cortona Drive, Suite 102, Goleta, CA 93117-5575, USA}
\altaffiltext{8}{Department of Particle Physics and Astrophysics, Weizmann Institute of Science, Rehovot 7610001, Israel}
\altaffiltext{9}{Tel Aviv University, ADDRESS PLEASE, Israel}
\altaffiltext{10}{Division of Physics, Math and Astronomy, California Institute of Technology, 1200 East California Boulevard, Pasadena, CA
91125, USA ; Cahill Center for Astrophysics, California Institute of Technology, Pasadena, CA 91125, USA}
\altaffiltext{11}{Astrophysics Science Division, NASA Goddard Space Flight Center, Mail Code 661, Greenbelt, MD 20771, USA}
\altaffiltext{12}{Joint Space-Science Institute, University of Maryland, College Park, MD 20742, USA}
\altaffiltext{13}{Department of Astronomy, University of California, Berkeley, CA 94720-3411, USA}
\altaffiltext{14}{Instituto de Astrof\'{i}sica, Facultad de F\'{i}sica, Pontificia Universidad Cat\'{o}lica de Chile, Vicu\~{n}a Mackenna 4860, 7820436 Macul, Santiago, Chile}
\altaffiltext{15}{Millennium Institute of Astrophysics, Vicu\~{n}a Mackenna 4860, 7820436 Macul, Santiago, Chile}

\slugcomment{submitted to ApJ in Nov 2018}
\begin{abstract}

{Unlike the ordinary supernovae (SNe) some of which are hydrogen and helium deficient (called Type Ic SNe), broad-lined Type Ic SNe (SNe Ic-bl) are very energetic events, and all SNe coincident with bona fide long duration gamma-ray bursts (LGRBs) are of Type Ic-bl. Understanding the progenitors and the mechanism driving SN Ic-bl explosions vs those of their SNe Ic cousins is key to understanding the SN-GRB relationship and jet production in massive stars. Here we present the largest set of host-galaxy spectra of 28 SNe Ic and 14 SN Ic-bl, all discovered before 2013 by the same untargeted survey, namely the Palomar Transient Factory (PTF). We carefully measure their gas-phase metallicities, stellar masses ($M_*$s) and star-formation rates (SFRs) by taking into account recent progress in the metallicity field and propagating uncertainties correctly. We further re-analyze the hosts of 10 literature SN-GRBs using the same methods and compare them to our PTF SN hosts with the goal of constraining their progenitors from their local environments by conducting a thorough statistical comparison, including upper limits. We find that the metallicities, SFRs and M$_*$s of our PTF SN Ic-bl hosts are statistically comparable to those of SN-GRBs, but significantly lower than those of the PTF SNe Ic. The mass-metallicity relations as defined by the SNe Ic-bl and SN-GRBs are not significantly different from the same relations as defined by the SDSS galaxies, in contrast to claims by earlier works. Our findings point towards low metallicity as a crucial ingredient for SN Ic-bl and SN-GRB production since we are able to break the degeneracy between high SFR and low metallicity. 
We suggest that the PTF SNe Ic-bl may have produced jets that were choked inside the star or were able break out of the star 
as unseen low-luminosity or off-axis GRBs.}

\end{abstract}

\keywords{supernovae --- 
galaxies: abundances}

\section{Introduction} \label{sec:intro}

Exploding massive stars in form of Supernovae (SNe) and long Gamma-ray Bursts (GRBs) are the most powerful explosions in the Universe. They are visible over large cosmological distances, but uncovering their progenitors is a difficult task. 
The origin of long GRBs has been conclusively shown to be connected to the death of some kind of massive stars in form of SNe (e.g., \citealt{galama98, stanek03, hjorth03}), also known as the SN-GRB connection, where the spectrum of all nearby bona-fide long GRB afterglow metamorphosed into that of a Type Ic supernova with broad lines (SN Ic-bl) (e.g., \citealt{modjaz06}, for review see, e.g., \citealt{woosleyjanka05, modjaz11-rev, cano17_obs_guide} ). SNe Ic are defined by the lack of H and He lines (for review, see \citealt{Filippenko97_review,Gal-Yam16}) indicating that the SN Ic progenitor has lost large amounts (if not all) of its H and He envelopes before explosion \citep{Branch06, Hachinger12, Liu16}. In addition, a subclass of SNe Ic show broadened lines in their spectra (thus called SNe Ic-bl), indicating high ejecta expansion velocities of order $\sim 15000\,-\,30000\,\rm km\,s^{-1}$\citep{modjaz16,Sahu18} and released kinetic energies as high as $\sim 10^{52} \rm \, erg\,s^{-1}$ according to some models \citep{mazzali17}, up to 10 times more than canonical SNe. While all bona-fide long GRBs have been associated with SNe Ic-bl, there are many SNe Ic-bl without observed GRBs and the big question is why.

Direct imaging of the progenitors of SNe Ic-bl with and without observed GRBs is either unsuccessful or impossible (e.g., \citealt{gal-yam05, Eldridge13}), so the progenitors and the explosions conditions remain unclear, and are the focus of current research. Although a better understanding of stellar evolution is an important goal in its own right, the question of SNe and GRB progenitors impacts a diverse group of research fields. For instance, GRBs are suspected sites for high-energy cosmic ray acceleration \citep{Abbasi12}. GRBs and SNe are also a fundamental part of the chemical history of the Universe, since particularly black-hole-forming SNe are thought to have made important contributions to the early chemical evolution of the Universe \citep{Nomoto06}. Finally, due to their high luminosity, long GRBs can be detected at very high redshift. With the most distant GRB at redshift $\rm z \sim 9.4$ \citep{cucchiara11}, their mere existence at such distances makes them ideal tools to probe the star formation rate (SFR) history of the universe and provide constraints on the properties of interstellar dust, the reionization history and star formation in the early Universe \citep{perley16c, wang11, robertson12}, and thus highly motivates the search for a unified model for GRBs and SNe Ic-bl.





SNe Ic-bl belong to the class of Stripped-Envelope core-collapse SNe \citep{Clocchiatti97, Modjaz14, Zapartas17}, for short Stripped SNe and constitute a small fraction of Stripped SNe ($\sim4\%$, \citealt{Shivvers17}\footnote{Note the caveat that this SN Ic-bl rate is based on one object in the LOSS sample.}). Since the mechanism behind the energetic outflows in SNe Ic-bl is not clear, the role of GRBs in SNe Ic-bl is particularly interesting. Although not all SNe Ic-bl have been observed with an associated GRB \citep{Berger02, soderberg10_09bb, margutti14_12ap, Milisavljevic15_12ap},
all SNe associated with GRBs are of type Ic-bl (for reviews, see e.g.\citealt{woosley06_rev, modjaz11-rev, cano17_obs_guide}). In fact, for all bona-fide long-duration GRBs at redshift $z \lessapprox 0.3$ a corresponding SN Ic-bl has been found (e.g. \citealt{mazzali14, cano17_obs_guide})\footnote{While there is a growing class of "SN-less" GRBs such as GRB060614 (e.g. \citealt{gal-yam06}), whether the GRBs are truly long-duration GRBs or short-duration ones with extended tail is debated (e.g., \citealt{yu18}).} For the SN Ic-bl accompanied by GRBs, GRB jets may be the reason for the broadened lines in SNe Ic-bl as shown by \citealt{Barnes18}. 

As mentioned, there are many SNe Ic-bl without associated GRBs, and the question is why.
In the following we state several possible explanations as to why there are SNe Ic-bl without associated GRBs, some of which will be qualitatively assessed in subsequent sections.

\begin{inparaenum}[1)]
	\item The progenitor of the SN Ic-bl made a GRB, however it was off axis. Off-axis GRBs are discussed further in section \ref{sec:off-axis-grbs}. \\
	\item The broad lines in SNe Ic-bl have nothing to do with an engine, and no GRB is produced in association with those SNe Ic-bls that are seen without GRB. \\
	\item An engine is present, but the jet did not punch through the stellar envelope as it was too low in energy, was choked or did not last long enough \citep{Milisavljevic15_12ap,modjaz16}. Understanding which CCSNe harbor choked jets would also impact identifying the astrophysical sources for the diffuse flux of high-energy Icecube neutrino \citep{Senno18}.
\end{inparaenum}

If an engine is present, be it either a collapsar \citep{Woosley93, MacFadyen99, MacFadyen01} or magnetar \citep{Thompson94, usov92, Wheeler00, Thompson04},  one could suggest that jets occur in nearly all SNe Ic-bl, accelerating the outflows and causing line broadening, though GRBs are not always observed because they are either off-axis or stifled. We discuss this possibility later in the context of our results. 
In fact, recent observations by \cite{soderberg10_09bb, margutti14_12ap, Chakraborti15} show that there may be SN events that populate the gap between energetic, but non-relativistic SNe Ic-bl and SN-GRBs by producing relativistic, engine driven outflows but no observable GRBs. With these new observations, the connection between SN-GRBs and SNe Ic-bl appears to become increasingly relevant.

 Many studies have concluded that GRBs strongly prefer low metallicity environments
(e.g. \citealt{fynbo03, Stanek06, modjaz08_grbz, levesque10_grbhosts, Graham17}). Furthermore, it is clear that LGRBs are linked to SNe Ic-bl, and their progenitors must be massive, rapidly rotating stars that have lost their hydrogen and helium envelope, with the ability to create significant quantities of nickel during the explosion (e.g. \citealt{mazzali01, Maeda09, cano17_obs_guide}).

Several different mechanisms have been proposed to achieve the stripping of the stellar envelope, and from those mechanisms, single Wolf-Rayet (WR) stars and binary star systems emerge as GRB/SN Ic-bl progenitor channels (for a recent reviews, see e.g.,  \citealt{smartt15,levan16}). Binary star systems and WR channels can also be at play at the same time: for instance \cite{Cantiello07} suggest that LGRB progenitors can be made through  quasi-chemically homogeneous evolution in low metallicity environments once a WR star has been spun up in a massive close binary.

Since massive stars and binary star systems are likely progenitor channels for stripped CCSNe, such as SNe Ic-bl which are linked to GRBs, it is interesting to look for environmental markers that favor the evolution of such channels. \cite{wang11} show that variations in the stellar initial mass function (IMF), such as those proposed by \cite{Dav08} could lead to higher rates of close-binary systems or more massive stars in galaxies, both of which are associated with the occurrence of GRBs. If GRBs were products of dynamical processes in young dense star clusters as suggested by \cite{vandenHeuvel13}, GRBs would clearly prefer host galaxies with the highest star formation rates (SFR). Although multiple studies have shown that GRBs prefer low metallicity environments, it is still somewhat debated whether low metallicity causes GRB progenitors to form, or whether it is a side-effect of high sSFR, since there is a galaxy relationship in which high sSFR galaxies are also metal-poor (e.g. \citealt{Mannucci11}). Here we compare the sSFR of PTF Stripped SNe with those of SN-GRBs, something not done before, as well as with those of the general population of star-forming galaxies, to test whether the hosts of SNe Ic-bl without GRBs as well as those of PTF SNe Ic are as highly starforming as GRB hosts and to conclusively test whether low metallicity is the necessary ingredient for producing jets.



Until recently, GRBs and SN were typically detected in different ways. GRBs are found in all-sky surveys with gamma-ray satellites such as BATSE, HETE or Swift, and there is little or no host-galaxy bias associated with their detection. However, traditional SN searches such as LOSS \citep{Filippenko01} or CHASE \citep{Pignata09}, with their small fields-of-view, specifically target luminous galaxies that contain many stars, in order to increase their odds of finding those that explode as SN. Because more luminous galaxies are more metal-rich \citep{Tremonti04}, including SNe that were found by targeted surveys introduces a bias towards finding SNe in high-metallicity regions \citep{modjaz08_grbz, Young08, sanders12}, complicating the comparison of the environments of the two kinds of stellar explosions. 
While \cite{modjaz08_grbz}, for the first time, pointed out that the targeted surveys are biased towards galaxies that are massive and thus more metal rich, they could only include half of their SNe Ic-bl sample from untargeted surveys, given the limited number of surveys at that time. Thus, a large sample of SNe from an untargeted survey has to be examined to truly compare the environments of SNe with and without GRBs.

Here we fulfill this requirement by exploiting the largest single-survey, untargeted, spectroscopically classified, and homogeneous collection of SNe Ic and SNe Ic-bl discovered by the Palomar Transient Factory (PTF) before 2013, and by conducting an unparalleled and thorough study of their host galaxies. PTF alleviates the aforementioned galaxy-bias (and thus, implied metallicity-bias) since the 1.2m telescope at Palomar Observatory observes a much larger patch of the sky with its 7.8-square-degree field of view \citep{Law09, Rau09}. On top of omitting bias, our sample with directly measured metallicities provides a factor of $\sim 2$ increase in SNe Ic and SNe Ic-bl from untargeted surveys compared to previous studies \citep{modjaz08_grbz, Arcavi10, sanders12, Kelly12}, and an increase of 1.5$-$2 for the SN-GRB hosts compared to other SN-GRB host compilation studies (e.g. \citealt{modjaz08_grbz, levesque10_grbhosts}).
We study the environments of a large sample of members of the SN Ic family, namely SNe Ic and SNe Ic-bl, in order to discern systematic trends that characterize their stellar populations. In order to pinpoint the physical process that give rise to observed jets during the explosions of some massive stripped stars in form of GRBs, we compare them to host galaxies of SN-GRBs, i.e., SN Ic-bl with an associated GRB, in order to uncover any differences in the host galaxies and environments of SNe Ic-bl with and without observed GRBs. Thus, we compare those derived galaxy properties such as metallicity and SFR with those of published host galaxies of 10 SN-GRB with spectroscopic IDs within $z \lessapprox $0.3, whose emission lines values we take from the literature and analyze in the same way as our SNe Ic-bl hosts in order to ensure a self-consistent comparison. We also report the error estimates for the SN-GRB host galaxy parameters that we measure from the literature spectra, which previous studies did not report (see e.g., \citealt{levesque10_grbhosts}), 


\citet{levesque10_grbhosts2_mz} suggested that SN-GRBs prefer special kinds of galaxies that follow a different mass-metallicity relationship than a comparison sample of SDSS galaxies, namely ones that exhibit lower metallicities for the same galaxy masses. However, the available host data set of SN-GRB was relatively small in 2010 (only five objects) and in the intervening 8 years, five more SN-GRB have been discovered, doubling the initial sample. We include the most recent SN-GRBs in our study, and revisit the question of whether SN-GRB prefer a special population of galaxies. In addition, progress has been made in the field of measuring chemical abundances and thus we are including new metallicity scales in our work here. 

In Section 2, we introduce the PTF SNe in our sample. In section 3 we give details of the spectroscopic and photometric data of the PTF SN host galaxies. In section 4 we introduce a sample of GRBs and describe how we derived host galaxy properties from the literature for comparison with our PTF SN host galaxies. In section 5 we explain how the SN host galaxy properties were derived, including the metallicities for the entire sample, and $M_{*}\rm{s}$ and SFRs for PTF SNe Ic/Ic-bl host galaxies only. In sections 6 and 7 we present our results and discuss implications for jet production and GRB progenitor models. Section 8 describes potential caveats in our work, and section 9 concludes our results. Throughout the paper, we adopt a Hubble constant of $H_0 = 70~{\rm km~s^{-1} Mpc^{-1}}$. 








\section{Discovery and Classification of PTF SNe Ic and Ic-bl in our Sample}\label{sec:sample}

Our study is based on the SN harvest of the Palomar Transient Factory \citep[PTF,][]{Law09} survey between 2009 and 2012, 
which is a galaxy-untargeted and wide-field (7.8 sq. deg field of view) survey, discovering transients up to a limiting magnitude of 20.5 mag in the Mould {\it R}-band, 
independent of their host galaxies. While here we concentrate specifically on PTF SNe Ic and SNe Ic-bl, more details about 
the discovery and analysis of PTF SNe Ib/c are given in \citet{Corsi16}, \citet{fremling18}, Taddia, et al. (in prep) and Barbarino et al. (in prep). Independent of \citet{fremling18}, we performed our own spectral classification on the PTF SN spectra by running the state-of-the-art code SNID 
\citep{Blondin07}, with the augmented SNID library of Stripped SNe and Superluminous SNe Ic published 
by the SNYU group \citep{Liu14, modjaz16, Liu16,liu17}. Our spectroscopic IDs are given in Table \ref{tab:obs}. 
For a number of PTF SNe, we have updated their spectroscopic IDs, while they had been announced or published with a different ID 
\footnote{PTF~09q (Ic $->$ Ia), PTF~10bip (Ic $->$ uncertain: Ic/Ic-bl), PTF10vgv (Ic $->$ Ic-bl), see also \citet{modjaz16}}. 
 We note that while our IDs for the PTF SNe Ic-bl in our sample are fully consistent with those in \citet{Corsi16}, some of our IDs are inconsistent with those in \citealt{fremling18} (F18); in particular: PTF10tqv and PTF11qcj are included as SNe Ic-bl in our sample, while F18 include them as SNe Ic, and all the SNe in our "uncertain ID" group, except 10gvb, are included as SNe Ic in F18.


While our sample consists of 14 SNe Ic-bl and 28 SNe Ic, we also include the host galaxy data of five SNe with uncertain or peculiar SN types - those were cases where the classification was not clear either based on low S/N spectra or their uncertain physical nature. In particular, we have two SNe, PTF 10gvb and 12hni, whose spectra match with those of both SNe Ic-bl and SLSNe Ic, since the spectra of SLSNe Ic contain broad lines, similar to  those of SNe Ic-bl \citep{liu17,quimby18} - and indeed \citet{quimby18} independently classifies them as possible SLSN Ic based on his own code.
However, for our final analysis and statistical tests on the host galaxy data, we only include SNe Ic-bl and Ic with clean IDs. 

While the final PTF sample includes two additional SNe Ic with clean IDs for which we were not able to obtain host galaxy spectra 
(PTF~09dh and PTF~11mnb, for the latter see \citet{Taddia17} for its peculiar spectra and spectral ID),
we are confident that their omission will not affect our results, 
as all our conclusions are limited by the small number statistics of PTF SNe Ic-bl and the comparison sample of SN-GRBs, not those of PTF Ic.

\section{PTF SN Host Data} \label{sec:data}

In this section, we describe data of the PTF SN host galaxies. 
For spectroscopic data, we give details on how we conduct the observations, reduce and extract spectra, 
and measure emission line fluxes from the spectra, 
as well as on how we query to the SDSS archive for emission line fluxes in supplement.  
For photometry data, we explain how we assemble reliable UV-to-optical broadband photometry from SDSS, Pan-STARRS, and {\it GALEX}. 
We briefly discuss the redshift and luminosity distributions of the full sample at the end of this section, 
as well as point out unique aspects of our sample, as well as check that there are no obvious biases introduced when comparing the different SN host galaxies in our sample.

Host spectra of a total of seven PTF SNe Ic and SNe Ic-bl in our sample have been obtained and published independently by \citet{sanders12}. We compare our data and analysis with theirs in detail in the Appendix section~\ref{app:allgalscomparison} - overall, their metallicity values agree with ours (except for PTF11hyg) while we have better data than they do, possess the most updated spectroscopic IDs for the PTF SNe and analyze the data in more recent metallicity scales.

\subsection{Spectroscopic Data}

\subsubsection{Spectroscopic Observations and Reduction}

Optical spectra were obtained with a variety of telescopes: the 10m Keck I and Keck II telescopes, 
the 200-inch Hale Telescope of Palomar Observatory (P200), and the 8.1m Gemini-North and 
Gemini-South telescopes (GN-2011A-Q-93 and GS-2011A-C-5, PI: Modjaz). 
The spectrographs employed were the Low Resolution Imaging Spectrometer \citep[LRIS,][]{Oke95} at Keck I 
and DEIMOS \citep{Faber03} at Keck II, the Double Spectrograph \citep[DBSP,][]{Oke82} on the P200, 
and the GMOS-North and GMOS-South \citep{Hook03} at Gemini.

The majority of our host spectra were obtained long after the SNe themselves had faded, 
including the Keck and P200 runs after 2013, and Gemini runs during 2011. 
We expect such host spectra to have minimum contamination from the SN emission. 
Exposure times were chosen to yield a signal-to-noise ratio (S/N) of at least 15 in H$\alpha$ line, 
in order to robustly determine the line ratios for metallicity diagnostics.  
We re-observed if the desired S/N is not reached from an earlier run, but only the spectrum 
of the best quality for a given host is presented in this paper. 

During our major observing runs with Keck I, the spectra for nine hosts were obtained on Nov 21, 2014 and 
for four hosts on June 15, 2015. During these runs, we used 600/4000 grism on the blue side and 400/8500 grating on the red side, 
yielding a FWHM resolution of $\sim 4{\rm \AA}$ and $\sim 7{\rm \AA}$, on the blue and red sides respectively, 
for a slit width of 1$^{\prime\prime}$. 
The spectrograph and dichroic setup allows continuous wavelength coverage 
of at least 3500 -- 8000 $\rm\AA$. Given the redshifts of our targets, 
all the major nebular emission lines required to derive oxygen abundances, 
including [O{\sc ii}] $\lambda\lambda3726,29$, are within this wavelength range. 
We used HgNeAr arclamps for the wavelength calibration. 
Standard bias frames and lamp flats were obtained for each night. 
Multiple standard stars were taken throughout nights for the flux calibration.


\begin{deluxetable*}{llllllllllll}
\tabletypesize{\tiny}
\tablenum{1}
\tablewidth{0pt}
\tablecaption{Observation Information.}
\label{tab:obs}
\tablehead{\colhead{PTF} & 
\colhead{$z$} & 
\colhead{UT Date} &
\colhead{Tel.} &
\colhead{Instr.} &
\colhead{Wave. Range} &
\colhead{Res. (r/b)} &
\colhead{P.A.} &
\colhead{Airmass} &
\colhead{Seeing (r/b)} &
\colhead{Slit} &
\colhead{Exp. (r/b } \\
\colhead{name} &
\colhead{} &
\colhead{} &
\colhead{} &
\colhead{} &
\colhead{(\AA)} &
\colhead{(\AA)} &
\colhead{($^\circ$)} &
\colhead{} &
\colhead{($^{\prime\prime}$)} &
\colhead{($^{\prime\prime}$)} &
\colhead{if different) (s)} }
\startdata
	\multicolumn{12}{c}{SN Ic-bl}\\
\hline
   09sk &  0.035 &    2016-02-09 & KECK & LRIS &     3200-10000 &    6.6/4.6 &  360.00 & 1.03 &    1.0/1.1 & 1.0 &            900/980 \\
  10aavz &  0.062 &    2014-11-21 & KECK & LRIS &     3075-10290 &    9.3/5.9 &   34.40 & 1.56 &    1.9/1.7 & 1.5 &         1200x2+900 \\
   10bzf\tablenotemark{a}  &  0.049 &    2016-06-06 & KECK & LRIS &     3200-10000 &    6.4/4.3 &    4.01 & 1.35 &    1.3/1.3 & 1.0 &      600x2/620+640 \\
   10ciw &  0.115 &    2015-06-15 & KECK & LRIS &      3040-9500 &    6.2/4.7 &   48.00 & 1.07 &   0.93/1.0 & 1.0 &              900x3 \\
   10qts &  0.09 &    2015-06-15 & KECK & LRIS &      3100-9500 &    6.2/4.7 &   35.00 & 1.10 &   0.93/1.0 & 1.0 &             1200x3 \\
   10tqv &  0.079 &    2016-06-10 & KECK & LRIS &     3100-10000 &    6.2/7.8 &   59.00 & 1.38 &   0.94/1.0 & 1.0 &        1200x2/1200 \\
   10vgv &  0.015 &    2015-06-15 & KECK & LRIS &      3100-9500 &    6.2/4.7 &  236.00 & 1.21 &   0.93/1.0 & 1.0 &              900x2 \\
   10xem &  0.056 &    2014-11-21 & KECK & LRIS &     3070-10260 &    6.1/4.8 &  256.60 & 1.02 &   0.86/1.1 & 1.0 &              900x2 \\
   11cmh &  0.1055 &    2016-06-06 & KECK & LRIS &     3200-10000 &    6.4/4.3 &   73.70 & 1.72 &    1.3/1.3 & 1.0 &      1000x2/1040x2 \\
   11gcj &  0.148 &    2017-06-25 & KECK & LRIS &     3200-10000 &    6.3/4.6 &  80.00 & 2.10 &    0.84/0.89 & 1.0 &   900x2/900+1020 \\  
   11img &  0.158 &    2016-09-09 & KECK & LRIS &     3200-10000 &    6.1/4.5 &  211.00 & 1.39 &    1.1/1.2 & 1.0 &   1200x2/1200+1340 \\
   11lbm &  0.039 &    2014-11-21 & KECK & LRIS &     3070-10260 &    6.1/4.8 &   98.49 & 1.17 &   0.86/1.1 & 1.0 &               1200 \\
   11qcj &  0.028 &    2011-12-31 & KECK & LRIS &     3100-10000 &    6.4/8.8 &  245.00 & 1.36 &   1.0/0.94 & 1.0 &          420x2/900 \\
    12as &  0.033 &    2012-04-27 & KECK & LRIS &      3200-7350 &    5.3/6.5 &   30.00 & 1.07 &   1.0/0.96 & 0.7 &         540x2/1200 \\
\hline
	\multicolumn{12}{c}{SN Ic}\\
\hline
   09iqd &     0.034 &    2010-01-09 & KECK & LRIS &      3200-8750 &    4.6/3.7 &  141.00 & 1.11 &  0.81/0.95 & 1.0 &          150x2/420 \\
   10bhu &     0.036 &    2015-06-13 & PALO & DBSP &     3300-10000 &    5.3/3.8 &  153.00 & 1.09 &    1.3/1.6 & 1.0 &              900x3 \\
   10fmx &     0.045 &    2015-06-13 & PALO & DBSP &     3300-10000 &    5.3/3.8 &  227.00 & 1.07 &    1.3/1.6 & 1.0 &              900x3 \\
   10hfe &     0.049 &    2010-06-08 & KECK & LRIS &     3105-10000 &    6.5/8.2 &  127.00 & 1.71 &  0.80/0.93 & 1.0 &          260x2/600 \\
   10hie &     0.067 &    2015-06-13 & PALO & DBSP &     3300-10000 &    6.8/4.2 &  265.00 & 1.13 &    1.3/1.6 & 1.5 &              900x2 \\
   10lbo &     0.053 &    2016-02-09 & KECK & LRIS &     3200-10000 &    6.6/4.6 &  289.60 & 1.33 &    1.0/1.1 & 1.0 &            600/680 \\
   10ood &     0.059 &    2010-09-05 & KECK & LRIS &     3100-10000 &    5.2/5.2 &  226.00 & 1.13 &  0.68/0.71 & 0.7 &          150x2/420 \\
   10osn &     0.038 &    2014-09-30 & PALO & DBSP &      3380-8240 &    3.6/4.5 &  101.51 & 1.04 &    1.3/1.7 & 1.5 &              900x2 \\
   10qqd &     0.08 &    2015-06-13 & PALO & DBSP &     3300-10000 &    6.8/4.2 &  201.00 & 1.14 &    1.3/1.6 & 1.5 &                900 \\
   10tqi &     0.038 &    2014-11-21 & KECK & LRIS &     3070-10260 &    6.1/4.8 &  262.74 & 1.16 &   0.86/1.1 & 1.0 &               1200 \\
   10wal &     0.028 &    2014-11-21 & KECK & LRIS &     3070-10260 &    6.1/4.8 &  214.60 & 1.22 &   0.86/1.1 & 1.0 &  1200x2+300/1200x2 \\
   10xik &     0.071 &    2014-11-21 & KECK & LRIS &     3070-10260 &    6.1/4.8 &  231.10 & 1.28 &   0.86/1.1 & 1.0 &             1200x2 \\
   10yow &     0.024 &    2015-06-13 & PALO & DBSP &     3300-10000 &    6.8/4.2 &  321.01 & 1.16 &    1.3/1.6 & 1.5 &              900x3 \\
   10ysd\tablenotemark{b}&     0.096 &    2014-09-30 & PALO & DBSP &      3380-8240 &    3.6/4.5 &  351.90 & 1.04 &    1.3/1.7 & 1.5 &              900x2 \\
   10zcn &     0.02 &    2014-09-30 & PALO & DBSP &      3380-8240 &    3.6/4.5 &  177.41 & 1.01 &    1.3/1.7 & 1.5 &              900x2 \\
   11bov\tablenotemark{c} &     0.022 &    2012-01-18 & PALO & DBSP &     3470-10300 &    6.4/4.4 &  330.01 & 1.03 &    1.9/2.4 & 1.5 &           600/1200 \\
   11hyg\tablenotemark{d}&     0.03 &    2014-09-30 & PALO & DBSP &      3380-8240 &    3.6/4.5 &  123.61 & 1.12 &    1.3/1.7 & 1.5 &              900x2 \\
   11ixk &     0.021 &    2015-06-12 & PALO & DBSP &     3300-10000 &    5.2/3.6 &  206.01 & 1.06 &    1.1/1.5 & 1.0 &              900x2 \\
   11jgj &     0.04 &    2015-06-13 & PALO & DBSP &     3300-10000 &    6.8/4.2 &   74.60 & 1.17 &    1.3/1.6 & 1.5 &              900x2 \\
   11klg &     0.026 &    2014-11-21 & KECK & LRIS &     3070-10260 &    6.1/4.8 &  274.79 & 1.03 &   0.86/1.1 & 1.0 &               1200 \\
   11rka &     0.074 &    2015-06-15 & KECK & LRIS &      3100-9500 &    6.2/4.7 &   55.00 & 1.04 &   0.93/1.0 & 1.0 &                900 \\
   12cjy &     0.044 &    2015-06-12 & PALO & DBSP &     3300-10000 &    6.6/4.3 &  290.00 & 1.07 &    1.1/1.5 & 1.5 &              900x2 \\
   12dcp &     0.031 &    2012-05-17 & KECK & LRIS &      3210-7350 &    1.9/3.6 &  231.00 & 1.05 &  0.94/0.92 & 1.0 &                600 \\
   12dtf &     0.061 &    2014-09-30 & PALO & DBSP &      3380-8240 &    3.6/4.5 &   95.00 & 1.15 &    1.3/1.7 & 1.5 &              900x2 \\
   12fgw &     0.055 &    2015-06-12 & PALO & DBSP &     3300-10000 &    6.6/4.3 &  297.01 & 1.00 &    1.1/1.5 & 1.5 &              900x2 \\
   12jxd &     0.025 &    2014-11-21 & KECK & LRIS &     3075-10295 &    9.3/5.9 &  357.85 & 1.13 &    1.9/1.7 & 1.5 &               1200 \\
   12ktu &     0.031 &    2014-09-30 & PALO & DBSP &      3380-8240 &    3.6/4.5 &    7.90 & 1.38 &    1.3/1.7 & 1.5 &              900x2 \\
\hline
	\multicolumn{12}{c}{weird/uncertain SN subtype}\\
\hline
    09ps\tablenotemark{e} &     0.106 &    2015-06-13 & PALO & DBSP &     3300-10000 &    6.8/4.2 &  141.00 & 1.10 &    1.3/1.6 & 1.5 &              900x2 \\
   10bip\tablenotemark{e}  &    0.051 &    2011-06-30 & Gemini-S & GMOS-S & 4000-7200 & 7.7 & 155.6 & 1.57 &    0.7        & 1.0 &           1200 \\
   10gvb\tablenotemark{f} &    0.1 &    2015-04-23 & KECK & LRIS &     3102-10000 &    6.1/6.6 &  270.00 & 1.07 &    1.1/1.3 & 1.0 &      500+580/600x2 \\
   10svt\tablenotemark{g} &     0.031 &    2014-11-21 & KECK & LRIS &     3070-10235 &    9.3/5.9 &  168.11 & 1.62 &    1.9/1.7 & 1.5 &               1200 \\
   12hni\tablenotemark{f} &     0.107   &   2016-10-25 & KECK & DEIMOS & 4500-9600 &   5 &  116.0 & 1.31  &  0.6 & 1.2  &            600x2 
\tablenotetext{a}{aka SN~2010ah}
\tablenotetext{b}{aka SN~2011bm}
\tablenotetext{c}{aka SN2011ee}
\tablenotetext{d}{SN~2004aw-like, thus possibly a transition object between SN Ic and SN Ic-bl} 
\tablenotetext{e}{Ic/Ic-bl}
\tablenotetext{f}{SLSN/SN Ic-bl - \citet{quimby18} independently ID it as a possible SLSN Ic with Superfit, a different SN classification code.}
\tablenotetext{g}{Ib/c}
\enddata
\end{deluxetable*}

We placed the slit center at the SN site to catch the immediate environment of the explosion, 
which can be significantly different from the galaxy nucleus due to metallicity gradients. 
LRIS is equipped with an Atmospheric Dispersion Corrector, which allowed us to orient the slit at an angle different from parallactic angle without any loss from atmospheric differential refraction. We chose position angles that covered both the SN position and the nucleus of the host galaxy along the radial direction. If the SN site is far away from the nucleus with little local host galaxy emission, 
we can still characterize host properties via the brighter regions 
being closer to the nucleus that fall within our slit. 
In fact, this only happens for a few cases amongst the full sample 
and they are distinguished by different symbols on the plots in our analysis. 
A slit width of either 1.0 or 1.5$^{\prime\prime}$ was used, depending on the seeing condition. 

In addition, 
nine host spectra presented here were acquired at Keck during 
2015$-$2017 (one with DEIMOS and all others with LRIS), 
with similar instrumental setups as our major Keck runs in 2014 and 2015, so as to deliver a homogeneous dataset. 
Note that the wavelength coverage of the DEIMOS observation starts at 4500 $\rm\AA$ and thus the [O{\sc ii}] $\lambda\lambda3726,29$ 
lines are outside of coverage for the host galaxy of PTF~12hni. 

The P200 runs were conducted on the nights of 
Sep 30, 2014 
(for six hosts) and Jun 12-13, 2015 
(for 10 hosts), following similar conventions as the Keck runs 
(e.g., the SN sites were placed at the slit center). 
We took FeAr arclamp exposures on the blue side and HeNeAr on the red side. 
The brighter hosts were preferentially observed during the P200 runs. 

One host spectrum presented here was obtained on Gemini-South and is that of PTF~10bip, with the [O{\sc ii}] $\lambda\lambda3726,29$ lines 
lying outside of the wavelength coverage. 

In addition, we gathered data for hosts of five PTF SNe Ic and two PTF SNe Ic-bl that were observed in the year 2012 or earlier when the SN was still present (one of them with P200 and rest with Keck I $-$ the SNe Ic spectra have been published in \citealt{fremling18}). Thus, we can easily locate the SN sites by the bright continuum of SN light in the two-dimensional frames. While the SN spectra were superimposed on the spectra of their host galaxies, we were able to remove them during the analysis since SN spectra have much broader lines than the HII regions (see section \ref{sec:line}). 


Details about our spectroscopic observations for 
46 host galaxies of the PTF SNe in our sample (not including two, namely PTF12gzk and 12hvv, for which we downloaded SDSS spectra, see Section~\ref{sec:sdssspec}) are shown in Table \ref{tab:obs}, including both the ones from our recent runs 
and earlier ones based on data with SN spectra superimposed. 
Column 1 lists the name of the PTF SN, with the first two digits being the year of SN detection. 
Column 2 indicates the most up-to-date SN classification performed by us. 
Column 3 lists UT Date of the observations. The ones based on data with SN spectra superimposed have a UT Date in the year of 2012 or earlier. 
Column 4 and 5 list the telescope and instrument used, being Keck I (LRIS), Keck II (DEIMOS), Gemini-S (GMOS-S), or P200 (DBSP). 
Column 6 lists the full wavelength coverage, whereas we only consider a range of 3500 -- 8000 $\rm\AA$ in our analysis. 
Column 7 lists the spectral resolution on red side and blue side respectively. 
They were measured from the width of night sky lines or arclamp lines if only a few night sky lines exist on the blue-side spectra. 
Column 8 lists position angle of the slit in degrees, counter-clockwise from north, Column 9 airmass, 
Column 10 seeing, Column 11 slit width. In Column 12 we list exposure time, red and blue side separated 
by slash in case they are different. 

\subsubsection{Spectroscopic Reduction and Analysis}

We followed standard procedures in IRAF to prepare the Keck/LRIS data obtained from our major runs in 2014 and 2015 for further reduction, 
including bias subtraction, trimming, and flat fielding. 
For the LRIS data obtained before 2013, as well as in the year of 2015--2017 
following our major runs, the pre-processing 
were performed by the automated reduction pipeline in IDL, {\it LPipe}\footnote{http://www.astro.caltech.edu/~dperley/programs/lpipe.html}, 
which delivers similar results 
as by the standard procedures in IRAF. 
For the P200/DBSP data, we pre-processed with a PyRAF-based reduction pipeline, {\it pyraf-dbsp} \citep{Bellm16}. 

Subsequent cosmic ray removal, spectrum extraction, and wavelength calibration were performed in the same manner for all datasets. 
If at least three exposures were taken on the same target, 
{\it imcombine} task was performed in IRAF using median image combine to remove cosmic rays. 
Otherwise, we ran the IDL routine {\it P-Zap}\footnote{http://www.astro.caltech.edu/~dperley/programs/pzap.pro} for cosmic ray removal, which has been improved for better  
treatment around bright emission line regions and absorption features in standard star spectra. 
We adopted optimal spectrum extraction from the IRAF task, {\it apall}, which by nature applies higher weights to brighter regions in aperture. 
The metallicity measurement as a result can be considered as a luminosity weighted average over the aperture. 
The final steps of flux calibration, atmospheric band removal, and refinement of wavelength calibration against night 
sky lines were performed by our customized IDL routines \citep{Matheson01}.


\begin{figure*}[ht!]
\epsscale{0.8}
\makebox[\textwidth][c]{\includegraphics[width=.8\textwidth]{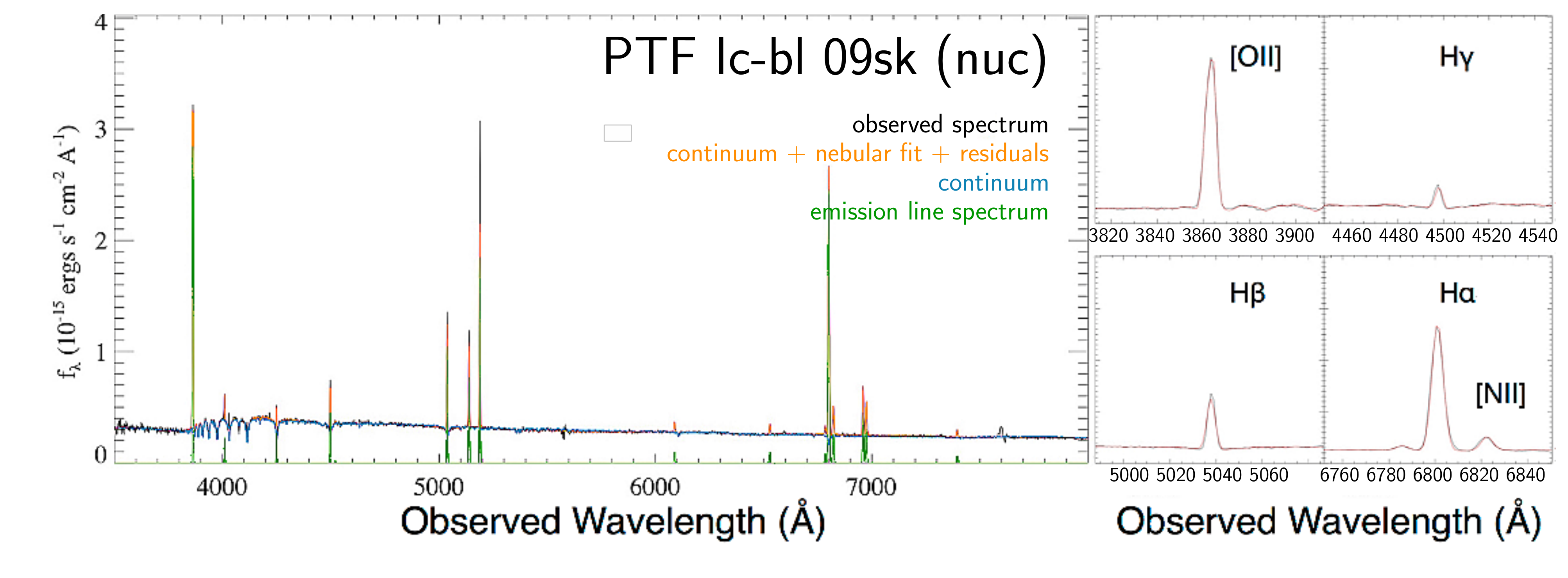}}%
\vspace{0.5cm}
\makebox[\textwidth][c]{\includegraphics[width=.8\textwidth]{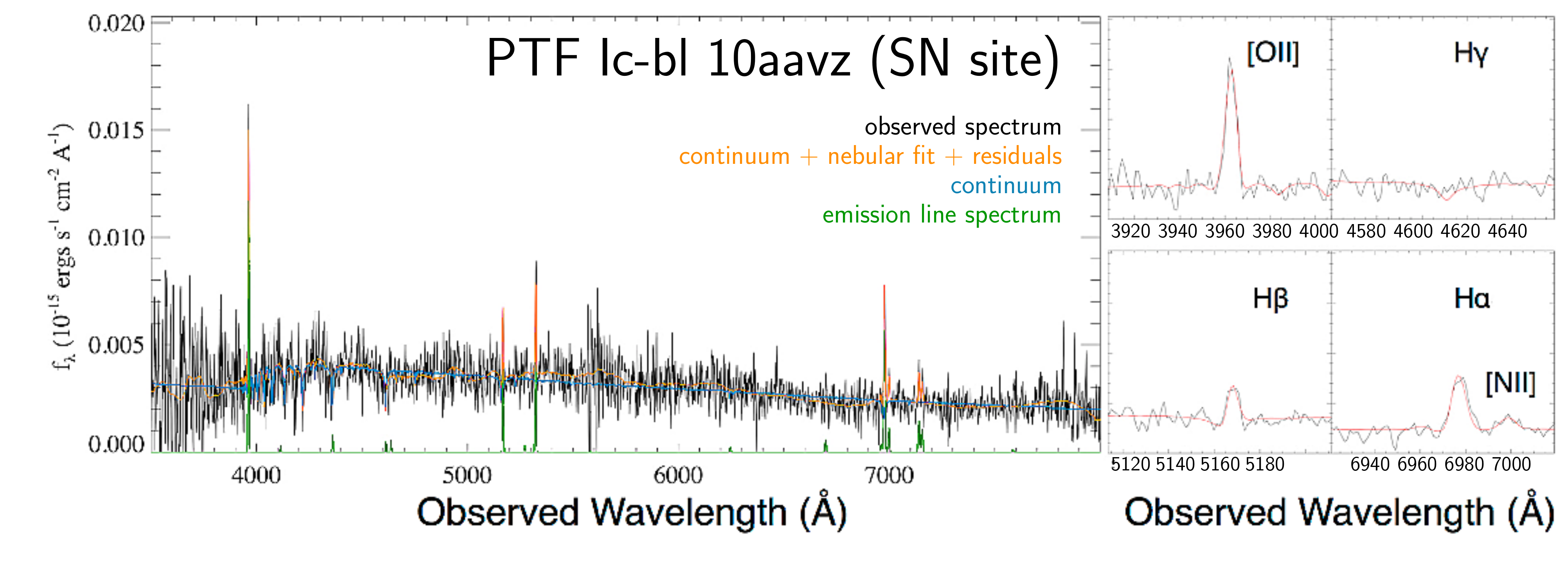}}%
    \qquad
\vspace*{-0.6cm} 
\caption{Two examples of the host spectra for SNe Ic-bl in our sample with spectral fits superimposed: the top panel shows a high S/N spectrum (PTF 09ks host, taken of the nucleus), while the lower panel shows an example of a low S/N spectrum (PTF 10aavz host, taken at the SN site) amongst the host spectra in our sample. The spectral fits, shown in different colors, are the outputs of {\it platefit}, the standard spectral fitting pipeline for SDSS \citep{Tremonti04, Brinchmann04}, decomposing the original spectrum into three components: a continuum (blue), an emission line spectrum (green), and the sum of continuum, nebular fit, and residuals shown in orange; see Section\ref{sec:line} for more details. On the right are zoom-ins onto a few of the important lines used for reddening correction and metallicity measurements. Note the importance of stellar absorption removal necessary for obtaining correct Balmer emission lines that will be used for the subsequent analysis. }
\label{fig:Example_Icbl}

\end{figure*}

In particular, we carefully select apertures for spectrum extraction to best probe the immediate environments of SNe. 
The spectra should also provide sufficient S/N in major nebular emission lines for metallicity calculation. 
As a result, we center the aperture at the peak of H$\alpha$ emission from the nearest star-forming region to the SN site in the slit. 
However, we caution that we use the term star-forming region here to infer a cluster of regions with significant H$\alpha$ emission. 
We cannot spatially resolve individual star-forming regions given the distances of our sources. 
Given that the progenitors of SN Ic/Ic-bl are likely to be massive stars with short lifetime, they generally haven't traveled 
far away from the star-forming regions since birth until explosion.  We confirmed by our data that star-forming regions with 
plenty of H$\alpha$ emission are usually found very close by to the SN site, with an offset usually comparable to the seeing. 

Observed long after explosion, most SNe themselves were no longer observable in our data. 
During out observations, we located the SN site by offsetting from a nearby bright star in the finding chart 
that was obtained when the SN was still present. 
In order to locate the SN site along the slit during spectrum extraction, we placed the SN site at slit center, 
as well as took standard star exposures at the slit center. 
The positional accuracy of the SN site in the slit is comparable to the seeing as determined in this way 
by matching to the position of standard stars. 
We therefore chose an aperture size for spectrum extraction to be twice of the seeing. 
Together with the fact that the aperture is centered at the nearest star-forming region which is usually very nearby, 
this ensures the SN site to fall within the aperture for most of our targets. 
We denote such apertures as 
the SN sites in subsequent analysis. 
In the remaining small number of cases (seven out of all 
46 host spectra that we extracted), 
SN sites have large offsets from the host nucleus or simply are in regions with little H$\alpha$ emission. 
 for these cases, the above described strategy of aperture placement leaves the SN site outside of the aperture. 
We denote such apertures as `H{\sc ii}' instead, or `nuc' if that peak of H$\alpha$ emission also happens to be at galaxy nucleus. 
In order to extract spectra with emission lines detected, we compromised to use a region further away from the SN site in such cases. 
They are less representative of the immediate environments of the SN explosions and are plotted with different symbols in the figures later on. 
However, they make up only a small fraction (7/48) 
of all the host spectra we extracted. 

It is also important to appropriately select sky background regions during spectrum extraction. 
For most projects aiming at the study of SNe per se as point sources, 
background regions closely bracketing the SN sites in the slit are chosen. 
However, the emission from host galaxy is extended so that such close background regions 
may still contain star forming regions from other parts within the same galaxy. 
Considering that the nebular emission line ratios vary throughout the galaxy, 
choosing close background regions would effectively alter the line ratios arising from the star forming region  
near the SN site from their intrinsic values. 
We instead choose to place the background regions far out enough to avoid 
the extended emission from host galaxies, such that they only contain sky, not galaxy light. 
Because the background regions are far away from the SN site in extended galaxies, 
the host spectra extracted at SN site can be very noisy at the wavelengths of night sky lines. 
Such artifacts are masked out during line flux measurements but 
lead to high uncertainty in line flux value if that nebular emission line 
coincides with a night sky line. 

\begin{figure*}[ht!]
\epsscale{0.8}
\makebox[\textwidth][c]{\includegraphics[width=.8\textwidth]{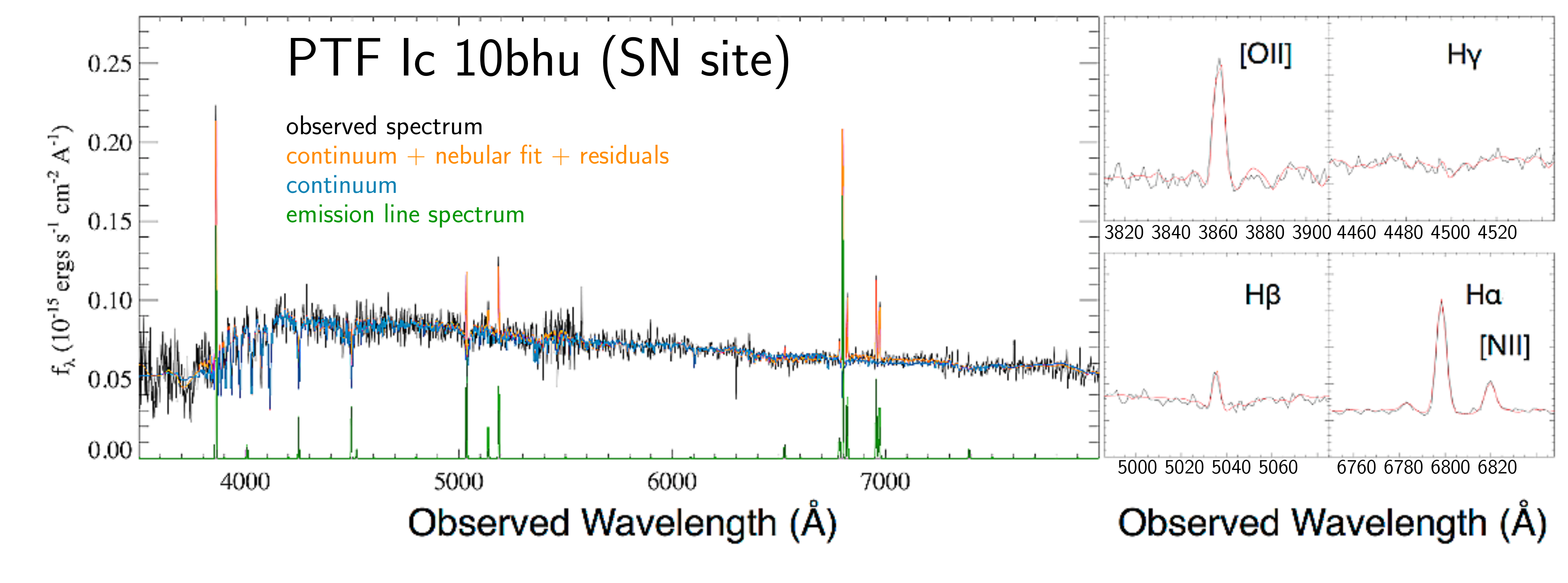}}%
\vspace{0.5cm}
\makebox[\textwidth][c]{\includegraphics[width=.8\textwidth]{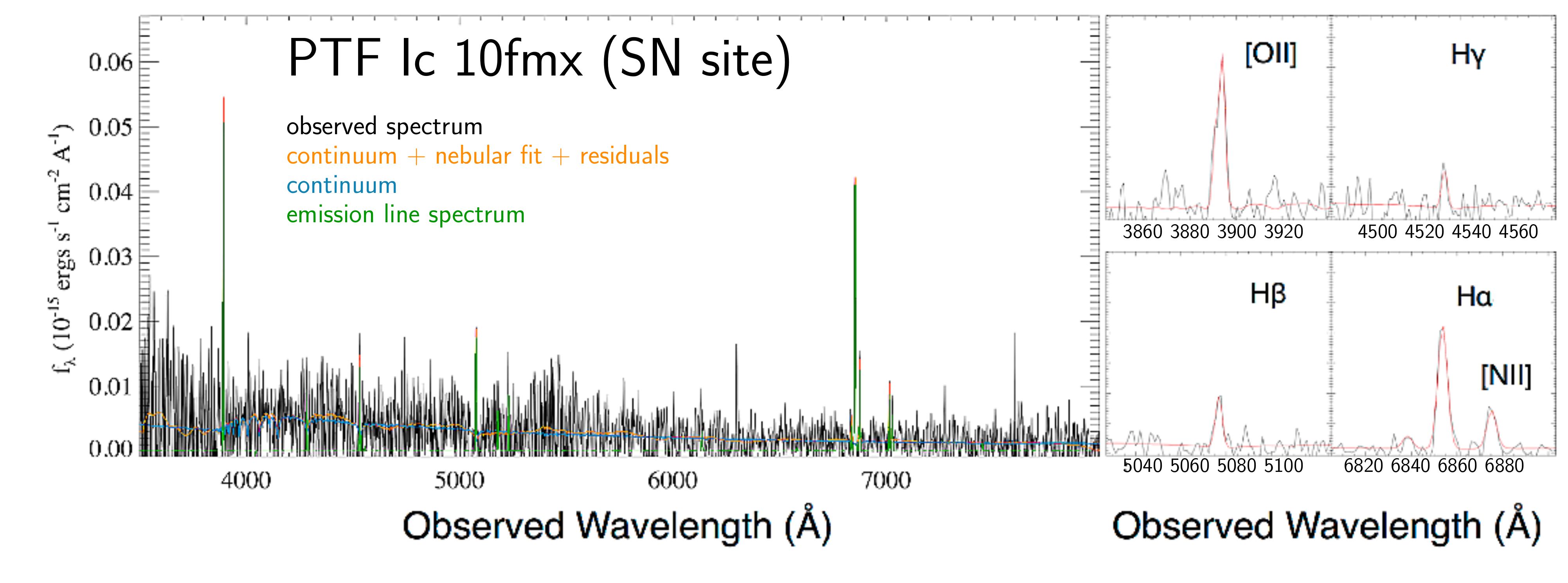}}%
    \qquad
\caption{Same as in Figure~\ref{fig:Example_Icbl}, but for two example hosts of PTF SNe Ic in our sample.} 
\label{fig:Example_Ic}
\end{figure*}

Our final host spectra are released on both the github page of our SNYU group\footnote{https://github.com/nyusngroup/PTFSNhosts/finalspectra} and on WISEREP\footnote{https://wiserep.weizmann.ac.il/} \citep{yaron12}. Figures~\ref{fig:Example_Icbl} and ~\ref{fig:Example_Ic} show examples of our spectra for SNe Ic-bl and SNe Ic hosts, respectively, that represent both high S/N and low S/N cases, with our spectral fits superimposed, which we describe in the next section.

\subsubsection{Line Flux Measurements}
\label{sec:line}

In this paper we focus on the analysis of the nebular emission lines of the host galaxy, 
since they encode physically important properties of the starforming regions at the SN sites. 
However, the presence of stellar absorption features can contaminate the emission components, especially the Balmer lines 
\citep[e.g.,][]{Tremonti04}. 
For both the Keck/LRIS and P200/DBSP data, 
in order to subtract stellar spectra and measure line fluxes from pure nebular emission spectra, 
we employed the SDSS standard spectral fitting pipeline, {\it platefit} \citep{Tremonti04, Brinchmann04}, 
which can be applied to non-SDSS data. 
Stellar absorption is usually non-negligible especially in H$\beta$ and H$\gamma$, even for the spectra 
extracted at the SN sites that are offset from galaxy nuclei, as shown in the example host spectra (Figures~\ref{fig:Example_Icbl} and~\ref{fig:Example_Ic}).

The {\it platefit} pipeline models the observed spectrum as a combination of three components: stellar continuum, 
nebular emission lines, and residual, as also shown in the example plots.
The continuum, including all stellar absorption features, is fit to the observed spectrum with the emission-line features masked, 
using stellar population synthesis templates from \citet{Bruzual03}. 
This approach ensures better constraints on the amounts of stellar absorption than on those derived by only fitting to the wings of absorption features 
for one Balmer line at a time. 
The redshift of emission lines is not tied to that of the continuum. It is determined by the fit of a sum of Gaussian functions 
to the emission line only spectrum, assuming the emission lines to have the same velocity offset. 
Visual inspection confirms that the final fit can very well reproduce the observed 
spectra, including the wings of the Balmer lines and the global slope of the continuum, 
with the residuals only being significant at the edges of spectra with unphysical trends, 
or if there are SN features superimposed on the host spectra (see below). 

The full spectra were separated into blue and red sides for both the Keck/LRIS and P200/DBSP data. 
We stitched the two spectra by applying a scaling factor on the blue side as to match the continuum level 
over the overlapping wavelength regions on both sides. 
Since transmission drops off rapidly at the edges of both spectra, small deviations in the wavelength calibration 
convert to large ones in the flux calibration. 
Unrealistic rising or falling trends at the edges make it hard to stitch by matching 
the observed continuum, whereas the {\it platefit}'s residual component characterizes such non-physical trends. 
To stitch the blue- and red-side spectra together,
we match the continuum fit as an output component from the {\it platefit}, free of the artificial rising or falling trends (included in the residual component). 
We therefore ran {\it platefit} twice: the first time to obtain the continuum fits for the two sides in order to then stitch them, 
and the second time on the stitched spectrum in order to obtain the spectrum with pure nebular emission lines that we use for line flux measurements. The line flux measurements that we present in Table\ref{tab:emission} are based on standard single Gaussian fits to the pure nebular emission lines spectrum and are corrected for Galactic reddening. 



To ensure that the automated pipeline, {\it platefit}, is working properly on our spectra, we compare the line flux measurements 
given by {\it platefit} with the values we measured by hand via the {\it splot} task in IRAF. 
The {\it platefit} derives uncertainties on line fluxes by 
the Levenberg-Marquardt least-squares fitting, 
and we follow \citet{Perez03} to derive statistical errors on the {\it splot} line fluxes. 
The uncertainty of scaling factor applied on the blue spectrum for stitching purpose is estimated by the 
standard deviation of ratios between the continuum from both sides over the whole overlapping wavelength range, 
being typically $\sim 10\%$ of the continuum level. 
We folded this into the variance spectrum to calculate flux uncertainties in both methods. 
We confirmed that the {\it splot} line fluxes generally agree with the {\it platefit} ones within the uncertainties, 
except for the Balmer lines for which only {\it platefit} properly corrects stellar absorption. 

For the P200 and Keck I data taken in the year 2012 or earlier, the imprints of SN spectra are superimposed on the spectra of host galaxies.  
We followed a similar approach to analyze these data, e.g., 
we run {\it platefit} to measure line fluxes and it eliminates the SN contribution at the same time. 
Being much broader than the nebular emission lines, the SN features are treated as residual by {\it platefit}. 
Various observational setups have been employed to produce these data, 
but the majority result in spectra with continuous wavelength coverage and spectral resolution comparable to our more recent data, 
being sufficient to resolve all the nebular emission lines of interest to us for metallicity derivation. 
As an exception, very different grating and grism setups were used to observe the host galaxy of PTF~12dcp 
that a $\sim 100~{\rm \AA}$ gap is present between the blue- and red-side spectra. 
Instead of matching the continuum level within the overlapping wavelength range, 
we stitch these spectra based on Balmer decrement. 
We predict the H$\alpha$ flux on the red side from the H$\gamma$ and H$\beta$ line fluxes on the blue side. 
To account for dust extinction, we assume case B recombination \citep{Osterbrock89} and the standard Galactic reddening law with 
$R_V = 3.1$ \citep{Cardelli89}. 
The scaling factor applied on the blue spectrum is thus determined by the ratio between the observed H$\alpha$ flux 
and this predicted H$\alpha$ flux. 

For the two spectra obtained at Gemini-South (PTF~10bip) and Keck II (PTF~12hni), we report the line fluxes 
measured by the {\it splot} task in IRAF. Both of them are star forming galaxies with little stellar absorption. 


We present line flux measurements for the full sample of 
48 hosts of PTF SNe Ic/Ic-bl in Table \ref{tab:emission} 
(including 
46 from new observations presented in this work and two from SDSS, see section \ref{sec:sdssspec}). 
Column 1 lists the name of the PTF SN. 
Column 2 indicates the type of aperture used for spectrum extraction. Values other than `SNsite' indicate 
that the SN site is outside of the aperture (see above). 
Column 3 lists emission line redshift measured within the aperture. 
Because the host redshifts reported by the PTF survey are sometimes extracted from the host nucleus, 
they can be slightly different from the redshifts presented here due to galaxy rotation. 
Column 4-11 list flux measurements for all eight of the nebular emission lines in need to derive oxygen abundances, 
including [O{\sc ii}] $\lambda\lambda3726,29$, H$\beta$ $\lambda4861$, [O{\sc iii}] $\lambda4959$, 
[O{\sc iii}] $\lambda5007$, H$\alpha$ $\lambda6563$, [N{\sc ii}] $\lambda6584$, [S{\sc ii}] $\lambda6717$, 
and [S{\sc ii}] $\lambda6731$ (in units of $10^{-15}{\rm~erg~s^{-1}~cm^{-2}}$, 
corrected for Galactic reddening and stellar absorption, 
but not internal extinction). 

\subsubsection{SDSS Spectra} 
\label{sec:sdssspec}

In supplement, we searched in the SDSS database spectra for the 
PTF SN Ic/Ic-bl host galaxies 
that are not observed by us (or with bad data quality) and found two of them, the hosts of 
PTF~12gzk (Ic-peculiar) and PTF~12hvv (Ic). 

For these two host galaxies, we make use of the emission line measurements from the MPA-JHU spectroscopic 
reanalysis of the SDSS DR8, which were also generated by {\it platefit}. 
We include these two hosts in addition to the other hosts observed by us 
in Table \ref{tab:emission}. 
The [O{\sc ii}] $\lambda\lambda3726,29$ lines are outside of the wavelength coverage for PTF~12gzk. 
However, PTF~12hvv has a redshift above 0.021 and thus 
its SDSS spectra cover all the major emission lines that we need for metallicity derivation, including the 
[O{\sc ii}] $\lambda\lambda3726,29$ lines on the blue end. 
Note that we queried SDSS for the combined [O{\sc ii}] $\lambda\lambda3726,29$ line fluxes from a free fit,
because the doublet is similarly un-resolved by most of our observations. 
For 
PTF~12hvv, the SN site is outside of the SDSS fiber area (3$^{\prime\prime}$ in diameter), 
so that we list their region type as `nuc' rather than `SNsite' in the table. 

\subsection{Photometry Data}
\label{sec:phot}

In order to estimate the $M_*$s and SFRs of the PTF SN host galaxies from SED fitting (see section \ref{sec:sed}), 
we retrieve their photometry in the optical bands for all of the PTF SN host galaxies: 
44 from SDSS and four from Pan-STARRS, 
and in the ultraviolet (UV) bands for 
36 of them from {\it GALEX}. 
In this section, we describe the sources of these photometry data which we argue provide reliable global magnitudes for SED fitting. 

\subsubsection{SDSS photometry}

All but four of the PTF SN host galaxies are covered by the SDSS imaging survey. We adopt their photometry in {\it u,g,r,i,z} bands 
from either the NASA-Sloan Atlas (NS-Atlas\footnote{www.nsatlas.org}) if available, or from the SDSS catalog otherwise. 
We ensure that these magnitudes are of good quality in both circumstances. 

Although 
about half of the PTF SN host galaxies are too faint to be included in the SDSS main spectroscopic galaxy sample, 
all of them are bright enough to be detected by the SDSS imaging survey. 
We retrieve model magnitudes from the SDSS catalog (DR8), except for the four outside of the SDSS footprint. 
The DR8 is chosen to be consistent with the other data products that we use, 
including those from the NS-Atlas and the MPA-JHU spectroscopic reanalysis. 
We note however that the four host galaxies outside of the SDSS DR8 footprint are still outside of the footprint in SDSS DR13. 
Robust colors are essential to constrain the shapes of the global SEDs. 
The model magnitudes are chosen because they usually provide the best available colors for extended sources like our low redshift galaxies, 
relative to the other magnitudes in the SDSS catalog, e.g., the Petrosian magnitudes. 
We correct for Galactic extinction using the far-IR map from \citet{Schlegel98} and standard Galactic reddening law with $R_V = 3.1$ \citep{Cardelli89}. 
We note that the conclusions are unchanged even if we use the new reddening map \citep{Schlafly11}. 

However, the SDSS photometry pipeline is optimized for small and high surface brightness objects, 
while it suffers from shredding for low redshift galaxies that are of large angular extent. 
The shredding happens during deblending, when the light from each ``parent'' object as an island of contiguous detected pixels is 
deblended into several different ``child'' objects. 
Deblending is necessary to eliminate light contamination from the foreground stars and background galaxies, 
and thus the magnitudes for ``parent'' objects as intermediate products are usually useless. 
However, when the deblending process is too aggressive, multiple star-forming sites that are resolved in the disks 
may be treated as separate child objects by the pipeline. 
In such cases, the magnitudes for the central child objects usually characterize the redder light from bulges, 
whereas they miss the bluer light emitted by the star forming regions at larger radii. 
If we use the SDSS catalog magnitudes from the central child objects to derive galaxy stellar properties, 
especially the sSFRs that are sensitive to color, we will systematically underestimate sSFRs when shredding happens. 
In the worst-case scenario, the shredding causes inconsistent deblending of light into child objects across bands, e.g., 
the fraction of light deblended into a certain child in $u$ band may be very different from that deblended into the same child in $r$ band.  
This sometimes gives rise to unrealistic colors for each child and thus completely fails the SED fitting. 

Visual inspection shows that the shredding of SDSS photometry pipeline is significant for 
eight PTF SN hosts. 
To alleviate the shredding problem, we utilize the NS-Atlas, 
as a reanalysis of all the galaxies bright enough to be included in the SDSS main galaxy sample (DR8) and with $z<0.05$. 
It creates image mosaics from SDSS and rephotometeres the {\it ugriz} bands with improved background subtraction. 
In particular, the NS-Atlas uses a better deblending technique 
\citep{Blanton11} compared to that of the standard SDSS pipeline, with the primary differences being: 
(a) the NS-Atlas uses constant templates across bands such that the subsequence colors are more robust; and 
(b) the NS-Atlas requires a much higher significance to deblend a child as a galaxy. 
These improvements make the NS-Atlas implementation more stable for the large galaxies. 
When available, the NS-Atlas reanalysis alleviates the shredding problem and avoids the failure in SED fitting that is caused by bad colors. 

For the 
eight hosts that are shredded by the SDSS pipeline, we search in the NS-Atlas and find 
seven of them (only PTF~10aavz is outside of NS-Atlas, see below). 
For these 
seven galaxies, we adopt their NS-Atlas Sersic fluxes, which are more robust compared to the NS-Atlas Petrosian fluxes. 
In addition, there are 
12 PTF SN hosts that are not shredded by the SDSS pipeline, but are within the NS-Atlas. 
We also adopt their NS-Atlas Sersic fluxes, even though their SDSS pipeline magnitudes are acceptable. For the ones outside of NS-Atlas, we use their model magnitudes from the SDSS pipeline. 
In fact, the ones outside of NS-Atlas are either faint or distant ($z>0.05$), usually with small angular extent and thus, are less likely to be shredded. 
The host galaxy of PTF~10aavz is the only one outside of the NS-Atlas that is shredded by the SDSS pipeline, even though it is a faint, distant and small galaxy. Usually, the parent objects of nearby bright galaxies with large angular extent include light contamination 
from foreground stars or background galaxies, so that the magnitudes for such parent objects are useless. 
Here however, the parent object of PTF~10aavz is deblended into only two children which both belong to the host galaxy itself. 
We know by inspection that the parent object does not include contamination, so that we 
easily recover its photometry by adopting the model magnitudes for the parent object of the host of PTF~10aavz derived by the SDSS standard pipeline. 

We also check that the NS-Atlas fluxes are generally consistent with the SDSS pipeline magnitudes that do not suffer from shredding. 
When compared to the SDSS colors derived from correct pipeline model magnitudes for a big sample of MPA-JHU galaxies, 
we confirm that the Sersic fluxes from NS-Atlas result in consistent colors. 
Therefore, we expect to obtain self consistent SEDs for the two subsets: 
25 of the PTF SN host galaxies with SDSS photometry from 
the pipeline model magnitudes, and 19 from the NS-Atlas Sersic fluxes. 

\subsubsection{Pan-STARRS photometry}

For the four PTF SN host galaxies outside of the SDSS footprint, 
we perform aperture photometry on Pan-STARRS images in the bands, {\it g,r,i,z,y}, using the python photometry package, {\it photutils}. 
The elliptical aperture is chosen by varying the semi-major axis value with a constant eccentricity 
so that a sufficient amount of the total galaxy flux is contained. For each fitting, we mask all the pixels of nearby sources that would 
contaminate the galaxy aperture. We calibrate the galaxy photometry using a star within each field and apply individual zero points to 
each aperture to correct the instrumental magnitude of the galaxy. Uncertainties in the measurements are derived by taking a standard 
deviation of the background measurements. 

\subsubsection{{\it GALEX} photometry}

The Far-Ultraviolet (FUV) luminosity is the most robust SFR indicator for individual galaxies with low total SFRs and low dust attenuation \citep{Lee11}, 
such as the PTF SN host galaxies. Thus, we downloaded images of the host galaxies in our sample from the Galaxy Evolution Explorer ({\it GALEX}) whose imaging mode surveyed the sky simultaneously in FUV 
(effective wavelength of 1516~$\rm \AA$) and NUV (2267 $\rm \AA$), with a field of view $\sim 1.2 ^\circ$ in diameter for each tile \citep{Morrissey07}. 
We draw the FUV and NUV magnitudes of the PTF SN host galaxies from {\it GALEX} GR6/7 data release. 

The PTF SN host galaxies are usually covered by several tiles generated from multiple satellite visits to the same area of sky, 
but not combined. 
Due to the failure of the FUV detector in 2009, many of these tiles have no FUV coverage. 
We consider only the tiles with FUV coverage, because the UV continuum at 
$\lambda < 2000 \rm \AA$ is used as an SFR indicator \citep[e.g.][]{Kennicutt98}. 
The {\it GALEX} mission includes several survey modes that differ in their exposure time per tile, ranging from 
All-sky Imaging Survey (AIS, $\sim$~100 sec to $\sim$~20.5~mag), Medium Imaging Survey (MIS, $\sim$~1,500 sec to $\sim$~23.5~mag) 
to Deep Imaging Survey (DIS, $\sim$~30,000 sec to $\sim$~25.0~mag). 
When adopting the {\it GALEX} magnitudes, we give preference to sources extracted from the tile with the longest exposure time. 
In a few cases, the FUV and NUV objects for the same host are not merged due to an astrometry error, but we match them by hand. 
For eight of the PTF SN host galaxies there were no {\it GALEX} images since the galaxies were outside of {\it GALEX} coverage.

Visual inspection shows that the {\it GALEX} pipeline magnitudes 
of only two PTF SN hosts (PTF~12dcp and PTF~11jgj) suffer from shredding,  
because of the poorer imaging resolution ($\sim 4.5^{\prime\prime}$) compared to that of the SDSS, as well as the lower UV source density. 
We exclude those  from our analysis. 
We further exclude {\it GALEX} magnitudes for 
two other PTF SN hosts from our analysis: PTF~10hie (contaminated by a UV bright star nearby) and 
PTF~11img (not detected in FUV from an MIS tile). 

In summary, {\it GALEX} magnitudes with good quality in both FUV and NUV bands are available for 
36 out of all 48 PTF SN host galaxies 
(23 from AIS and the rest from deeper surveys). 
We note that the lack of {\it GALEX} magnitudes only results in a poorer constraint on the SFR by SED fitting, 
not a lower limit on the SFR estimate (see section \ref{sec:sed}). 

\subsection{Sample Properties of the PTF SN Hosts}

Our final sample of PTF SN Ic and Ic-bl host galaxies consists of 
48 sources (14 SNe Ic-bl, 28 SNe Ic, and six weird/uncertain SN subtype transients). 
Thus, our sample of SN Ic-bl hosts is almost twice as large as the ones in the earlier studies of \citet{sanders12} and  \citet{Kelly12}
for untargeted Stripped SN hosts. 
The additional crucial difference is that their samples were taken from a heterogeneous set of SN surveys, 
while ours are all from the same single, and thus more homogeneous, untargetted survey. 

\begin{figure*}[ht!]
\epsscale{0.8}
\plotone{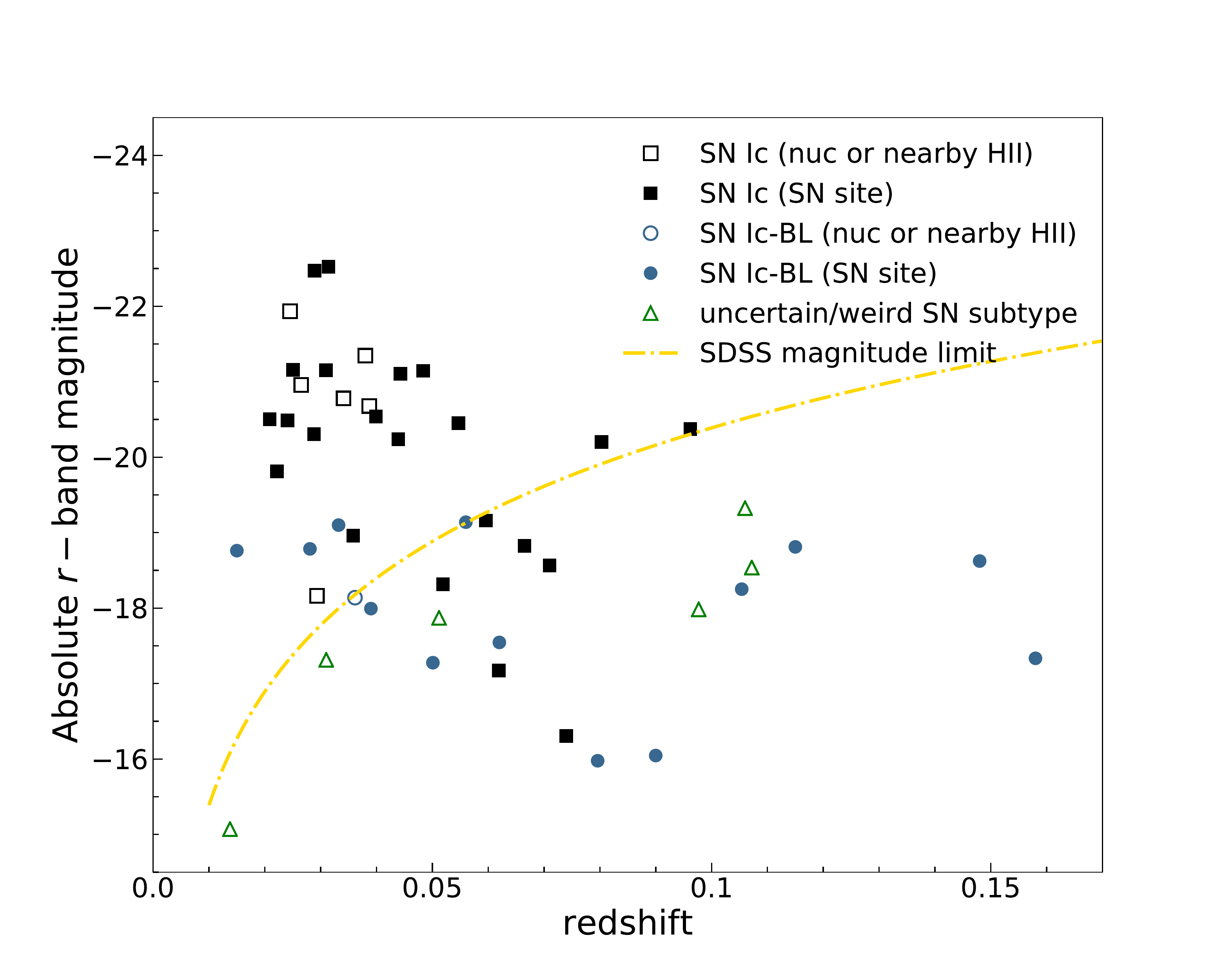}
\caption{Absolute {\it r-}band magnitudes as a function of redshift for the host galaxies of PTF SNe Ic (black squares), SNe Ic-bl (blue circles), and weird/uncertain SN subtype transients (green triangles). For the PTF SN Ic and Ic-bl hosts: if the SN site is within the aperture then closed symbols are used, otherwise open symbols (for both subsets). The {\it r-}band magnitudes are global values derived from Pan-STARRS and SDSS photometry and are corrected for shredding if applicable (see text for details). Redshifts are re-measured from the spectra in this work. The SDSS legacy galaxy redshift sample has an apparent {\it r-}band magnitude limit of 17.77~mag, which is denoted by the yellow dash-dotted curve. Sitting below the dashed curve, 2/3 of the SN Ic-bl hosts are too faint to be covered by the SDSS legacy galaxy redshift survey. 
}
\label{fig:LD}
\end{figure*}

As an untargeted survey, the PTF is not biased towards massive galaxies or nearby ones that are bright 
and extended in angular sizes.  
In Figure \ref{fig:LD}, we show the absolute {\it r}-band magnitude versus redshift for 
the full sample of 
48 PTF SN Ic/Ic-bl hosts (including six weird/uncertain SN subtype transients that are represented by green triangles). 
The {\it r}-band magnitudes are refined global values derived from the SDSS or Pan-STARRS images 
and redshifts are re-measured in this work. 
No uncertainty is plotted here, but it is usually smaller than the symbol size. 
The PTF SN Ic hosts have $z_{\rm median} = 0.038$ and $z_{\rm average} = 0.044$ with a standard deviation of 0.019.
The PTF SN Ic-bl hosts have $z_{\rm median} = 0.059$ and $z_{\rm average} = 0.073$ with a standard deviation of 0.043. 
Galaxy evolutionary effects on metallicity or star formation properties have no impact on the comparison between the two subsets 
within such a small redshift range, for which, e.g., the metallicity content of the universe does not change significantly \citep{Tremonti04}. 

Selected to be roughly twice the seeing, the aperture size used for spectrum extraction usually stays 
constant in angular size for all data from a given night. 
Since the sample of host galaxies spans a large range in luminosity and distances, 
this angular size corresponds to various physical scales. 
We check here if this will have a significant impact on the physical sizes that we are actually probing 
for the SN Ic hosts as a population in comparison with the Ic-bl hosts, 
considering the fact that the hosts of SN Ic-bl overall lie further away (see Figure \ref{fig:LD}). 
The aperture size varies from 1.2 to 5.9 kpc for the SN Ic hosts with a median of 2.1 kpc, 
and varies from 0.7 to 7.1 kpc for the Ic-bl hosts with a median of 3.0 kpc. 
The physical scales of the environments that we probe are 
on average more extended for the hosts of SNe Ic-bl than those for the SNe Ic. 
However, both the K--S test and Anderson Darling test show that this difference 
is not statistically significant, assuming a significance level, $\alpha=0.05$ (see section \ref{sec:bpt}). 
Because the hosts of SNe Ic-bl are less luminous intrinsically and are at higher redshifts, 
they are fainter and thus, they were all observed with Keck. The Keck runs have on average better seeing than the P200 ones, 
giving rise to the similar aperture sizes in physical scales for the SN Ic-bl hosts 
as for the nearby, more luminous SN Ic hosts observed predominantly with P200. 

The SDSS main spectroscopic galaxy sample has an apparent {\it r}-band magnitude limit of 17.77~mag \citep{Strauss02}, 
which is denoted by the dashed curve for different redshifts in Figure \ref{fig:LD}. 
Sitting below the dashed curve, $\sim$2/3 of the SN Ic-bl hosts are too faint to be covered 
by the SDSS legacy galaxy redshift survey. 
For the rest of our sample, presumably the hosts are bright enough so that SDSS spectra exist. 
However, especially the SN Ic hosts occupying the upper left part of this diagram 
are bright, nearby, and thus appear large on the sky. 
For these galaxies, the SDSS fiber, which is 3$^{\prime\prime}$ in diameter, is generally centered on the galaxy nucleus 
and thus, misses the SN site. 
Most isolated late-type spiral galaxies display strong metallicity gradients, being more metal-rich at the center. 
The spectra we present here cover the SN sites and thus are crucial to probe the immediate environments of SN explosions. 

\section{Comparison Samples}
\label{sec:comparison}
In this section, we define three control samples: (1) SN-GRB hosts, (2) local galaxies 
from the SDSS and (3) the Local Volume Legacy Survey (LVL). 
We also describe how we compile relevant observables and derived galaxy properties from the literature, 
which include the emission line fluxes for the SN-GRB hosts and the SDSS galaxies, metallicities for the LVL galaxies, 
and global $M_*$s and SFRs for all the samples. 
At the end of this section, we summarize the redshift distributions of the control samples, 
relative to that of the PTF SN hosts. 
We show that their very different redshift ranges have little impact on the intercomparison between the samples, 
specifically for their star formation and metallicity properties. 

\subsection{Hosts of SN-GRBs}
\label{sec:sngrbhosts}

Here we construct a sample of host galaxies of SN-GRBs, i.e., SN Ic-bl with an associated GRB, 
in order to uncover any differences in the host galaxies and environments of SNe Ic-bl with and without observed GRBs, 
and thus to pinpoint the physical process that gives rise to observed jets during the explosions of some massive stripped stars. 
If our PTF SNe Ic-bl are associated with off-axis GRBs, we expect to see no significant difference of host environments 
between the PTF SNe Ic-bl and SN-GRBs since viewing angle effects are a random process.
If the PTF SNe Ic-bl in our sample have intrinsically no GRB associated with them, 
then there needs to be some unique property that forces GRBs to occur in some massive stripped stars 
in contrast to the usual SN Ic-bl without GRB - and our environmental study could reveal what that unique property is. 
To distinguish between these two scenarios, we compare the host galaxies of 
SN-GRBs with the PTF SNe Ic-bl without GRBs.  In section \ref{sec:discussion} we discuss whether the PTF SNe Ic-bl harbored off-axis GRBs.

For our comparison sample of SN-GRBs, we are including spectroscopically ID'ed SN-GRBs at z$<$0.3 (in order to mitigate any significant cosmic evolution for a fair comparison with the PTF SN sample) with published host galaxy data before August 2018.
There are 10 such SN-GRB hosts in the literature, which we list in Table \ref{tab:GRBspec}, along with their nebular emission line fluxes that we adopt in order 
to compute their line ratios and metallicities in the same calibrations as we do for the PTF SN hosts (see section \ref{sec:meta}). 
If multiple sets of flux measurements exist in the literature, we adopt the ones providing flux uncertainties 
since the code we are using requires flux uncertainties for computing metallicity uncertainties (see section \ref{sec:meta}).
If multiple spectra are extracted from the same host at different sites, we include only the one from the SN site. 
We list notes for individual objects in the Appendix, section~\ref{app:grbsn} and compare the metallicities that we compute to those reported in the literature, based on the same data, in the Appendix, section~\ref{app:allgalscomparison}. In section~\ref{sec:caveats}, we discuss the caveats that arise from our criteria and future work that can address them.
While there are GRBs within this redshift volume with no observed SNe (e.g., \citet{dado18}), with the most famous being GRB 060614 with very deep SN limits (e.g., \citealt{galyam06}), their classification as bona fide long-duration bursts is debated, as they may be short-duration GRBs with extended tails \citep{ofek07_grb060505,caito09,perley09} or altogether another type of GRB \citep{gehrels06,lu08}. Indeed their host properties are very different from those of confirmed GRB-SNe, and closer to those of short GRBs \citep{levesque07,stanway15}.


We further compiled the $M_*$ and SFR values for these SN-GRBs from the database of GRB Host Studies 
(GHostS\footnote{www.grbhosts.org}), converted to be consistent with the stellar initial mass function (IMF) that we adopt (see section \ref{sec:sed}). 
Except for the host of GRB161219B/SN2016jca \citep{cano17_obs_guide}, the $M_*$s and SFRs are available for all the other 10 SN-GRB hosts. 
The $M_*$s were derived from spectral energy distribution (SED) fitting to the ``pseudophotometry'', 
as a homogeneous photometry reconstructed by sampling the observed SEDs from the literature in a reduced set of filters \citep{Savaglio09}. 
In particular, the SED is modeled as a young burst component superimposed on a population of older stars, which is more general 
than assuming a single simple star formation history. It yields more realistic $M_*$ estimates for the low mass galaxies 
with bursty star formation behavior \citep{Huang12a}. 
We applied a similar approach to obtain $M_*$ estimates for the host galaxies of PTF SNe (see section \ref{sec:sed}), 
but we used different stellar population synthesis codes and different grids of dust extinction, stellar metallicity, etc., to generate the SED models. 
One of the SN-GRB hosts, GRB/XRF060218/SN2006aj, is detected in SDSS. 
We confirmed that our SED fitting process making use of the SDSS photometry yields a $M_*$ 
estimate consistent with the GHostS value for this source within 2$\sigma$, with our value being slightly lower. 
Thus, we expect no significant systematic differences in the $M_*$ estimates between our PTF SN hosts and the GRB-SN hosts. 

In order to estimate the SFRs, the GHostS prefers the H$\alpha$ luminosity over UV luminosity 
or over the luminosity of other emission lines (e.g., H$\beta$ or [O{\sc ii}]). 
All of the SN-GRB hosts have H$\alpha$ detected and thus their SFRs from the GHostS are based on H$\alpha$ luminosity, 
corrected for aperture-slit loss and dust extinction. 
While the computed metalllcities for both SN-GRBs and our PTF SN host sample 
reflect local values in cases where the SN site is distinct from the galaxy nucleus, 
the $M_*$s and SFRs reflect global values for both the GHostS and our sample (see section \ref{sec:sed}). 


\subsection{Galaxy Comparison Samples}

In order to understand whether the SNe Ic/Ic-bl or SN-GRBs preferentially occur in certain types of galaxies over others, 
we select two control samples of low redshift galaxies to set the baseline of comparison. 
The first sample is selected from the SDSS, which is frequently invoked in previous works of SN Ic-bl and GRB hosts 
to represent the overall population of star forming galaxies \citep[e.g.,][]{modjaz08_grbz, levesque10_grbhosts2_mz}. 
However, the SDSS legacy galaxy redshift survey is highly incomplete below $M_* \sim 10^{8.5} M_\odot$ and has a fiber bias. 
Following \citet{Perley16b}, the second sample is selected from the LVL survey, which provides a volume complete sample within 11 Mpc. 
However, limited to such a small volume, the LVL survey suffers from cosmic variance and is 
biased against the most massive galaxies that are rare in the local universe. 
Most importantly, there are no homogenous metallicity measurements for the LVL galaxies. 

An ideal control sample should have representative 
distributions in all physical parameters of interest to us, e.g., metallicity, $M_*$, and SFR. 
In particular, Figure \ref{fig:LD} shows that the hosts of SN Ic-bl are generally of low luminosity. 
Inspection of their SDSS images demonstrates that many of them are dwarf irregulars barely resolved by the survey. 
It is therefore important to employ a control sample that is inclusive of low mass galaxies 
and has star formation behavior typical of that in the local universe. 
Thus, the two samples that we define here are not ideal, but are complementary for our purpose of comparison. 

\subsubsection{SDSS galaxies}
\label{sec:sdss}

We select the SDSS galaxies from DR8, which is provided by the MPA-JHU catalog of galaxies, 
including all the main physical parameters of interest to us: $M_*$s, SFRs, and emission line fluxes for metallicity calculation.

In order to self-consistently compare to our PTF SN Ic/Ic-bl hosts, 
we retrieve redshifts from the `galSpecInfo' table, line fluxes from the `galSpecLine' table, and from the `galSpecExtra' table: 
$M_*$s, SFRs, and metallicities \citep[calculated by][]{Tremonti04}.  
The three tables contain the MPA-JHU reanalysis of all the SDSS spectra that are classified as galaxies. 
Note that for the PTF SN Ic/Ic-bl host galaxies, we follow the same approach as adopted by the MPA-JHU group to 
obtain line flux measurements (section \ref{sec:line}) and $M_*$ estimates (section \ref{sec:sed}). 
Most importantly, we derive metallicities in four calibrations from the line fluxes for the SDSS galaxies, 
following what we do for the PTF SN Ic/Ic-bl host galaxies (section \ref{sec:meta}). 
Some of the calibrations are very recent and thus are not considered by the previous works that 
compare the SN Ic-bl and GRB hosts with the SDSS galaxies. 
For the SFR estimates, the MPA-JHU values that we adopt have been improved 
from their original values as presented in \citet{Brinchmann04}, 
so that the out-of-fiber SFRs are derived from SED fitting to {\it ugriz} photometry. 
For the population of star-forming galaxies in particular, 
these MPA-JHU SFRs are consistent with the SFRs that are derived from the SED fitting involving the UV bands \citep{Salim16}, 
similar to what we do for the PTF SN Ic/Ic-bl hosts (section \ref{sec:sed}). 
Therefore, these $M_*$s, metallicities, and SFRs for the SDSS galaxies are consistent with those for the PTF SN hosts. 

In particular, the SDSS sample serves the purpose of assessing if the PTF SN Ic/Ic-bl and SN-GRB hosts follow the same mass--metallicity ($M-Z$) 
relation as defined by the local galaxies (section \ref{sec:mz}). Therefore, to select a parent sample of the SDSS galaxies, 
we adopt the following criteria, which are similar to \citet{Kewley08}, who derived the $M-Z$ relation for the overall SDSS galaxies: 

\begin{itemize}
\item For reliable metallicity estimates, an S/N$>8$ in the strong emission lines ([O{\sc ii}] $\lambda\lambda3726,29$, H$\beta$, [O{\sc iii}] $\lambda5007$, H$\alpha$, 
[N{\sc ii}] $\lambda6584$, [S{\sc ii}] $\lambda6717$, and [S{\sc ii}] $\lambda6731$), following \citet{Kewley08}, 
and an S/N$>3$ in  [O{\sc iii}] $\lambda4959$. 

\item A lower redshift limit $z > 0.04$, which corresponds to a flux covering fractions of $> 20\%$ on average for normal star forming galaxies 
observed through the 3$^{\prime\prime}$ SDSS fibers \citep{Kewley08}. Such a covering fraction is required for metallicities to begin 
to approximate global values \citep{Kewley05}. Plus, the [O{\sc ii}] $\lambda\lambda3726,29$ lines are not measured at $z < 0.03$. 

\item The $M_*$ derived from inside the fiber being $> 20\%$ of that derived from the global photometry, to eliminate the large 
luminous galaxies that require a higher redshift to satisfy the covering fraction requirement. 

\item An upper redshift limit $z < 0.1$, above which the SDSS star-forming sample becomes highly incomplete \citep{Kewley06_sdss}. 

\item A bptclass of star-forming from the `galSpecExtra' table, so that the non-starforming galaxies and 
the galaxies containing active galactic nuclei (AGNs) are removed. 



\item The $M_*$ and metallicity estimates \citep[calculated by][]{Tremonti04} are both available from the MPA-JHU catalog. 

\end{itemize}

These selection criteria result in a parent sample of 40879 SDSS galaxies, with a median redshift $z \sim 0.067$. 
To select a subset that is reasonable for metallicity calculation by our code, {\it pyMCZ} (section \ref{sec:meta}), 
we randomly sample 500 galaxies from the parent sample. We ensure that the differences are not statistically significant 
between the distributions of metallicities \citep[calculated by][]{Tremonti04}, $M_*$s, and SFRs for this subset of 500 galaxies 
and those for the parent sample, i.e., that the subset is representative of the general SDSS population. 
In support of this, we can reproduce the \citet{Kewley08} $M-Z$ relation with this subset, based on the \citet{Tremonti04} metallicities.

The SDSS legacy galaxy redshift survey is highly incomplete below $M_* \sim 10^{8.5} M_\odot$. 
While one could correct for incompleteness by applying a volume weight \citep[e.g.,][]{Huang12b}, 
it is not sufficient for our purpose here, because we have applied further selection criteria on emission line fluxes, etc. 
The volume weight only corrects for the fact that the faint galaxies below the flux limit are missed, 
but not that the additional galaxies are missed due to further selection criteria.

We also note that more recent IFU-based M-Z calibrations have been published, including based on the CALIFA survey \citep{sanchez18} that mitigate the fiber bias of SDSS - however since they do not extend to the low galaxy masses in which our PTF SNe Ic-bl and the SN-GRBs are found and are not calculated in majority of the metallicity scales in which we present our work, we do not include them here, but mention them for completeness.

\subsubsection{LVL galaxies}

In contrast to the SDSS, the LVL survey provides a sample of galaxies that is complete in volume 
within 11 Mpc \citep{Dale09}, dominated by dwarfs. 
Thus, the LVL galaxies form a better control sample that the SDSS galaxies \citep{Perley16b}, 
for the comparison with the PTF SN host galaxies, especially with the hosts of SNe Ic-bl that have overall low $M_*$s.  
Furthermore, the LVL catalog includes photometry from UV to NIR, yielding $M_*$ estimates from SED fitting. 
We adopt UV-based SFRs for the LVL sample because they are known to be more reliable 
than the H$\alpha$-based ones due to stochastic star formation in the regime of low mass galaxies \citep{Lee09}. 

However, the LVL Survey is only limited to the local volume and thus suffers from cosmic variance. 
Plus, the most massive galaxies which are rare in the universe are under-represented in the local volume. 
Most importantly, the LVL is a photometric survey without spectroscopy and thus, there are no metallicity measurements as part 
of the initial survey. Metallicity measurements exist in the literature for about 2/3 of the galaxies 
in that sample and we use the compilation in \citet{Perley16b} (see also other references therein). 
We caution that these metallicities for the LVL galaxies are literature values in a variety of calibrations 
(typically electron temperature $T_e$-based at low metallicities and various strong-line diagnostics at higher metallicities), 
whereas for the hosts of PTF SNe and  SN-GRBs, we re-calculated the metallicities in various calibrations  (see section \ref{sec:meta}). 
The line fluxes for the LVL galaxies were not published, and thus we cannot run them through {\it pyMCZ}, as we do for the SN-GRB hosts. 
Moreover, these literature values were derived from spectroscopic observations that 
covered various fractions of the entire galaxies. As a result, the average trends as defined by the LVL sample involving metallicities 
serve the purpose of only a qualitative comparison. 

\subsection{Redshift Distributions of the Comparison Samples}

For the 10 SN-GRB hosts, their redshift distribution has a $z_{\rm median} = 0.146$, $z_{\rm average} = 0.154$, 
with a standard deviation of  0.104. 
Among them, GRB130427A/SN2013cq has the highest redshift ($z = 0.3399$). 
On the contrary, all the hosts of PTF SNe Ic/Ic-bl are at 
$z < 0.158$, with a $z_{\rm median} = 0.059$, $z_{\rm average} = 0.072$, 
with a standard deviation of  0.044. 
For the 500 SDSS galaxies, $z_{\rm median} = 0.066$, $z_{\rm average} = 0.067$. 
Furthermore, all the LVL galaxies lie within 11 Mpc and thus have much lower redshifts than the PTF SN or SN-GRB hosts do. 
We fully acknowledge that the redshift distributions of the four samples are vastly different. 

However, to be able to compare these samples self-consistently, we still expect the evolutionary effect in metallicity or star-formation properties to be negligible. 
For example, \citet{Whitaker12} studied the redshift evolution of the SFR--mass sequence of star forming galaxies, 
and the lowest redshift bin is defined to be $0.0<z<0.5$ in that work, i.e., a redshift range than encompasses those of all of our samples.
\citet{Lamareille09} defined the same lowest redshift bin in their study of the evolution of $M-Z$ relation. 
The evolution in the metallicity of star-forming galaxies between $z\simeq0.08$ and 0.29 is less than 0.1~dex at $M_*\sim 10^9{M_\odot}$
\citep{Zahid14}. 
In conclusion, the evolution of the average trends in SFR--mass sequence or $M-Z$ relation within $z\lesssim0.3$ 
is negligible relative to the observed scatter in these correlations. 
Although the SN-GRB hosts have overall higher redshifts than the PTF SN Ic/Ic-bl hosts and LVL members do, 
all these galaxies are usually considered to be local in the observational studies of redshift evolution of galaxy properties. 

\section{Derived Host Galaxy Properties} \label{sec:prop}

In this section, we explain how we derive galaxy properties from observables, 
including the metallicities from emission line fluxes for the SN-GRB, PTF SN Ic/Ic-bl hosts and the SDSS galaxies, 
as well as the $M_*$s and SFRs from broadband photometry for the hosts of PTF SNe Ic/Ic-bl only. 

\subsection{Metallicity}
\label{sec:meta}

Theoretically, the metallicity is expected to influence the outcome of the deaths of massive stars as SNe or as GRBs \citep{Heger03}. 
For example, many GRB models favor rapidly rotating massive stars at low metallicities as likely progenitors. 
Massive stars at low metallicities are likely to avoid losing angular momentum from mass loss, 
if the mass loss rate is set by line-driven, and thus, metallicity-driven winds. 
Observationally, the metallicities of SN Ic/Ic-bl or SN-GRB progenitors can be traced 
by the oxygen abundance of the H{\sc ii} regions at the SN sites, 
given that the massive progenitors have short lifetimes and thus should not move far from their birth H{\sc ii} regions 
\citep[see the review by][for detailed discussion and caveats]{modjaz11-rev}. 
Our observations of the metallicities of SN host galaxies, measured in many cases at the SN sites, 
will therefore help to test such SN and GRB progenitor models. 

\subsubsection{Metallicity Estimation}

One of the main purposes of this work is to understand the metallicity dependence 
in the production of different subtypes of the large family of SNe Ic: SNe Ic, SNe Ic-bl, and SN-GRBs. 
The nebular oxygen abundance is the canonical choice of metallicity indicator for the studies of interstellar medium (ISM) 
and we use these two terms, metallicity and oxygen abundance, interchangeably in this work. 
We re-calculate the metallicities of the SN-GRB host galaxies and the subset of 500 SDSS galaxies from literature emission line fluxes by following the same approach we applied to the PTF SN hosts in order to make a self-consistent comparison. 

Two main methods of metallicity estimates can be applied to extragalactic studies: 
the $T_e$-based method and the strong-line method 
\citep[see a summary in][and the references therein]{Bianco16}. 
The $T_e$-based method requires 
auroral lines such as [O{\sc iii}] $\lambda4363$. 
However, the auroral lines are generally very weak, and saturate above solar metallicity \citep{Stasinska02}. 
The strong-line methods estimate metallicities from ratios of strong nebular emission lines emitted from H{\sc ii} regions. 

Commonly used line ratios include:
\begin{eqnarray*}
\left.\begin{aligned}
N2O2 & = {\rm \frac{[NII] \lambda6584}{[OII] \lambda\lambda3726,29}}, \\
R_{23} & = {\rm \frac{[OII] \lambda\lambda3726,29 + [OIII] \lambda4959 + [OIII] \lambda5007}{H\beta}}, \\
O3N2 & = {\rm \frac{[OIII] \lambda4959 + [OIII] \lambda5007}{H\beta} \times \frac{H\alpha}{[NII]\lambda6584}}, \\
N2 & = {\rm \frac{[NII] \lambda6584}{H\alpha}}.
\end{aligned}\right.
\end{eqnarray*}
Depending on the calibration of the observed emission line ratios, 
the strong-line methods can be further categorized into three types: empirical methods that are calibrated on observed $T_e$-based metallicities; 
theoretical methods that are calibrated on theoretically simulated line ratios using stellar population and photoionization models; the hybrid methods that are calibrated on a mixture of the above two. 

The auroral line that is needed by the $T_e$-based method is too weak to be detected in the host galaxies of PTF SNe. 
Thus, we derive the metallicities from strong emission lines using 
an open-source Python package developed by our group, {\it pyMCZ} \citep{Bianco16}. 
It is based on the original IDL code of \citet{Kewley02} with updates from 
\citet{Kewley08}, and expanded to include more recent calibrations. 
Specifically, the {\it pyMCZ} requires the emission line fluxes and its uncertainties as inputs for Monte Carlo sampling, 
and calculates the probability distributions (and their median and confidence intervals) of the metallicity 
in various calibrations. 
The code also calculates from Balmer decrement the probability distribution of host galaxy extinction values, $E(B-V)_{\rm host}$, 
and applies internal extinction correction to the emission line fluxes in the calculation of metallicities. 
In order to derive the Balmer decrement, H$\alpha$ and H$\beta$ fluxes are available for all the host galaxies of PTF SNe and SN-GRBs. 

The {\it pyMCZ} code samples the synthetic line flux measurements from a Gaussian distribution with 
standard deviation equal to the measurement error, and centered on the measured flux value. 
Therefore, \citet{Bianco16} suggest that the input lines should have a S/N of at least three, 
assuring that fewer than $\sim 1\%$ of the sampled fluxes fall below zero (and are thus invalid). 
The lines with $S/N < 3$ may be eliminated by setting their values to $NaN$ in the input files to {\it pyMCZ}. 
The metallicity will be calculated only for the calibrations that use valid, non-$NaN$, line fluxes. 
By observational design, all of our spectra have $S/N > 15$ in H$\alpha$ and $S/N > 3$ in H$\beta$, with the only exception being PTF~10aavz, 
that has a $S/N = 13.5$ in H$\alpha$ at the SN site. However, this does not ensure that the $S/N$ is above three 
for all the other strong lines that we need, e.g., the $N2$ or $R_{23}$ ratios are exceptionally low in some host spectra. 

We may choose not to input the line fluxes with $S/N < 3$, following the suggestion in \citet{Bianco16}. 
However, only in a few cases is the low S/N due to noise in the spectra (e.g., the [O{\sc iii}] $\lambda5007$ 
line of PTF~10ciw falls on a night sky line). 
For the rest, the low S/N is caused by that line being particularly weak 
relative to the Balmer lines. For example, the $N2$ ratio is sensitive to ionization parameter, 
but to first order, a low $N2$ ratio indicates a low metallicity. 
As a result, eliminating the [N{\sc ii}] line with $S/N < 3$ generally causes a bias 
against the most metal poor systems in the calibrations that require [N{\sc ii}] flux for calculation. 
Four of the PTF SN hosts have $S/N < 3$ in [NII] $\lambda$6584, two of  SNe Ic-bl (10qts, 10tqv) and the other two of SNe Ic (10hie, 11rka),  and all are computed to be metal poor. Similarly, in the metal-rich regime of $\log ({\rm O/H}) + 12 \gtrsim 8.7$, the lower the $R_{23}$ ratio, 
the higher the metallicity \citep[e.g.][]{Kewley04}. 
Eliminating the [O{\sc ii}] or [O{\sc iii}] lines with $S/N < 3$ generally causes a bias against the most 
metal rich systems.

Therefore, we choose to input $NaN$ to {\it pyMCZ} only if the lines are outside of the wavelength coverage 
(including the [O{\sc ii}] lines for PTF~10bip, PTF~12gzk, and PTF~12hni) or are not detected at all
(including both [O{\sc iii}] lines for PTF~10qqd and PTF~10yow). 
They appear as missing values in Table \ref{tab:emission}. 
For the lines detected with $S/N < 3$, we keep their flux measurements as derived by {\it platefit} in the input files to {\it pyMCZ}. 
The {\it pyMCZ} code has been modified to properly deal with input lines that have $S/N < 3$ since it
sets invalid negative synthetic line fluxes to zeros during sampling  - and we check if a sufficient sample size is retained from 2000 Monte-Carlo trials to build the final distribution of the metallicity. 

\subsubsection{Metallicity Calibrations}

Different metallicity calibrations give systematically different metallicity values, even based on the same line fluxes \citep[e.g.,][]{Kewley08} 
and there is no consensus which calibration to use. In order to make sure our results are independent of the chosen metallicity calibration, 
we present our analysis in four different calibrations that are relatively independent. To decide which calibrations should we present, 
we examine the 
intercomparisons between the metallicity values based on our data for all the calibrations implemented by {\it pyMCZ} 
\citep[over 23 of them, for a full definition of the calibrations see][and references therein]{Bianco16}.

Although available through {\it pyMCZ}, some older calibrations, such as M91 \citep{McGaugh91} and Z94 \citep{Zaritsky94} 
are obsolete. We evaluate the performance of the remaining metallicity calibrations in a data-driven fashion 
and we do not report metallicities for the calibrations that generally produce measurements with large scatters compared to 
the results in most of the other calibrations for our 
48 PTF SN hosts: 
that includes M08\_O3O2 \citep{Maiolino08}, P10\_ON and P10\_ONS \citep{Pilyugin10}. 
In addition, we chose a single calibration for each metallicity diagnostic (set of line ratios). 
For example, several calibrations are based on $N2$: 
D02 \citep{Denicolo02}, PP04\_N2H$\alpha$ \citep{Pettini04}, KK04\_N2H$\alpha$ \citep{Kobulnicky04}, and M08\_N2H$\alpha$ \citep{Maiolino08}. 
We only present the result in one of them, M08\_N2H$\alpha$. 
Similarly, where there are more than one calibration based on the same theoretical model, which thus 
should deliver self consistent metallicities, we present only one. 
For example, eight calibrations from \citet{Dopita13} can be calculated with pyMCZ and our line inputs, 
but we only present the result for one of them, D13\_N2S2\_O3S2.

In summary, we choose to present the median values, 16th and 84th percentiles of our metallicity estimates 
in these four calibrations: KD02comb, D13\_N2S2\_O3S2, PP04\_O3N2, and M08\_N2H$\alpha$, together with 
the $E(B-V)_{\rm host}$ estimates. Table \ref{tab:PTFZ} contains these values for the hosts of PTF SNe Ic/Ic-bl (including the six weird/uncertain SN subtype transients). 
Table \ref{tab:GRBZ} contains the same values for the hosts of SN-GRBs. 
We calculate them using {\it pyMCZ} in the same manner for both datasetes to make a fair comparison. 
Amongst these four calibrations, KD02comb and D13\_N2S2\_O3S2 are theoretical; 
PP04\_O3N2 and M08\_N2H$\alpha$ are hybrid. 
We note that none of them is an empirical calibration calibrated purely on the $T_e$-based metallicities. 
Historically, there have been large systematic offsets between the $T_e$-based and theoretical calibrations.

In the Appendix section~\ref{app:zcalib} we briefly describe pertinent details of each of the four calibrations we adopt.

There is a long-standing debate about which metallicity calibration should be used, and there are known systematic offsets 
between the various calibrations \citep[see][]{Kewley08}. Having chosen four calibrations that best fit our needs, we are not 
generally advocating for the use of these over others. 
Moreover, it should be noted that the {\it pyMCZ} code does not include systematic errors between the calibrations, 
so the reader should refrain from absolute comparisons of calibrations and especially from comparing galaxies across different 
calibrations. 
The metallicity uncertainties presented here are statistical ones and account for the flux uncertainties and calibration uncertainties if provided. 
The four calibrations are chosen to represent a variety of the calibrations, so as to test if any trends that we see 
are calibration-dependent. 
Although the absolute metallicity values vary depending on the calibration used, 
the relative metallicity trends can be robust and thus should be seen across different calibrations if the analysis 
is performed self-consistently for the same calibration \citep{Kewley08}.

\subsection{Stellar Masses and SFRs}
\label{sec:sed}

As described in section \ref{sec:phot}, we compile broadband UV-to-optical photometry with good quality 
for all 
48 PTF SN host galaxies (SDSS {\it u,g,r,i,z} for 
44 hosts, Pan-STARRS {\it g,r,i,z} for four hosts, and {\it GALEX} FUV, NUV bands for 
36 hosts). 
We present these magnitudes with uncertainties in Table \ref{tab:SED} (including the six weird/uncertain SN subtype transients), corrected for Galactic extinction. 
In this section, we derive the global $M_*$s and SFRs of these host galaxies via SED fitting 
and list them also in Table \ref{tab:SED}. 
More details of the SED fitting process can be found in \citet{Huang12a} and \citet{Huang12b}, 
which follows the method of \citet{Salim07}. 
We use the same SED models as the ones used by the MPA-JHU group, 
so that the $M_*$ values of the PTF SNe Ic/Ic-bl hosts are fully consistent with 
those that we retrieve from the MPA-JHU catalog for the SDSS galaxies (see \ref{sec:sdss}). 
We also expect no significant systematic offsets of the $M_*$ and SFR values between those derived in this work for the PTF SNe Ic/Ic-bl hosts 
and those compiled from the literature for SN-GRB hosts (see below). 

We use a library of 100,000 SED models generated using the \citet{Bruzual03} stellar population synthesis code \citep{Gallazzi05}, 
assuming a \citet{Chabrier03} IMF. 
Dust is accounted for with the \citet{Charlot00} two-component model to include attenuation from both the diffuse ISM 
and short-lived (10~Myr) giant molecular clouds. 
A grid of models with an extensive range of stellar metallicity (correlated but different from the gas phase metallicity as derived in section \ref{sec:meta}), 
internal extinction, and star formation histories (SFHs) are considered. 
The stellar metallicity ranges between 0.1 -- 2.0 Solar and the effective optical depth in the $V$ band, $\tau_V$, ranges between 0 -- 6. 
The prior on the distribution of the stellar metallicity is uniform, and that of $\tau_V$ is Gaussian with a standard deviation of 0.55~dex and a mean predicted by \citet{Giovanelli95}, 
being higher in more massive galaxies and in more inclined galaxies.

A common way to model the SFH is to parameterize it as a smoothly declining exponential, ${\rm SFH}(t; T) = e^{-t/T}$, 
where $T$ is the timescale of the decay, and then calculate a grid in $T$. 
The library of SED models that we use is generated 
following an alternative approach that mimics stochastic processes in which 
the SFHs are drawn randomly according to some prior assumptions. 
This technique was introduced for galaxy parameter estimation by \citet{Kauffmann03} 
and is widely used by e.g., \citet{Gallazzi05} and \citet{Salim07}, as well as by the MPA-JHU group to estimate the $M_*$ based on the fit 
to the SDSS photometry. 
The stochastic effect in star formation may be significant in low mass galaxies \citep{Lee09}, and hence 
the stochastic model, which allows for a wide range of SFHs, is more realistic. 
In particular, it allows random bursts in SFH that are superimposed on a continuous component. 
Similarly, for the $M_*$ estimates of SN-GRB hosts, we adopt the GHostS values derived by \citet{Savaglio09}, 
who also model the SFH as a young burst component superimposed on a population of older stars. 
However, we note that such methods that include bursts in the SFHs generally 
result in systematically smaller $M_*$ estimates than if assuming a smooth exponential SFH \citep[e.g.,][]{Bell03}. 
The difference increases with decreasing $M_*$ to be $\sim$0.4~dex on average at $M_* = 10^8~M_\odot$ \citep{Wyder07}, 
because the difference is more significant in the galaxies with higher ratios of current to past-averaged SFRs, 
which many low-mass galaxies have. 

Given the SED models, we follow a Bayesian approach to estimate galaxy parameters, e.g., $M_*$ and SFR, by deriving 
posterior probability density functions (PDFs) from the PDFs that are computed based on priors 
(the galaxy parameters are known for the models) and the likelihood functions. 
Assuming Gaussian errors in the flux, the likelihood of data given each model is calculated by 
minimizing $\chi^2$ of the fit. 
By convention, all the models are normalized to have a total of 1~$M_\odot$ in $M_*$ formed by the present day. 
For each model, the $M_*$ is essentially derived from a scaling factor to 
minimize the $\chi^2$ between the model fluxes and the observed magnitudes \citep[see e.g.,][]{Salim07}. 
Due to the stochastic nature of the star formation behavior in our adopted models, 
the physical parameters associated with a single best-fit 
model with the maximum likelihood (the best $\chi^2$) might lead to peculiar results at times. 
We therefore generate a full 
marginalized posterior PDF for each physical parameter. 
For the $M_*$ and SFR estimates of all 
48 PTF SN hosts, we report the median values from such distributions in Table \ref{tab:SED}, 
along with the uncertainties denoting the 16-84 percentiles from distributions. 

We now describe how we calculate SFRs for the models of galaxies, the physical meaning of this definition, 
as well as why the SFRs derived in this way (SED fitting) are consistent with those of comparison samples. 
For each model of a galaxy, the SFR is derived from averaging the SFH over the last 100~Myr. 
A UV bright star has a lifetime of $\sim$100~Myr and thus the UV luminosity as an SFR indicator also probes a timescale of $\sim$100~Myr. 
For 
36 of the PTF SN host galaxies with {\it GALEX} magnitudes, the SED fitting is applied to UV-to-optical bands 
and therefore their SFRs averaged over the last 100~Myr are UV luminosity based to first order. 
Similarly, the SFRs of LVL galaxies that we adopted are also UV luminosity based. 
We note however that the SFRs of SN-GRB hosts are H$\alpha$ based, which traces a more instantaneous timescale ($\sim$10~Myr). 
It was found amongst low mass galaxies from the LVL survey that H$\alpha$ tends to increasingly under-predict the total SFR relative to the FUV, 
probably due to stochasticity in the formation of high-mass stars \citep{Lee09}. 
However, this trend only becomes evident for galaxies with SFR $\lesssim 0.01~M_\odot~{\rm yr}^{-1}$, 
whereas all the host galaxies we consider in this work have SFR $> 0.01~M_\odot~{\rm yr}^{-1}$, and most of them have SFR $> 0.1~M_\odot~{\rm yr}^{-1}$. 
As a result, it is fair to compare the H$\alpha$-based SFRs of SN-GRB hosts with the FUV-based SFRs of PTF SN hosts and LVL galaxies.  

For the PTF SN hosts without {\it GALEX} magnitudes, our SED fitting process is applied to optical-only bands. 
Their SFRs are poorly constrained, especially for the two hosts outside of both SDSS and {\it GALEX} (PTF~09iqd and 10svt with neither {\it u} nor FUV/NUV magnitudes). 
This effect is reflected in the broader posterior PDFs for their SFR estimates and thus give rise to higher SFR uncertainties. 
In the following figures that involve SFRs, we plot the PTF SN Ic/Ic-bl hosts without {\it GALEX} magnitudes by open symbols, 
indicating that their SFRs are less reliable. 
In contrast, the $M_*$ derived from a scaling factor is well constrained by SED fitting to optical-only bands, and is insensitive to the extra data in UV. 

\section{Results} \label{sec:result}

In this section, we compare the distributions of various physical parameters derived in previous section for the PTF SN Ic/Ic-bl hosts 
with those for the SN-GRB hosts and LVL galaxies, in order to address the question whether the SNe Ic-bl or SN-GRBs preferentially 
occur in certain types of galaxy over others from comparison samples and what these preferences are if any. 
Such constraints from the environment will help to test the formation models of SNe Ic-bl and SN-GRBs 
which we undertake in section \ref{sec:discussion}. 

\subsection{BPT Diagram}
\label{sec:bpt}
\begin{figure*}[ht!]
\epsscale{0.8}
\plotone{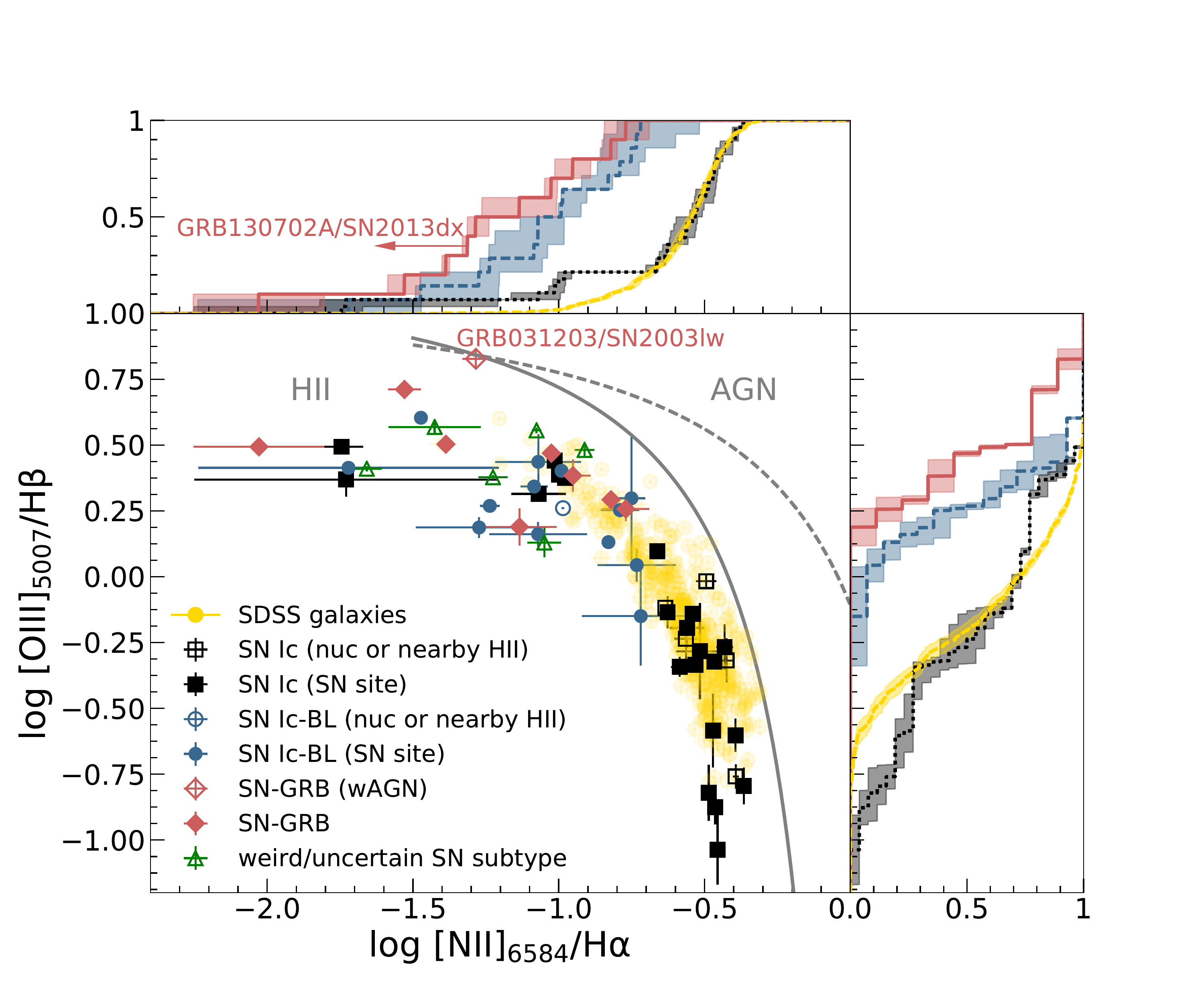}
\caption{
BPT diagram for the hosts of PTF SNe Ic (black squares), 
Ic-bl (blue circles), SN-GRBs (red diamonds), and weird/uncertain SN subtype transients (green triangles), as well as for a representative sample of SDSS galaxies (yellow crossed circles). The top and right panels show cumulative distributions of 
the two line ratios, respectively, for the hosts of PTF SNe Ic (black dotted line), Ic-bl (blue dashed line), and SN-GRBs (red solid line) and for the SDSS galaxies (yellow solid line). 
The flux uncertainties are denoted by error bars in the diagram and by colored bands in the side and top panels. 
The gray curves in the BPT diagram show the separation between star-forming galaxies (below the solid curve), composite galaxies (in between the solid and dashed curves), and galaxies whose spectra have a significant AGN contribution (above the dashed curve). All the host galaxies in our sample can be safely classified as star-forming, except for  GRB031203/SN2003lw 
which is at the intersection of the three regions (open diamond, see discussion in text).
}
\label{fig:BPT}
\end{figure*}

We explore the distributions of host galaxies across a Baldwin, Phillips \& Terlevich (BPT) diagram, 
in order to (i) identify the ones with AGN contribution so that the quantities derived from emission lines 
should be taken as approximate; (ii) compare the line ratio distributions between the hosts of different SN populations; 
and (iii)  study the implications in metallicity, ionization, and age conditions, 
which altogether predict a certain location in the diagram \citep[e.g.][]{Dopita00}, 
given the different parameter spaces occupied by samples. 

\subsubsection{Identifying AGNs}

The BPT diagram was first proposed by \citet{Baldwin81} to separate the main excitation mechanism of 
emission-line objects via the line ratios of [O{\sc iii}]$\lambda5007$/H$\beta$ versus [N{\sc ii}]$\lambda6584$/H$\alpha$. 
The lower left panel in Figure \ref{fig:BPT} shows a BPT diagram for the hosts of PTF SNe Ic/Ic-bl and SN-GRBs. 
The lower solid line in the BPT diagram is an empirical division derived from SDSS galaxies in \citet{Kauffmann03} and 
the upper dashed line is a theoretical upper limit for pure star-burst models from \citet{Kewley01}. 
They are usually invoked to distinguish between classical star-forming objects (below the solid line), AGN powered sources 
(above the dashed line), and composite objects (in between the two lines). 

The ionization source for all the host galaxies in our PTF SN and SN-GRB samples is star formation rather than AGN activity, 
except potentially for the host of GRB031203/SN2003lw. 
Thus, the strong line methods that we adopt for metallicity calculations are valid for almost all of the hosts in our sample, 
so are the H$\alpha$-based SFRs for SN-GRB hosts in particular. 
However, the AGN contamination may render the metallicity and SFR measurements incorrect for the host of GRB031203/SN2003lw.
We present our analysis results with and without this object and plot it with a different symbol, in order to assess its impact. 
See Appendix \ref{app:bpt} for more details on the classification of all host galaxies, 
especially why we cannot rule out AGN contamination for the host of GRB031203/SN2003lw. 

\subsubsection{Line Ratios}
\label{sec:liner}

The analysis of line-ratios is a model independent diagnostic, thus line-ratio plots are an important diagnostic tool.
Before we study what the line ratios imply in terms of the metallicity, ionization, and age conditions, which are all model-dependent quantities, 
we can make a direct comparison of line ratio distribution between our samples to assess statistical differences.

In the lower left panel of Figure \ref{fig:BPT}, the hosts of PTF SNe Ic are represented by black squares, 
SNe Ic-bl by blue circles, and SN-GRBs by red diamonds. 
For the hosts of SNe Ic and Ic-bl, open (not filled) symbols indicate that the SN sites are outside of the apertures for spectra extraction. 
Amongst both samples, the open symbols do not systematically deviate from the closed ones, i.e., the metallicities of 
the regions that are less representative of the SN local environments are not significantly different from those that are representative. 
The hosts of SNe Ic occupy the full range of the sequence, 
whereas the hosts of SNe Ic-bl and SN-GRBs belong exclusively to the upper left half of the sequence. 
The difference between the hosts of SNe Ic-bl and SN-GRBs is more subtle. 
We therefore perform a quantitative comparison between the samples by investigating the two line ratios separately. 

In the top and right panels of Figure \ref{fig:BPT}, 
we compare the empirical cumulative distribution functions (CDFs) of the three samples, respectively for these two line ratios. 
The colored bands denote 1$\sigma$ flux measurement errors. 
For each sample, the CDF includes both open and close symbols from the plot in the lower left panel. 
Note that the two SN Ic hosts with no detections in [O{\sc iii}]$\lambda5007$ 
show up only in the top panel for [N{\sc ii}]$\lambda6584$/H$\alpha$. 
The host of GRB130702A/SN2013dx, which has no detection in [O{\sc iii}]$\lambda5007$ 
and an upper limit in [N{\sc ii}]$\lambda6584$, also shows up only in the top panel: 
it appears as a value associated with an arrow pointing to the left. 

Considering these upper limits, we should describe the CDF of [N{\sc ii}]$\lambda6584$/H$\alpha$ for SN-GRBs by 
the Kaplan-Meier estimator, which uniformly distributes the weight of each upper limit among the detections with lower values. 
In absence of censored data (lower or upper limits), the Kaplan-Meier estimator reduces to the usual CDF. 
We use the ASURV package for survival analysis \citep{Lavalley92} to calculate the Kaplan-Meier estimator. 
However, new methods are still needed that combine the virtues of survival analysis and measurement errors \citep{Lavalley92}. 
In order to demonstrate how the differences in distributions compare with the typical measurement errors, 
we choose to plot the usual CDF rather than Kaplan-Meier estimator for SN-GRBs in the upper panel by treating the upper limit as uncensored data 
but indicating it by an arrow. 

In the upper panel, the overall [N{\sc ii}]$\lambda6584$/H$\alpha$ ratios for the three samples follow the order of SN-GRB $<$ SN Ic-bl $<$ SN Ic, though this trend has not yet been statistically tested for actual significance. 
In the right panel, the overall [O{\sc iii}]$\lambda5007$/H$\beta$ ratios for the three samples follow the order of SN Ic $<$ SN Ic-bl $<$ SN-GRB , though this trend has not yet been statistically tested for actual significance.
For both line ratios, the differences between the CDFs of SNe Ic and Ic-bl hosts are much higher than the $1\sigma$ measurement errors 
(colored bands), but those between the CDFs of SN Ic-bl and SN-GRB hosts are less so. 
For the [N{\sc ii}]$\lambda6584$/H$\alpha$ ratio in particular, the CDF offset between the hosts of SNe Ic-bl and SN-GRBs 
can be largely explained by measurement errors. 

Thus, in order to assess if the differences in line ratio distributions are significant, we perform statistical tests. 
The tests account for the fact that the CDFs 
are based on our data, which do not sample the full underlying populations, but do not account for the measurement errors. 
For each pair of the samples that are adjacent in the sequence of SN Ic, Ic-bl, and SN-GRB, 
we test the null hypothesis that their line ratios are drawn from the same underlying distribution. 
The ASURV package applies the Logrank test and various generalized Wilcoxon tests on censored data, 
but the Logrank test can only be used when observations are censored. 
To be consistent, we report results of the Peto \& Peto generalized Wilcoxon test for censored data, 
and apply the Wilcoxon rank sum test with {\it scipy} in absence of censored data. 
Our conclusion 
is based on computing the {\it p}-value which is the probability of obtaining an effect at least as extreme as the one observed 
assuming the truth of the null hypothesis. 
Throughout this work, we choose a significance level, $\alpha=0.05$, 
which indicates a 5\% risk of concluding that a difference exists when there is no actual difference (i.e., type I error). 
If the {\it p}-value we obtain is less than or equal to this significance level, 
we reject the null hypothesis, i.e., we conclude that the difference is statistically significant at the 2$\sigma$ level. 

Between the samples of SN Ic and Ic-bl hosts, 
the {\it p}-value of the Wilcoxon rank sum test on the distributions of [N{\sc ii}]$\lambda6584$/H$\alpha$ ratios is very 
close to zero and that on the distributions of [O{\sc iii}]$\lambda5007$/H$\beta$ ratios is 
0.002 ($<0.05$). 
Now we turn to the important case of comparison between the SN Ic-bl and SN-GRB hosts. 
If GRB031203/SN2003lw 
with potential AGN contribution is included in the SN-GRB sample, 
the {\it p}-value of Peto \& Peto generalized Wilcoxon test on the distributions of [N{\sc ii}]$\lambda6584$/H$\alpha$ ratios is 
0.210 ($>0.05$), 
but that of the Wilcoxon rank sum test on the distributions of [O{\sc iii}]$\lambda5007$/H$\beta$ ratios is 
0.022 ($<0.05$). 
However, if GRB031203/SN2003lw 
is excluded, the same {\it p}-values are 0.26 and 0.08 (both $>0.05$), 
respectively for the [N{\sc ii}]$\lambda6584$/H$\alpha$ and [O{\sc iii}]$\lambda5007$/H$\beta$ ratios. 
Due to the fact that the GRB031203/SN2003lw host 
has a high [O{\sc iii}]$\lambda5007$/H$\beta$ ratio, 
and that the Wilcoxon rank sum test is sensitive to the difference in the tails of the distributions, 
the null result is reliable for the comparison between the [O{\sc iii}]$\lambda5007$/H$\beta$ ratios of the SN Ic-bl and SN-GRB hosts 
excluding potential AGNs.

To conclude, we infer from our data that 
{\it the differences in line ratio distributions between the populations of SN Ic and SN Ic-bl hosts are statistically significant, 
but not those between the SN Ic-bl and SN-GRB hosts, especially if the host that may harbor potential AGNs is excluded}. 
We have applied several other hypothesis tests including Logrank test and various Generalized Wilcoxon tests on censored data, 
as well as K--S test and Anderson Darling test on uncensored data. All of them reach the same conclusion, 
except for the 
tests on the [O{\sc iii}]$\lambda5007$/H$\beta$ ratios between the SN Ic-bl and SN-GRB hosts, if potential AGNs are included. 
However, we note that the small sample sizes of both the SN Ic-bl and SN-GRB hosts may affect the power of these tests, 
leading to higher type II error rate in presence of inherently different populations. 

Even if additional correction for stellar absorption is applied to the SN-GRB hosts, 
it has little impact on the CDF of [N{\sc ii}]$\lambda6584$/H$\alpha$, since the correction is low for H$\alpha$. 
Given a sample of 46 GRB hosts (only five of them are SN-GRB hosts), 
\citet{Savaglio09} derive the mean correction for H$\alpha$ to be 4\%, 
which, if applied to all our SN-GRBs hosts, would suggest 
a shift in CDF for SN-GRBs by only 0.02~dex to the left in the upper panel of Figure \ref{fig:BPT}. 
However, as the overall correction for H$\beta$ is higher, 
the CDF of [O{\sc iii}]$\lambda5007$/H$\beta$ for SN-GRB hosts would shift more in the right panel 
\citep[downwards by 0.08~dex, assuming a 20\% correction for H$\beta$ from][derived from the full sample of GRB hosts]{Savaglio09}, 
and therefore would appear even closer to that of SN Ic-bl hosts. 
This consideration strengthens our result that we find no significant difference in the line ratio distributions 
between SN Ic-bl and SN-GRB hosts. 


\subsubsection{Implications for Physical Parameters of galaxies hosting SNe}

In order to interpret observed line ratios, various theoretical works have combined stellar population synthesis with photoionization models to predict 
line ratios, given variations in the metallicity, ionization parameter, electron density and SFH, \citep[e.g.][]{Dopita00, Kewley01}. 
In particular, expected grids of constant metallicities and ionization parameters have been overlaid on a BPT diagram, 
with all the other parameters fixed, e.g., grids calculated based on instantaneous zero-age starburst models or on continuous starburst models. 
In these works, a well-known degeneracy exists such that a given location in the BPT diagram can be explained by 
different combinations of metallicity and ionization parameters, especially for continuous starburst models \citep[e.g.][]{Kewley01}. 
Therefore, it appears that only when fixing one of the two parameters involved in the grid (metallicity or ionization parameter) 
is it possible to derive a trend with the other parameter across the BPT diagram. 

On the other hand, observations show that the more metal-poor H{\sc ii} regions are generally located in the upper left corner 
of the diagram, and the more metal-rich ones are located towards the lower right corner \citep[e.g.][]{Sanchez15}, 
which cannot be explained purely by the theoretical grids. 
In fact, the metallicity is observed to anti-correlate with ionization parameter 
\citep[e.g.][with metallicities derived from $T_e$]{Dopita86, Perez14}, 
so that not any arbitrary combination of the two is allowed by nature. 
For example, given the model grids calculated by \citet{Kewley01} that assume continuous star bursts, 
the upper left corner (-1, 0.5) can be explained either by a model with a super-solar metallicity but extremely high ionization parameter, 
or by a model with a sub-solar metallicity and moderate ionization parameter. The former model is unrealistic due to the anti-correlation 
between the metallicity and ionization parameter, and thus the upper left corner is generally associated with metal-poor H{\sc ii} regions. 
In a BPT diagram, the parameter space populated by real H{\sc ii} regions is therefore much more restricted than that predicted by model grids 
(see the narrow sequence in Figure \ref{fig:BPT}, though the exact location depends on redshift; e.g., \citealt{erb06,shapley15}). 
More importantly, after the models with unrealistic combinations of the two parameters are excluded, 
the rest define a cleaner trend of the increasing metallicity from the upper left to the lower right corner along the sequence. 
Such trend sets the foundation of some strong line methods, 
which derive metallicities from the line ratios like $O3N2$ and $N2$ in certain ranges monotonically. 
Similarly, an overall trend of decreasing ionization parameter from the upper left to the lower right along the sequence is observed \citep[e.g.][]{Sanchez15}. 

Based on this understanding, we can infer the average trends of gas-phase metallicities and ionization parameters for the hosts of 
PTF SNe Ic, Ic-bl, and SN-GRBs as populations by comparing their distributions on a BPT diagram in Figure \ref{fig:BPT}, 
regardless of the apparent degeneracy in model grids. 
Furthermore, \citet{Sanchez15} link a location on the BPT diagram to the average age and metallicity 
of the underlying stellar populations derived from the CALIFA observations. 
They find that the [N{\sc ii}]$\lambda6584$/H$\alpha$ increases with increasing stellar age and stellar metallicity, 
whereas the [O{\sc iii}]$\lambda5007$/H$\beta$ decreases with increasing stellar age and stellar metallicity. 
We can therefore infer the properties of underlying stellar populations of the hosts as well by comparing their distributions of line ratios. 

We find that the hosts of SNe Ic occupy the full range of the sequence, 
whereas the hosts of SNe Ic-bl and SN-GRBs belong exclusively to the upper left half of the sequence. 
In our samples, there seems to be an upper limit in the metallicity of host environment for the formation of SNe Ic-bl and SN-GRBs. 
In the side panels, the CDFs indicate that the overall metallicities for the three samples appear to follow the order of 
$Z_{\rm GRB} < Z_{\rm Ic-bl} < Z_{\rm Ic}$, and that the overall ionization parameters appear follow the opposite order of 
$q_{\rm Ic} < q_{\rm Ic-bl} < q_{\rm GRB}$ $-$ regarding the underlying stellar populations, the overall metallicities and ages follow the order of SN-GRB $<$ SN Ic-bl $<$ SN Ic. However, these trends have not yet been statistically tested for actual significance, which we perform below.
We note that such qualitative trends are implied by the overall physical conditions of H{\sc ii} regions observed along the sequence, 
even without having computed model-dependent values for the quantities of interest, such as metallicity and ionization parameter. 
We will present direct comparisons of the derived quantities like metallicities in the following sections, 
as well as test for the significance of the differences between the populations. 





\subsection{Metallicity Distribution}
\label{sec:ohdist}

In this section, we compare the metallicity distributions in four calibrations for the samples of PTF SN Ic, Ic-bl and SN-GRB hosts, 
in order to uncover any differences between the samples. 
We also compare the metallicity distributions for the same sample across different calibrations, 
in order to uncover any biases in these calibrations and their possible impact on our conclusions. 

\subsubsection{Statistics of the Distributions}


\begin{deluxetable*}{lcccccc}[ht!]
\tabletypesize{\small}
\tablenum{7}
\tablewidth{0pt}
\tablecaption{Summary of the metallicity distributions in various calibrations for the different samples}
\label{tab:OHdist}
\tablehead{
\colhead{SN type} & \colhead{$N$} & \colhead{Mean} & \colhead{SEM} & \colhead{25th} & \colhead{Median} & \colhead{75th}
}
\startdata
	&\multicolumn{6}{c}{D13\_N2S2\_O3S2}\\
\hline
PTF SN Ic & 26&8.782&0.08&8.677&8.867&9.103\\
PTF SN Ic-BL & 13&8.373&0.067&8.152&8.363&8.545\\
SN-GRB (excluding potential AGN) & 7&8.211&0.137&8.163&8.343&8.375\\
\hline&\multicolumn{6}{c}{KD02\_COMB}\\
\hline
PTF SN Ic & 28&8.779&0.055&8.73&8.828&8.938\\
PTF SN Ic-BL & 14&8.421&0.062&8.242&8.48&8.57\\
SN-GRB (including potential AGN)\tablenotemark{a} & 10&8.339& 0.071&8.115&8.183&8.515\\
SN-GRB (excluding potential AGN)\tablenotemark{a} & 9&8.304&0.070&8.107&8.160&8.479\\
\hline&\multicolumn{6}{c}{PP04\_O3N2}\\
\hline
PTF SN Ic & 26&8.594&0.048&8.515&8.642&8.774\\
PTF SN Ic-BL & 14&8.318&0.036&8.266&8.308&8.400\\
SN-GRB (including potential AGN) & 9&8.202&0.058&8.066&8.259&8.329\\
SN-GRB (excluding potential AGN) & 8&8.219&0.063&8.098&8.282&8.344\\
\hline&\multicolumn{6}{c}{M08\_N2H$\alpha$}\\
\hline
PTF SN Ic & 28&8.810&0.058&8.781&8.898&8.972\\
PTF SN Ic-BL & 14&8.480&0.057&8.353&8.486&8.669\\
SN-GRB (including potential AGN)\tablenotemark{a}& 10&8.295&0.110&8.068&8.282&8.521\\
SN-GRB (excluding potential AGN)\tablenotemark{a} & 9&8.287&0.124&7.917&8.298&8.535
\enddata
\tablenotetext{a}{containing an upper limit (GRB130702A/SN2013dx)}
\end{deluxetable*}

The statistics of the metallicity distributions are summarized in Table \ref{tab:OHdist} for each sample and in the four calibrations: 
D13\_N2S2\_O3S2, KD02\_COMB, PP04\_O3N2, and M08\_N2H$\alpha$, 
which were chosen for reasons given in section \ref{sec:meta}. 
Due to possible AGN contamination to the hosts of GRB031203/SN2003lw and GRB091127/SN2009nz, 
we consider both the cases with and without them in the SN-GRB sample. 
For the KD02comb, PP04\_O3N2 (if potential AGNs are included), and M08\_N2H$\alpha$ calibrations, 
the SN-GRB sample contains one upper limits (due to GRB130702A/SN2013dx) 
so that these statistics are calculated using the ASURV package Kaplan-Meier estimators. 

The sample sizes (column $N$ in Table \ref{tab:OHdist}) are calibration-dependent. 
Among these four calibrations, KD02comb and M08\_N2H$\alpha$ are the ones computable for all the hosts in our samples: 
for the hosts of 28 SNe Ic, 14 SNe Ic-bl, and 10 SN-GRBs (including potential AGNs).
Assuming a solar metallicity of $\log({\rm O/H}) + 12 = 8.69$ \citep{Asplund09_rev}, 
the mean metallicities for the hosts of SNe Ic, Ic-bl, and SN-GRB excluding potential AGNs range between 
0.80 -- 1.32, 0.42 -- 0.62, 
and 0.33 -- 0.41$Z_\odot$, respectively, across the four scales. 
Alternatively, assuming a solar metallicity of $\log({\rm O/H}) + 12 = 8.76 \pm 0.07$ \citep{Caffau11}, 
the same values range between 
0.68 -- 1.12, 0.36 -- 0.52, 
and 0.28 -- 0.35$Z_\odot$, respectively. 
Both the mean and median values show that the metallicities of the three samples follow the sequence of 
$Z_{\rm GRB} < Z_{\rm Ic-bl} < Z_{\rm Ic}$, which agrees with the implication from line ratios, though these trends have not yet been statistically tested for actual significance.
The mean and median metallicities for the SNe Ic hosts are about solar but those for the SNe Ic-bl and SN-GRB hosts are well below solar. 
The average differences between the SN Ic and Ic-bl hosts range from 
0.28~dex (PP04\_O3N2) to 0.41~dex (D13\_N2S2\_O3S2). 

The average differences between the SN Ic-bl and SN-GRB hosts (excluding potential AGNs) range from only 
0.10~dex (in PP04\_O3N2) to 0.19~dex (in M08\_N2H$\alpha$), 
which are comparable to the standard errors of sample means (SEMs), as listed in Table \ref{tab:OHdist}. 
With the smallest size, the sample of SN-GRB hosts always has the highest SEM among the three 
(from 0.06~dex for PP04\_O3N2 to 0.14~dex for D13\_N2S2\_O3S2). 
The columns of `25th' and `75th' list the percentiles and their difference characterizes the standard deviation of the distribution. 
For the only calibration without upper limits (D13\_N2S2\_O3S2), the SN Ic sample has the highest standard deviation, 
i.e., SNe Ic occur in galaxies with a large range of metallicities, whereas the SNe Ic-bl and SN-GRBs from our samples 
occur exclusively in galaxies with low metallicities. 

\subsubsection{Statistical Tests on Distributions}
\label{sec:ohtest}

\begin{centering}
\begin{deluxetable*}{lcccc}[ht!]
\tabletypesize{\small}
\tablenum{8}
\tablewidth{0pt}
\tablecaption{{\it p}-values of Wilcoxon tests on metallicity distributions for the different samples}
\label{tab:Htest}
\tablehead{
\colhead{} & \multicolumn{4}{c}{PTF SN Ic-BL hosts} \\
\colhead{} &
\colhead{D13\_N2S2\_O3S2} &
\colhead{KD02\_COMB} &
\colhead{PP04\_O3N2} &
\colhead{M08\_N2H$\alpha$} }
\startdata

	PTF SN Ic hosts & 0.001 & 0.0003 & 0.0004 & 0.0001 \\
SN-GRB hosts (including potential AGNs) & ... & 0.438\tablenotemark{a} & 0.114 & 0.168\tablenotemark{a} \\

SN-GRB hosts (excluding potential AGNs) & 0.452 & 0.225\tablenotemark{a}& 0.219 & 0.310\tablenotemark{a} \\
\enddata
\tablecomments{Throughout this work, we choose a significance level, $\alpha=0.05$. 
If the {\it p}-value we obtain is less than or equal to this significance level, 
we reject the null hypothesis that the two samples were drawn from the same parent distribution, i.e., we conclude that the difference is statistically significant at the 2$\sigma$ level.}
\tablenotetext{a}{containing an upper limit (GRB130702A/SN2013dx)}

\end{deluxetable*}
\end{centering}

In this section, we apply hypothesis tests on the metallicity distributions to determine if the observed differences between the 
hosts of SNe Ic, SNe Ic-bl, and SN-GRBs are statistically significant.

For each pair of the samples that are adjacent in the sequence of SN Ic, Ic-bl, and SN-GRB, we perform the 
same tests as we did in section \ref{sec:liner} on the line ratio distributions and adopt the same criteria - we perform
the Wilcoxon rank sum test 
for the null hypothesis that the metallicities are drawn from the same underlying distributions. 
If an upper limit is involved, we perform the Peto \& Peto generalized Wilcoxon test instead. 
The $p$-values from these tests are listed in Table \ref{tab:Htest}. 
In all calibrations, the differences in metallicities between the samples of SN Ic and Ic-bl hosts are statistically significant 
($p < 0.05$), whereas those between the samples of SN Ic-bl and SN-GRB hosts are not 
($p > 0.05$, with the lowest value being 
0.081 in the PP04\_O3N2 calibration if potential AGNs are included). 
These are consistent with the test results on the lines ratio distributions. 
We have applied several other hypothesis tests, including K--S test and Anderson Darling tests, which all suggest the same results. 

Our results agree with the previous findings from K--S tests on host galaxies of untargeted SNe Ic and Ic-bl. 
For example, \cite{sanders12} and \citet{Galbany16} both find statistically significant differences 
in metallicity distributions between the samples of SN Ic and Ic-bl hosts, in the PP04\_N2 and M13\_O3N2 calibrations, respectively. 
However, the PP04\_N2 calibration suffers from saturation above solar metallicity. 
The M13\_O3N2 calibration gives systematically lower metallicities for more metal-rich systems compared to all the other calibrations that we consider here, 
so that the M13\_O3N2 calibration results in the smallest variance in metallicities, given the same galaxies. 
Together with the fact that our sample sizes are larger than theirs by a factor of 3 (for \citealp{sanders12}) , we find a difference between the SN Ic and Ic-bl hosts that is 
more significant (i.e., a lower $p$-value from the K--S test) than those found by \cite{sanders12} or \citet{Galbany16}. 
We note that those two works are not independent, since the untargeted hosts in \citet{Galbany16} are from the literature, 
including those presented in \cite{sanders12}. 

However, our results are somewhat at odds with previous works that compare SN Ic-bl hosts with SN-GRB hosts. 
Based on the metallicities in 
the KD02 calibration \citep[updated by][]{Kobulnicky04}, \citet{modjaz08_grbz} derived a K--S test 
{\it p} value of 0.03 ($< 0.05$) from samples of 
six SN Ic-bl hosts found in an untargeted fashion and five SN-GRB hosts. 
\citet{modjaz08_grbz}, for the first time, pointed out that the targeted surveys are biased towards galaxies that are massive and thus more metal rich, 
and they tried to include as many SNe Ic-bl from untargeted surveys as possible. 
However, due to the very limited data from untargeted surveys at that time, 
only half of the SNe Ic-bl in the sample of \citet{modjaz08_grbz} were found in an untargeted fashion, whereas the SN-GRBs are always found in an untargeted fashion. 
Moreover, their SN-GRB sample at that time was also small and contained GRB031203/SN2003lw with potential AGN contamination. 
\citet{levesque10_grbhosts} expanded the sample size of GRB hosts to ten, including higher-redshift GRBs without observed SN-associations, and compared them with a sample of 
eight SN Ic-bl hosts selected from \citet{modjaz08_grbz} (six of the SNe Ic-bl detected by untargeted surveys). 
They derived a K--S test {\it p} value of 0.03 ($< 0.05$), but of 0.06 if GRB031203/SN2003lw is excluded ($> 0.05$), which in turn agrees with our null results. 
The conclusion of \citet{levesque10_grbhosts} was based on the metallicities in the KK04\_R$_{23}$ calibration (see its caveat below). 
Given that our test results are based on twice larger samples of untargeted SN Ic-bl and SN-GRB hosts, which are all detected in the same untargeted fashion, together with the fact that the metallicity calibrations used by us are more appropriate for such a study (see below), we think our null results are more reliable. 

 Indeed \citet{japelj18} also obtain a null result (see their first line in the Anderson Darling test results in their Table 5 which gives a {\it p} value of 0.43 and of 0.23 (both $>> 0.05$) for two metallicity calibrations), even though they claim the opposite in their conclusions since they do not consider their own statistical evidence. Their SNe Ic-bl metallicity sample consists mostly of the untargetted SNe Ic-bl from \citet{modjaz08_grbz}, \citet{modjaz11}, with the remaining few from \citet{sanders12}, and they compare them to a smaller subset of six SN-GRBS, also included in our sample.

Specifically, 
the KK04\_R$_{23}$ calibration used by \citet{levesque10_grbhosts} 
results in a gap in the metallicity distribution among all the SDSS galaxies, which is unphysical. 
According to Figure 1 from \citet{Kewley08}, the gap 
resides at 
$\sim 8.4 - 8.6$ for KK04\_R$_{23}$. 
It is caused by the relation between $R_{23}$ and metallicity, which is double valued 
and has a turn around point around $\log({\rm O/H}) + 12 \sim 8.4$. 
Other calibrations that rely on $R_{23}$ are also affected, e.g., M91 \citep{McGaugh91} and KD02 \citep{Kewley02}. 
The KD02 calibration is updated to KD02comb in \citet{Kewley08} (the latter one is used by this work), with modifications that include
alleviating the artificial lack of objects with metallicities $\sim 8.4$. 
It may be appropriate to use the KK04\_R$_{23}$ calibration for a study of metal rich galaxies from surveys like the SDSS, 
but not for a study of metal poor galaxies like the SN Ic-bl and SN-GRB hosts. 
Indeed, the samples of SN Ic-bl hosts and SN-GRB hosts are revealed to be roughly separated 
by a gap at $\sim 8.4 - 8.6$ from \citet{levesque10_grbhosts}. 
The difference between the two samples (GRBs vs SNe Ic-bl) may be therefore exaggerated by this gap, 
which is artificially introduced by the KK04\_R$_{23}$ calibration. 
Thus, as we stressed in \citet{modjaz08_grbz, modjaz11} and \citet{Modjaz12_proc}, 
it is always advisable to present metallicity trends in many different, independent calibrations 
so as to ensure that any observed trends are not due to artifacts in a particular calibration. 

\subsubsection{Plots of Metallicity Distributions}

\begin{figure*}[ht!]
\epsscale{1.17}
\plottwo{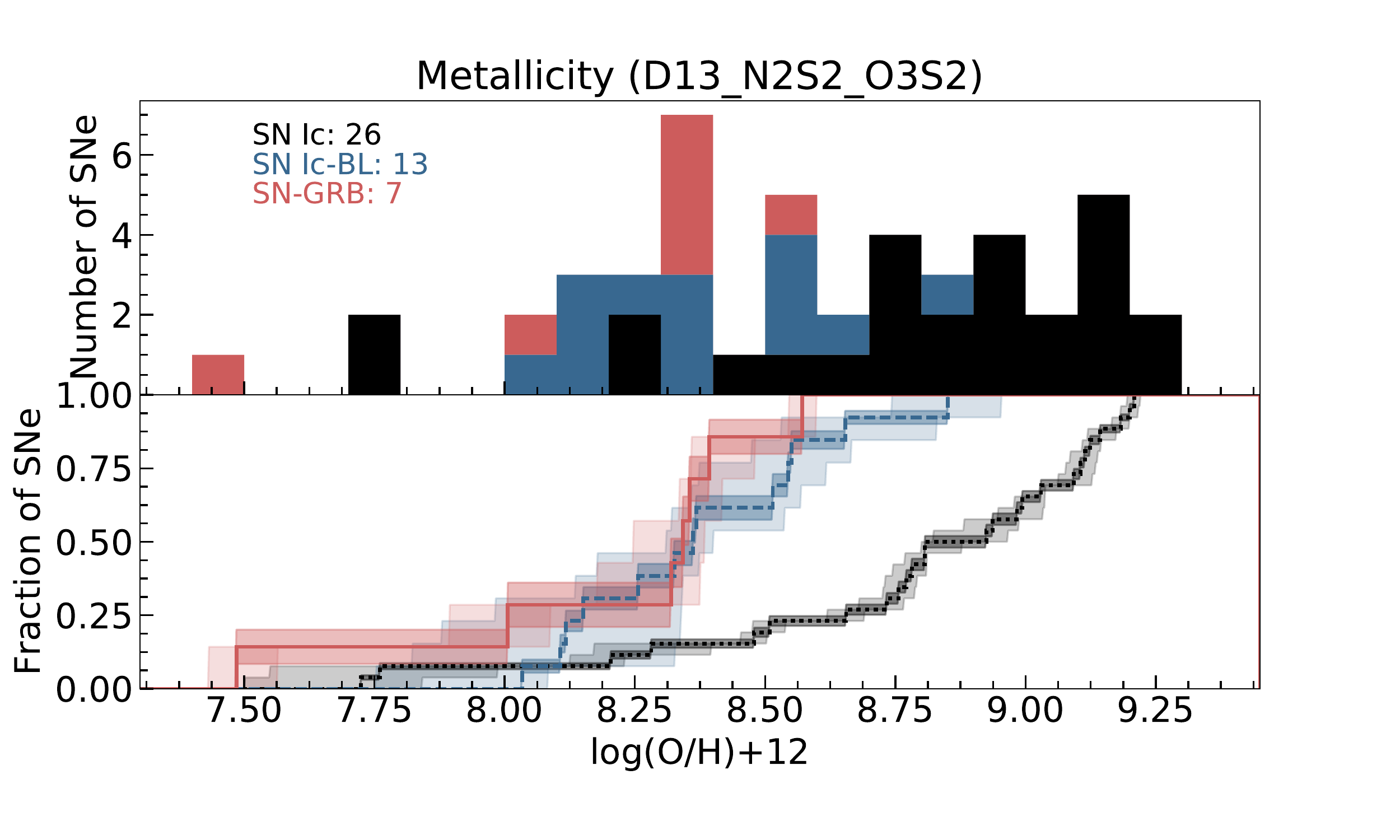}{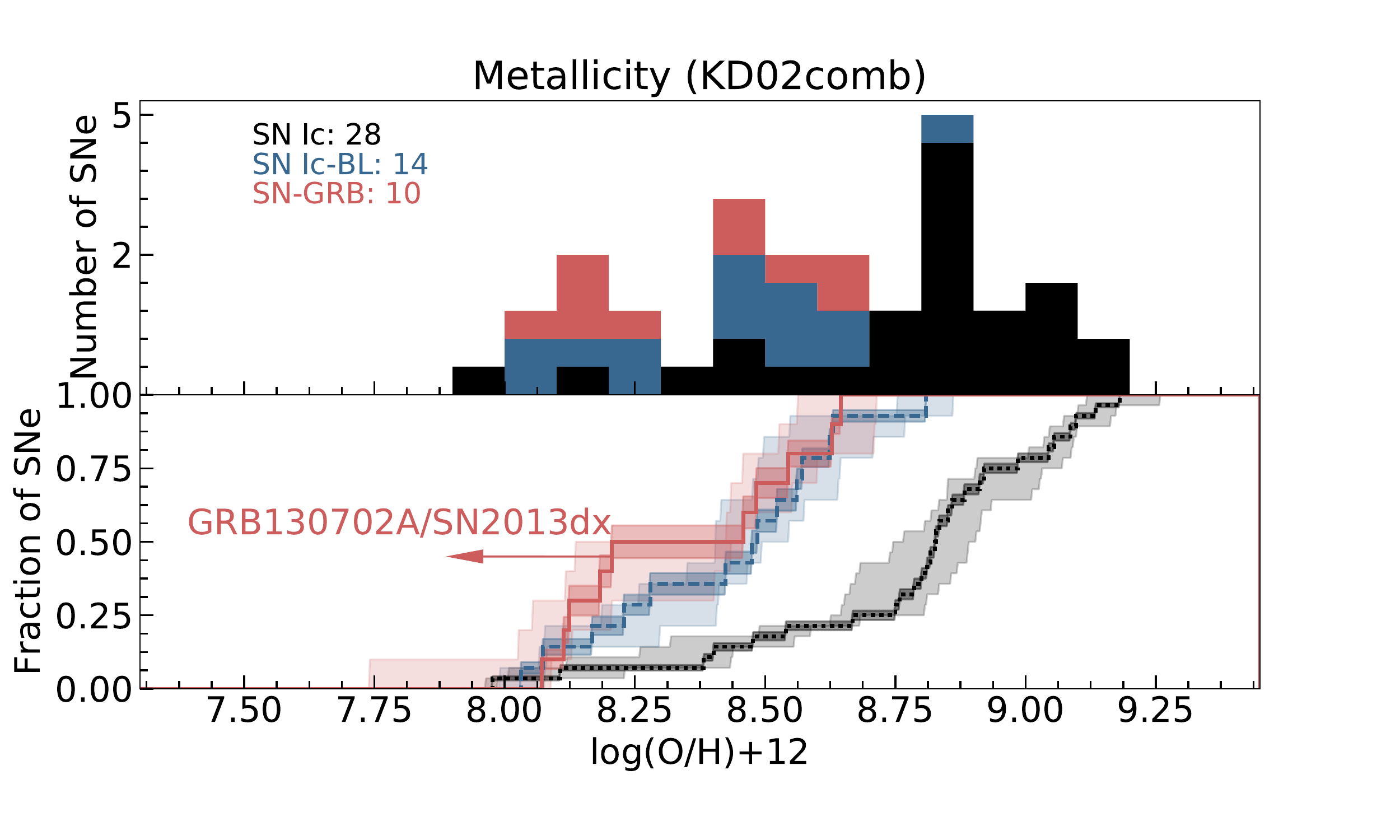}
\plottwo{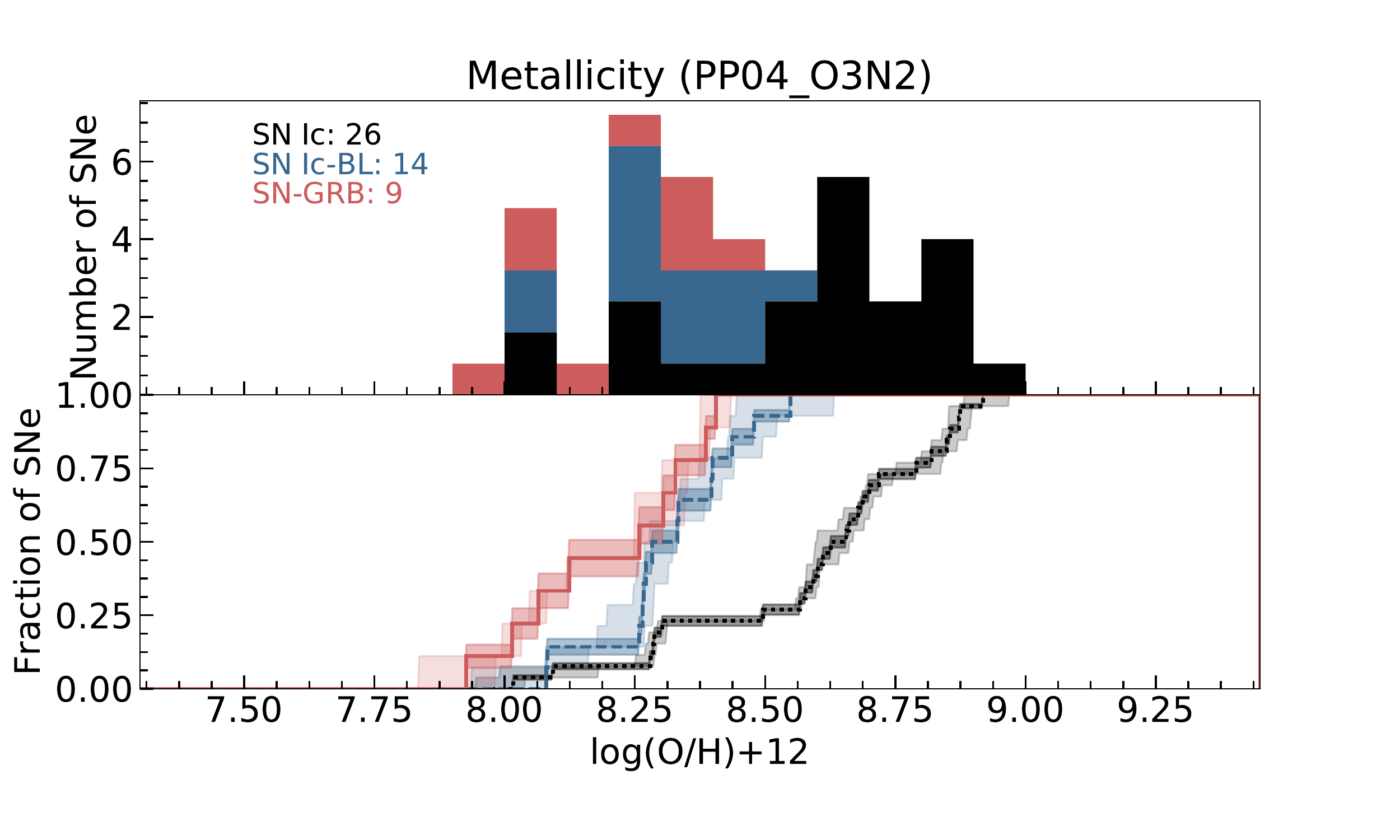}{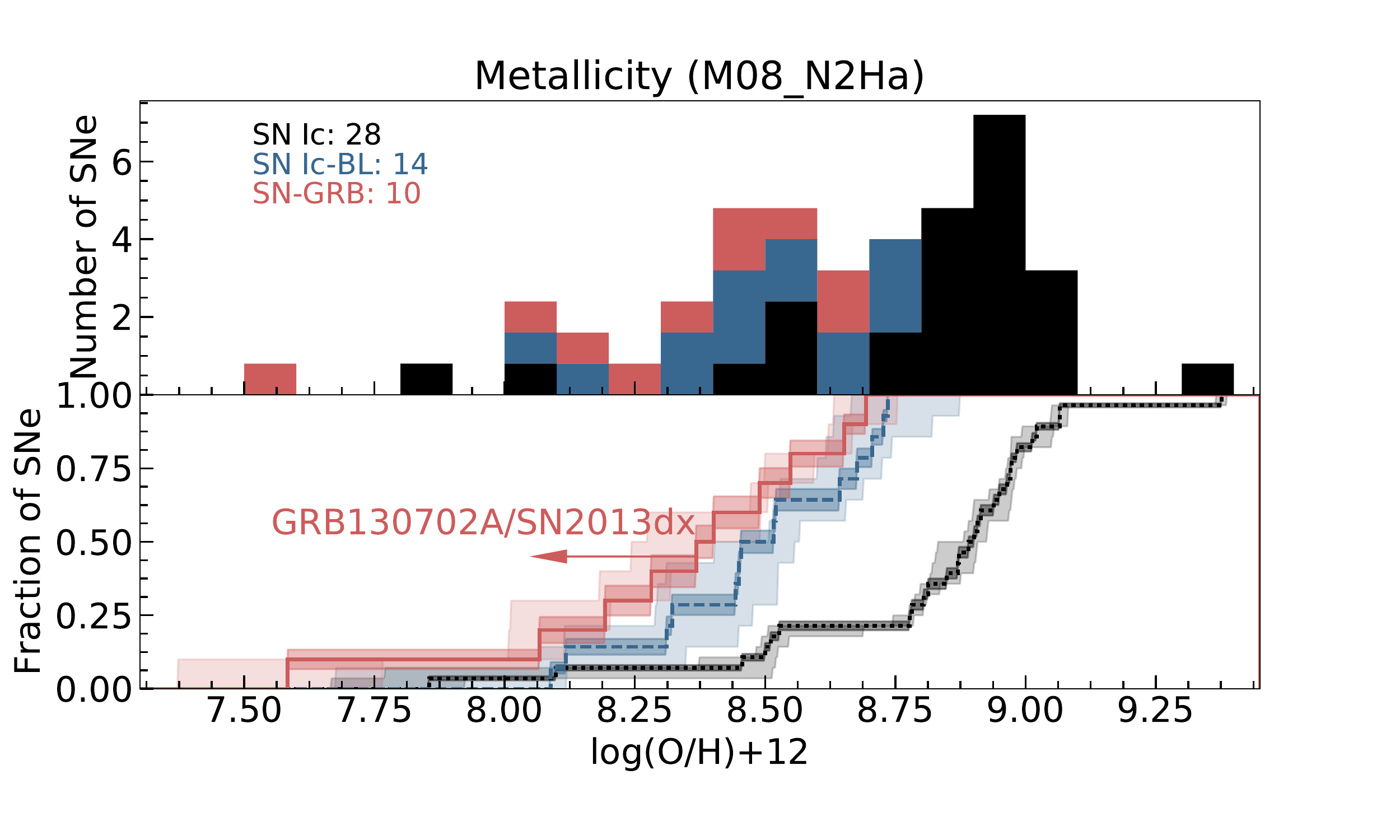}
\caption{
Stacked histogram (upper panels) and normalized cumulative distribution (lower panels) of metallicities in four different calibrations as calculated by {\it pyMCZ}. 
Color codes for PTF SN Ic/Ic-bl and SN-GRB follow the same convention as in Figure \ref{fig:BPT}. 
In the lower panels, the darker shaded area denotes the confidence band calculated by  {\it pyMCZ}, accounting for line measurement uncertainties, and the lighter shaded bands represent the uncertainty induced by the limited sample size, obtained by bootstrapping over the metallicity observations (see discussion in the text). The arrows indicate the contribution to the cumulative distributions of the upper limit measurement for GRB130702A/SN2013dx.
The host environment of SN Ic is more metal-rich than that of the SN Ic-bl and SN-GRBs and the difference is statistically significant. 
On the contrary, the SN-GRBs are found in environments with similar metallicities to those of the SNe Ic-bl. 
These trends are seen in all panels, indicating that they are robust to the calibration adopted. 
}
\label{fig:OHdist}
\end{figure*}

The statistical tests in the previous section account for the fact that data do not sample the full underlying populations, 
but they do not include the metallicity uncertainties. 
In this section, we show the plots of metallicity distributions with confidence bands that are derived from the metallicity uncertainties. 
In addition, we compare the combined sample across different metallicity calibrations, 
in order to understand the characteristics intrinsic to these calibrations and the possible impact on our conclusions. 

The plots of metallicity distributions are presented in Figure \ref{fig:OHdist}. 
These distributions include objects with host spectra extracted from both the apertures that cover the SN sites 
(filled markers in Figure \ref{fig:BPT}) and those that do not (unfilled markers in Figure \ref{fig:BPT}). 
Plotting both together is valid given the fact that the open symbols do not systematically deviate from the closed ones in Figure \ref{fig:BPT} 
(except for the open diamond that denotes GRB031203/SN2003lw which may have AGN contamination). 


In Figure \ref{fig:OHdist}, the upper panels show stacked histograms for the metallicity distributions in four calibrations. 
We compare the combined distributions from all samples across the calibrations, in order to understand the characteristic of each calibration. 
The well-known systematic offsets between the calibrations \citep[e.g.][]{Kewley08} are evident in the plots. 
For the most metal rich galaxies in the SN Ic host sample, the metallicities in the KD02comb calibration are on average 
$\sim 0.2$~dex higher than those in the PP04\_O3N2 calibration, which reinforced our statement that the comparison between samples 
should be performed with metallicities calculated in the same calibration. 

Amongst these four calibrations, the distribution in PP04\_O3N2 is the narrowest, i.e., it has the lowest standard deviation 
(0.26~dex for the combined sample of 50 hosts), even though the host galaxies are over a large range of masses and thus should have a large range of metallicities.
While \citet{Marino13} claimed that the M13\_O3N2 calibration (not shown) is superior to PP04\_O3N2, 
we think that the M13 calibration is not appropriate for the study of metal-poor galaxies such as those in our sample.
Together with the fact that the line ratios of several low metallicity SN-GRB hosts are outside of the range 
for the M13\_O3N2 calibration, the M13\_O3N2 calibration results in an even lower standard deviation than PP04\_O3N2 does 
(0.14~dex for the combined sample of 44 hosts), which is unrealistic considering that the host galaxies are over a large range of masses. 
The small difference in metallicity distributions between the SN Ic-bl and SN-GRB 
samples becomes completely indistinguishable if the M13\_O3N2 calibration is used.


The lower panels in Figure \ref{fig:OHdist} show CDFs of metallicities in the four calibrations, respectively. 
The upper limits of GRB130702A/SN2013dx in the KD02comb and M08\_N2H$\alpha$ calibrations
are denoted by arrows. 
The statistical tests account for the fact that data do not sample the full underlying populations, but 
do not include the metallicity uncertainties. 
We hereby derive the confidence bands for the CDFs from metallicity uncertainties via Monte Carlo simulations, 
which are shown by colored bands around the associated CDFs (central lines). 

For each CDF we compute two confidence intervals: one induced by the uncertainty in the line measurements as they propagate into the metallicity calculation, and the other as induced by the limited sample size. The former is an output of the {\it pyMCZ} code calculated via Monte Carlo as the 16th and 84th
percentiles of the metallicity distribution. The latter is computed by drop-1 (Jack Knife) bootstrapping over the SNe with the upper and lower intervals representing the standard deviation of the bootstrapped sample.



With these confidence intervals the differences between the SN Ic-bl and SN-GRB samples are not statistically significant for all metallicity calibrations, but those between the SN Ic and Ic-bl samples are. 
This is also indicated by our results from the statistical tests in section \ref{sec:ohtest}.

Finally, 
we now show that stellar absorption corrections will not change our results significantly, 
though we do not consider the second order effect on extinction corrections. 
While the stellar absorption correction is applied to all the SN Ic and Ic-bl hosts, this is the case for only four out of 10 SN-GRB hosts. 
Assuming typical corrections of 4\% for H$\alpha$ and 20\% for H$\beta$, 
which are the mean values derived from a sample of GRB hosts by \citet{Savaglio09}, 
only the calibrations that directly depend on H$\beta$ are highly affected by stellar absorption. 
Among the four calibrations that we choose, those are KD02comb (it depends on $R_{23}$ for $\log ({\rm O/H}) + 12 < 8.4$) and PP04\_O3N2. 
Given the line ratios and the lower branch relation between the metallicity and $\log R_{23}$ from \citet{Kobulnicky04}, 
applying stellar absorption correction would result in a decrease of $\lesssim$0.08~dex in $\log ({\rm O/H}) + 12$ 
for only one object, GRB161219B/SN2016jca. 
Considering the small difference between the CDFs of SN Ic-bl and SN-GRB hosts in the KD02comb calibration from Figure \ref{fig:OHdist}, 
a tiny shift of $\lesssim$0.08~dex to the left for only one object is unlikely to change our result. 
Likewise, if H$\alpha$ increases by 4\% and H$\beta$ flux increases by 20\%, the $\log O3N2$ value decreases by 0.06~dex, 
which corresponds to an increase of only 0.02~dex in $\log ({\rm O/H}) + 12$ \citep{Pettini04}. 
A tiny shift of 0.02~dex to the right for five 
SN-GRB hosts only slightly strengthens our result that 
the difference between the SN Ic-bl and SN-GRB hosts is not statistically significant. 
However, this probably explains the fact that the difference between the CDFs of SN Ic-bl and SN-GRB hosts 
is the most evident in the PP04\_O3N2 calibration among the four, although the PP04\_O3N2 calibration results 
in the smallest standard deviation in the metallicity distribution for each sample. 

{\it To conclude, the PTF SN Ic-bl host metallicity distribution is statistically consistent with that of the SN-GRB hosts, 
but inconsistent with the PTF SN Ic hosts, which are generally found at higher metallicities.} 
While no GRBs were observed in conjunction with the PTF SNe Ic-bl, we will discuss in section \ref{sec:off-axis-grbs} whether our null result could be due to the fact that the PTF SNe Ic-bl actually hosted off-axis GRBs or on-axis low-luminosity GRBs like GRB060218.


\subsection{Mass--Metallicity Relation}
\label{sec:mz}

\begin{figure*}[ht!]
\epsscale{1.2}
\plotone{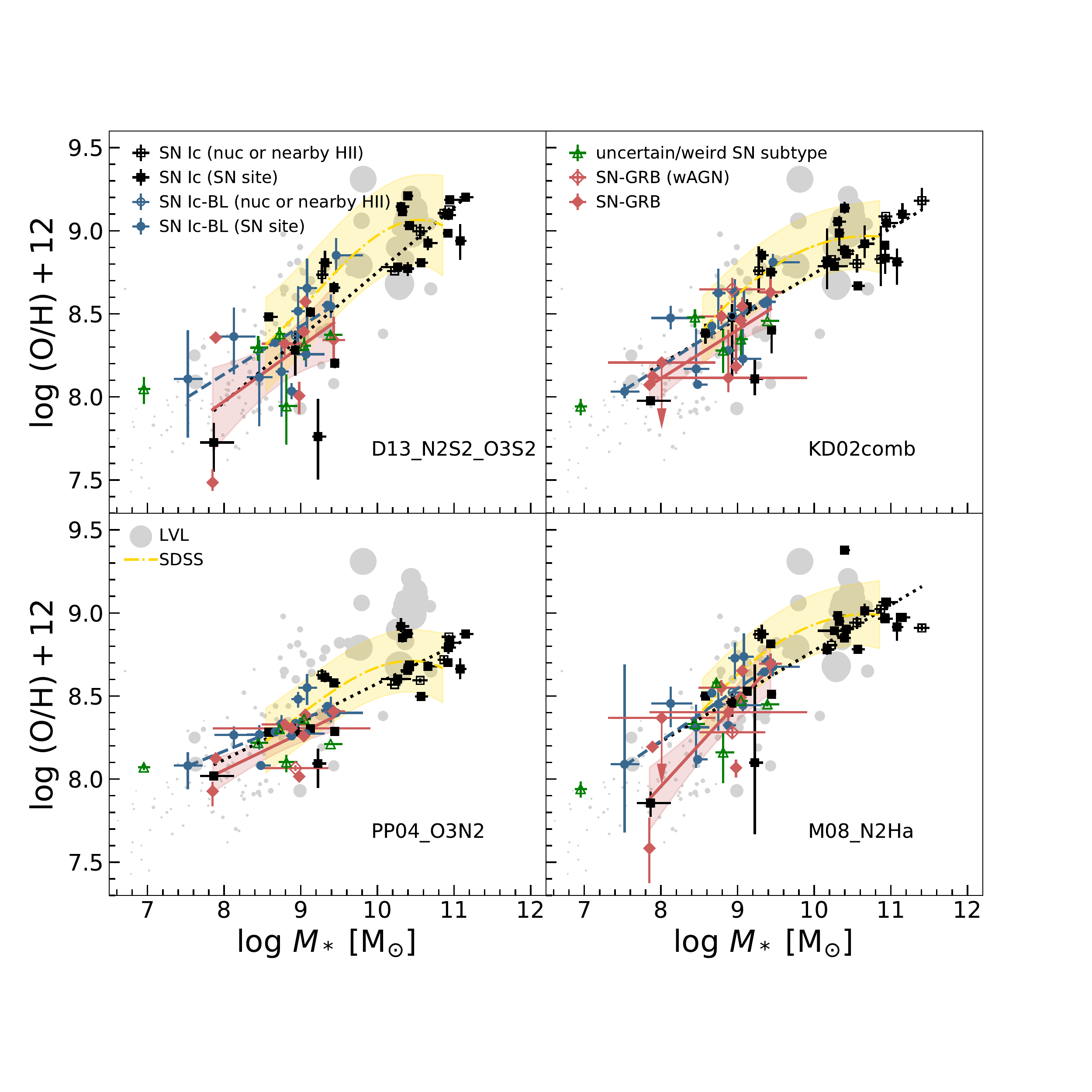}
\vspace{-1.8cm}
\caption{
 $M-Z$ relation for PTF SN Ic/Ic-bl hosts (black squares and blue circles respectively), compared to the SN-GRB hosts (red diamonds) and LVL galaxies (gray circles with symbol size proportional to SFR). 
Linear fits that characterize the average $M-Z$ relations for the PTF SN Ic, Ic-bl and SN-GRB hosts are denoted by 
black dotted lines, blue dashed lines, and red solid lines, respectively, and the corresponding 68\%  prediction intervals are denoted by the shaded regions for SN-GRB hosts. The average $M-Z$ relation for SDSS galaxies and its 68\% prediction interval is also shown, in yellow (see text on how the prediction interval was computed).
Metallicities of the PTF SN Ic/Ic-bl, SN-GRB hosts and the SDSS galaxies are calculated by {\it pyMCZ} for the four different calibrations across panels. 
Metallicities of the LVL galaxies are compiled from the literature in various calibrations and stay the same across panels. 
The metallicity of GRB130702A/SN2013dx can be calculated only as an upper limit and only for the KD02 combined and M08-N2H$\alpha$ scales, and is denoted by arrows pointing downwards. These upper limits are excluded from the linear fits, as are the hosts with potential AGN contamination (open red diamonds).
The PTF SN Ic-bl and SN-GRB tend to inhabit galaxies with low masses and metallicities, compared to those of PTF SN Ic. 
However, all the hosts generally follow the same $M-Z$ relation as discussed in the text. }
\label{fig:MZ}
\end{figure*}

As shown in section \ref{sec:ohdist}, the PTF SN Ic hosts have overall higher metallicities than the SN Ic-bl or SN-GRB hosts have.

Given the various galaxy correlations between metallicity, mass and sSFR, it is hard to pinpoint the physical driver for the preference of a specific SN type towards certain galaxies. Galaxy mass is a fundamental measurement and we use it to to elucidate whether the metallicity preference is intrinsic or a side-effect.


In this section, we take out this effect (i.e., that the SN Ic hosts are more massive) by deriving the $M-Z$ relations for the three SN host samples 
(SN Ic, Ic-bl, and SN-GRB) and assess if they are consistent with each other. We also compare these $M-Z$ relations 
with the ones derived for the SDSS or LVL galaxies, in order to answer the question whether the SN hosts fall 
significantly below the standard $M-Z$ relations of local galaxies. 
The $M-Z$ relation has been well-known and may be attributed to the larger neutral gas fractions and more efficient stripping 
of heavy elements by galactic winds in lower-mass galaxies \citep[e.g.,][]{Tremonti04}. 
If the SN hosts do not obey the $M-Z$ relation as defined by the standard local galaxies, 
they may have experienced a different regulation between the physical processes of metal enrichment 
(e.g., SN explosion) and metal depletion (e.g., SF and gas removal by feedback), 
which might suggest that they are a unique galaxy population. 

\subsubsection{$M-Z$ Relations for the SN Hosts}

In Figure \ref{fig:MZ}, we plot the metallicities in the four calibrations versus $M_*$ for the SN host galaxies. 
To quantify the average $M-Z$ relations, we fit a linear model to each SN host sample and show the results as lines in Figure \ref{fig:MZ}, 
which share the same color codes as the symbols. 

We are fully aware that the $M-Z$ relation is not perfectly linear. 
As a result, \citet{Kewley08} fit the $M-Z$ relation by a third-order polynomial, and \citet{Tremonti04} by a second-order one. 
Specifically, a gradual flattening occurs above $M_* \sim 10^{10.5} M_\odot$ \citep{Tremonti04}. 
Below $M_* \sim 10^{8.5} M_\odot$, the $M-Z$ relation is not constrained by the SDSS main spectroscopic galaxy sample, 
which becomes highly incomplete at the low mass end. 
However, the correlation is roughly linear from $10^{8.5}$ to $10^{10} M_\odot$ \citep{Tremonti04}. 
All the PTF SN Ic-bl and SN-GRB hosts have $M_* < 10^{9.5} M_\odot$ 
and thus are not affected by the obvious flattening observed over the range for massive galaxies. 
Together with the fact that the hosts from these two samples (PTF SN Ic-bl and SN-GRB hosts) span a similar $M_*$ range (from $10^{7.5}$ to $10^{9.5} M_\odot$), 
we expect the linear fits to provide a fair comparison of the average $M-Z$ trends between the two samples. 
We note that the linear fit to the PTF SN Ic hosts is less representative of the average $M-Z$ trend, 
because 
11 out of all 28 PTF SN Ic hosts have $M_* > 10^{10.5} M_\odot$. 
However, we still fit linear models to all three SN host samples to make fair comparisons. 
Plus, fitting higher order polynomials may 
fail to catch the intrinsic trends, given the small sample sizes of SN hosts. 

In all the calibrations, the linear fit to the $M-Z$ relation for the SN-GRB hosts falls slightly below that for the 
PTF SN Ic-bl hosts (red solid lines fall below blue dashed lines in Figure \ref{fig:MZ}). 
However, the relative position between the $M-Z$ relation 
for the PTF SN Ic hosts and those for the other two samples depends on the calibration (black dotted lines versus the others). 
Not only does the $y$-intercept of the $M-Z$ relation vary with the metallicity calibration, but so do its slope and even its shape \citep{Kewley08}. 
Due to the caveat of non-linearity in the intrinsic $M-Z$ relation over a large $M_*$ range, a linear fit 
to the $M-Z$ relation of PTF SN Ic hosts is not robust and varies with the calibration. 

To assess if these differences are significant between the $M-Z$ relations as obeyed by different SN host samples 
(with the emphasis of the difference between the SN Ic-bl and SN-GRB hosts, due to the caveat of a linear fit to the 
SN Ic $M-Z$ relation as discussed above), we fit a linear model 
for galaxies from each pair of the samples adjacent in the sequence of SN Ic, Ic-bl, and SN-GRB: 
\begin{equation}
\label{eq:mz1}
	\log ({\rm O/H}) + 12 = \beta_0 + \beta_1\log M_* + \beta_2 R + \beta_3 R  \log M_*, 
\end{equation} 
where $\beta_i$ ($i = 0, ..., 3$) are coefficients to be fit and $R$ is a variable that denotes the SN Ic-bl sample as the reference level: 
\begin{equation}
\label{eq:mz2}
	R = 
    \begin{cases} 
      0, & \text{if a galaxy belongs to the SN Ic-bl sample} \\
      1, & \text{otherwise.}
    \end{cases}
\end{equation} 
This multivariant model allows for different $y$-intercepts and slopes of the average trends as defined by different samples. 
For example, in the comparison between the SN Ic-bl and SN-GRB hosts, equation (\ref{eq:mz1}) reduces to 
\begin{equation}
\label{eq:mz3}
	\log ({\rm O/H}) + 12 = \beta_0 + \beta_1\log M_*, 
\end{equation} 
for the SN Ic-bl hosts (blue dashed lines in Figure \ref{fig:MZ}), and it reduces to 
\begin{equation}
\label{eq:mz3}
	\log ({\rm O/H}) + 12 = (\beta_0+\beta_2) + (\beta_1+\beta_3)\log M_*, 
\end{equation} 
for the SN-GRB hosts (red solid lines in Figure \ref{fig:MZ}). 
We exclude GRB031203/SN2003lw (with potential AGN contamination) and GRB130702A/SN2013dx 
(the metallicity value is an upper limit) from the fit. 

The estimates of the coefficients in equation (\ref{eq:mz1}) follow the Student's {\it t} distribution, and thus 
we apply Student's {\it t}-tests on the hypothesis that $\beta_2=0$ or $\beta_3=0$, 
which is equivalent to testing if the two average trends are the same as defined by different host samples. 
Between the SN Ic-bl and SN-GRB hosts, 
we find that the $p$-values of the {\it t}-tests on the hypothesis that $\beta_2=0$ or $\beta_3=0$ 
are much higher than 0.05 for all four calibrations, 
i.e., the difference between the average $M-Z$ relations as defined 
by the SN Ic-bl and SN-GRB hosts is not statistically significant. 
The tests on coefficients from the linear fits to the SN Ic-bl versus Ic hosts show similar results. 
Note that the statistical tests account for the fact that our data do not sample the full underlying populations, 
but do not include our data's measurement uncertainties. 
Considering the uncertainties, the average trends between the samples become even more indistinguishable. 

{\it To conclude, in all the calibrations, although the average $M-Z$ relation for the SN-GRB hosts lies slightly 
below that for the SN Ic-bl hosts, we find this difference to be not statistically significant. 
Similarly, the difference between the average $M-Z$ relations for the SN Ic-bl and Ic hosts is 
not statistically significant, while the range of SNe Ic hosts extends to higher metallicities and higher masses than the other two SN host samples, with the caveat in mind that the $M-Z$ relation becomes non-linear at this high masses. } 



\subsubsection{Comparisons with Local Galaxies}

\citet{modjaz08_grbz}, \citet{levesque10_grbhosts2_mz}, \citet{han10} and \citet{Graham13} argue that the hosts of SN-GRBs lie significantly below the 
standard luminosity--metallicity ($L-Z$) relations of local galaxies. 
In contrast, \citet{leloudas15} conclude only that the GRB hosts do not occupy the same region as the SDSS galaxies, 
which are more massive and more metal rich. 
\citet{leloudas15} suggest that the GRB hosts form a possible extension towards lower masses, 
as there is only a small number of low mass galaxies ($M_* < 10^{8.5}M_\odot$) in the SDSS. 
Similarly, \citet{Savaglio09} argued that GRB hosts lie on the same $M-Z$ relation as regular galaxies.
Now with a larger sample size, we examine here whether SN-GRBs occur in host galaxies below the standard $M-Z$ relation. 
Due to a positive correlation between the galaxy $M_*/L$ and optical color, 
the $M-Z$ relation is intrinsically tighter than the $L-Z$ relation, so that the former is more appropriate 
for this investigation. 

We first compare the SN host galaxies with 500 galaxies that are representative of the overall SDSS population (section \ref{sec:sdss}), 
in terms of the average $M-Z$ relations. 
We keep in mind that the SDSS main spectroscopic galaxy sample becomes highly incomplete at the low mass end. 
The average SDSS $M-Z$ relations are overlaid in Figure \ref{fig:MZ} by yellow dash-dotted lines extending between $10^{8.5} < M_* < 10^{11}M_\odot$, 
as a third-order polynomial fit to the SDSS galaxies. 
Indeed, the majority of the SN-GRB hosts with $M_* > 10^{8.5}M_\odot$ appear to fall below these average SDSS relations. 
Likewise, all three average $M-Z$ relations for the SN hosts (SN Ic in black dotted line, SN Ic-bl in blue dashed line, and 
SN-GRB in red solid line) fall below this SDSS curve above $10^{8.5} M_\odot$ in all four calibrations, 
with the caveat in mind of the non-linearity for SN Ic $M-Z$ relations. 
The confidence intervals plotted as shaded regions around the line fit represent the uncertainty on the line fit calculated as a two-sided 68\%confidence interval, i.e, for $\alpha = 0.32$: $\pm t_{1-\alpha/2, n-2} \times s $ where $t$ is the value of the $t$-distribution with n-2 degrees of freedom for the $\alpha$ confidence threshold,  and $s$ is the estimated standard deviation of the fit. 

However, we argue that the difference between the average SDSS and the average SN-GRB $M-Z$ relations, 
in particular, is not statistically significant, except for perhaps the metallicities in the KD02comb calibration. 
In Figure \ref{fig:MZ}, we overlay the 68\% confidence intervals 
of the linear fits to the SN-GRB $M-Z$ relations as red shaded bands. 
Due to small sample sizes of the SN-GRB hosts, as well as large scatter of data points around the average relations, 
the confidence intervals associated with the linear fits are generally wide at a fixed $M_*$. 
The yellow dash-dotted curves for the SDSS $M-Z$ relations therefore always fall within the confidence intervals of the SN-GRB $M-Z$ relations 
(i.e., the differences are not statistically significant at the 2$\sigma$ level), except for the KD02comb calibration 
(the yellow dash-dotted curve falls outside of the red region). 
We note that the KD02comb calibration is mildly effected by the discontinuity in the metallicity distribution around $\log ({\rm O/H}) + 12 \sim 8.4$. 
Furthermore, we do not consider the confidence intervals for the SDSS relations here, and 
the confidence intervals should become even wider if measurement uncertainties are included. 
Meanwhile, the average SDSS relation is derived from the metallicities measured within the fibers, 
which preferentially target the nuclear regions, i.e, regions with overall higher metallicities 
compared to the rest of the galaxy due to negative metallicity gradients, from metal-rich centers to metal-poor outskirts. 
Thus, the average SDSS relation should shift downwards if the
metallicities were to be measured at similar galactocentric radii as
those for the SN hosts, that is, be even closer to the SN-GRB $M-Z$
relation.  If we take into account all of these factors, our
conclusion is strengthened that the difference between the average
SDSS and the average SN-GRB $M-Z$ relation is not significant.

The above analysis, however, is a 1-dimensional test which answers the question:
given a mass value, is the $Z$ value of an SN-GRB host consistent with
that of an SDSS galaxy at the same mass on average? One could conversely
ask if the SDSS galaxies and the SN-GRB hosts occupy the $M-Z$ space
in a similar way - this would be a 2-dimensional test. 



To answer this question we performed a
\emph{cross-match} test \citep{rosenbaum05} which compares two
multivariate distributions by measuring distances between
observations. This is a non-parametric test. The distance
between points in the 2D space is measured and observations are paired
according to \emph{optimal non-bipartite matching}. Optimal
  non-bipartite marching algorithms find the set of matches that
  minimize the sum of distances based on a given distance matrix (for
  a review on optimal non-bipartite matching see \citealt{lu11}). In
the presence of observations from two samples, the test measures the
\emph{cross-match statistics}, which is the number of pairs containing
one observation for the first sample and one from the second (in this
case, the SDSS galaxies and SN-GRB hosts are the two samples). The
fraction of cross-matches has a well-defined probability distribution
if the two samples are identically distributed in that space. Thus a
given cross-match fraction is associated to a $p$-value for the null
hypothesis that the two samples are identically distributed. Since our SN-GRB hosts have lower masses (with a maximum mass of $10^{9.5} M_\odot$) than SDSS galaxies, we restrict the comparison to the same mass range of SN-GRB host and SDSS hosts, i.e. we impose a mass cut off at $10^{9.5} M_\odot$ for the SDSS galaxies. 


The cross-matching algorithm uses the Mahalanobis distance, which
takes into account the variance of each variable and the covariance
between samples by measuring the Euclidean distance of the
decorrelated standardized samples, as a distance metric to assign
pairs. This is necessary since the samples are obviously covariant with
positive $Z$ \emph{vs} $M$ relations
\citep{mahalanobis1936generalized}.

Since we have more SDSS galaxies than SN-GRB hosts, to minimize the
computational burden of the matching algorithm\footnote{Optimal
  non-bipartite matching algorithms perform naively as $O(N^4)$,
  though they can be optimized to $O(N^3)$, \emph{e.g.}
  \citealt{papadimitriou82}} we randomly select $N$ SDSS galaxies at
$M<10^{9.5} M_\odot$ for each Z diagnostic, where $N$ is the number of SN-GRB
hosts (which varies between 7 and 10 for different metallicity calibrations), and then
measure the cross-match statistics for this subset of SDSS galaxies and
the SN-GRB hosts, repeating this procedure 500 times. We take the mean over the 500 tests to be the value for the cross-match statistics for each $Z$ diagnostics.

The $p$-value associated with the null hypothesis is: 0.028 for
KD02comb, 0.23 for D13\_N2S2\_O3S2, 0.16 for PP04\_O3N2, and 0.07 for
M08\_N2H$\alpha$. Three out of the four values are above the 0.05 threshold,
and thus the two samples, SN-GRB hosts and the SDSS galaxies, are consistent with each other at the $2-\sigma$ level based on this second test.

Unfortunately, the power of the test is limited by the small number of
objects in the SN-GRB sample, which varies between seven and ten
depending on the metallicity diagnostic. The power of the test would
be much increased with a larger SN-GRB sample, since with such a small
number of objects, the $p$-value is high for the smallest
possible number of cross-matches. For example, D13\_N2S2\_O3S2 only
has 7 SN-GRB hosts with measured $Z$, and for samples of $N ~=~7$ the
minimum number of cross-matches, 1 cross-match, already has a probability of 0.08, which
is above the $2-\sigma$ threshold.


We are aware that the average SN-GRB $M-Z$ relations are not well constrained. 
The small sample size of SN-GRB hosts may weaken the power of our test 
to distinguish the small difference between the SDSS and SN-GRB $M-Z$ relations.

We also compare the SN hosts with the LVL galaxies, which are represented by light gray circles 
in Figure \ref{fig:MZ} with symbol sizes proportional to SFRs, following \citealt{Perley16b}. The LVL survey is expected to provide a volume complete sample 
in the local universe. 
However, it suffers from cosmic variance and the metallicities for LVL galaxies are compiled from the literature in heterogeneous calibrations, 
which are typically based on electron temperature $T_e$ at low metallicities and on various strong-line diagnostics at higher metallicities \citep{Perley16b}. 
As a result, the $M-Z$ relation as defined by the LVL galaxies serves the purpose of only a qualitative comparison. 
Qualitatively, the data points for the LVL galaxies, whose metallicity values stay constant across panels, 
occupy a similar parameter space as the overall SN hosts do. 
In the low mass regime, the LVL galaxies provide nice supplement to the SDSS galaxies. 
The SN Ic-bl or SN-GRB hosts give no strong indication of an offset relative to the LVL galaxies. 
At the high mass end, the offset between the LVL galaxies and the SN Ic hosts is also inconclusive, 
given that the LVL survey is biased against the most massive galaxies, and that the metallicity values are 
in heterogeneous calibrations. 
Only in the PP04\_O3N2 calibration do LVL galaxies have systematically higher metallicities at a fixed $M_*$ 
than the SN Ic hosts. 

In conclusion, {\it the average $M-Z$ relations as defined by the SN Ic, Ic-bl, and SN-GRB hosts fall slightly below the same relations as defined by the SDSS galaxies above $M_* = 10^{8.5} M_\odot$. 
However, the differences between these relations are not statistically significant at the 2 $\sigma$ level. 
}
In contrast to \citet{levesque10_grbhosts2_mz}, we do not find a significant offset between SN-GRB host galaxies and those in SDSS, 
having used a larger dataset (by a factor of two larger for SN-GRBs) and a more thorough statistical analysis. When comparing in detail the five SN-GRB hosts we have in common, we find that not only are our metallicities different (though we are using different scales), but also our host masses, which we took from \citet{Savaglio09}. Indeed, \citet{levesque10_grbhosts2_mz} use a different code to compute host masses than \citet{Savaglio09} and their error bars for the host masses are larger than in \citet{Savaglio09}, such that for most SN-GRBs in common (4 out of the 5), the host masses are formally consistent with each other between the two works within one standard deviation.

\subsection{$M_*$ and SFR}
\label{sec:sfsq}

\begin{figure*}[ht!]
\epsscale{1.25}
\plotone{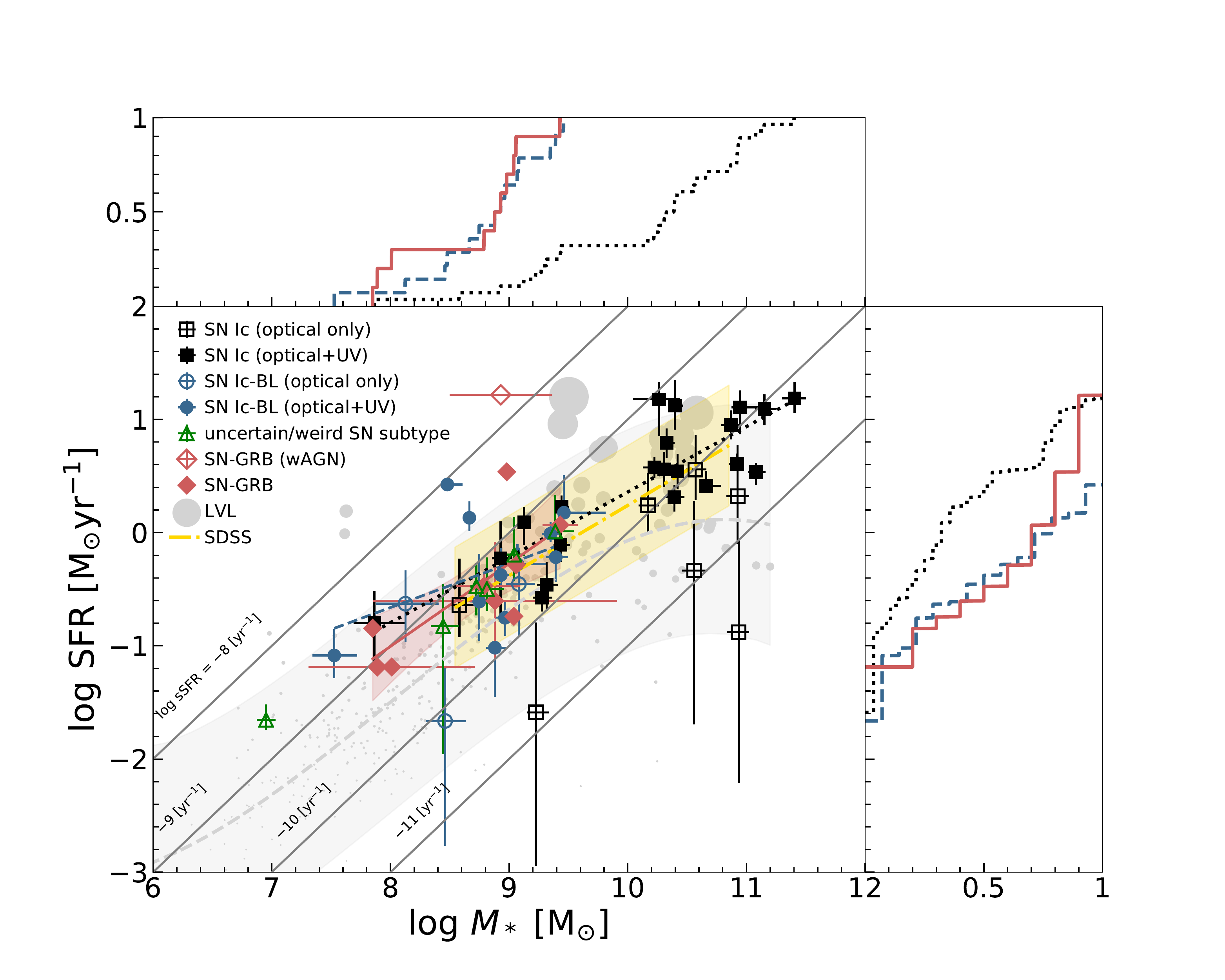}
\caption{
SFRs vs. $M_*$s of the PTF SN Ic/Ic-bl host galaxies compared to the SN-GRB hosts (symbols as in \ref{fig:BPT} and \ref{fig:MZ}) and local galaxies from the LVL (gray circles along with an average trend  denoted by a dashed line in gray, and with the corresponding prediction interval by the gray shaded area).
The SDSS average trend is denoted by the yellow dash-dotted line and the corresponding 68\% prediction interval by the shaded region.  
In this plot, open symbols denote the hosts with less reliable SFR estimates (see text). 
As in \ref{fig:MZ}, linear fits to the PTF SN Ic, Ic-bl and SN-GRB hosts are denoted by black dotted lines, blue dashed lines, and red solid lines, respectively, 
with a red shaded area that denotes the confidence interval for only the SN-GRB fit. 
Gray diagonals lines indicate lines of constant sSFR. 
Most of the SN host galaxies lie along the main sequence of star formation, i.e., falling in between the diagonal lines with sSFR of 10$^{-10}$ and 10$^{-9}$ yr$^{-1}$. 
Only one of the SN-GRB is found in a starburst galaxy with sSFR above 10$^{-8}$ yr$^{-1}$, but is potentially contaminated by AGN activity. 
The top and right panels show cumulative distributions of $M_*$s and SFRs, respectively, for the hosts of PTF SNe Ic (black dotted line), 
Ic-bl (blue dashed line), and SN-GRBs (red solid line). 
The PTF SN Ic-bl and SN-GRB tend to inhabit similar galaxies with low absolute levels of $M_*$s and SFRs, compared to PTF SN Ic. 
However, all hosts generally follow the same main sequence of star formation. 
}
\label{fig:SFR}
\end{figure*}

SNe Ic, SNe /Ic-bl and SN-GRBs mark the deaths of massive stars with short lifetimes and thus 
they are generally found in galaxies with active star formation. In this section, we compare the global SFRs of the 
PTF SN Ic/Ic-bl with those of the SN-GRB host galaxies, to see if they show preferences to galaxies 
with very different star formation levels. 
Moreover, the SFR is strongly correlated with the $M_*$: the star forming galaxies form a sequence on the 
SFR vs. $M_*$ diagram \citep[e.g.][]{Salim07}, known as the main sequence of star formation. 
In addition to the absolute levels of SFRs, we compare the relative positions of PTF SN Ic/Ic-bl and SN-GRB 
host galaxies to local star forming galaxies from the SDSS and LVL surveys, 
in order to quantify the relative effect of enhanced star formation. 

Figure \ref{fig:SFR} shows the PTF SN Ic/Ic-bl and SN-GRB host galaxies on and off this main sequence of star formation. 
The diagonal lines mark constant sSFRs. 
The majority of the PTF SN Ic/Ic-bl and SN-GRB host galaxies lie within the band 
between $10^{-10} < {\rm sSFR} < 10^{-9}~{\rm yr}^{-1}$, which roughly coincides with the main sequence of star formation. 
For the PTF SN Ic and Ic-bl hosts, open symbols denote that the SFRs are derived from SED fittings without {\it GALEX} bands. 
For the SN-GRB hosts, the SFRs are all derived from H$\alpha$ luminosity and open symbols denote the objects with potential AGNs. 
In both cases, the open symbols represent objects with unreliable SFR estimates. 
In fact, all the extreme outliers from the sequence are open symbols: 
above the sequence, only the host of GRB031203/SN2003lw with a potential AGN has ${\rm sSFR} > 10^{-8}~{\rm yr}^{-1}$; 
far below the sequence, all the hosts that are found in non-star-forming galaxies are likely due to poor SFR estimates. 
Among the solid cases, there are three moderate outliers above the sequence (starbursts with high sSFRs by definition): 
two SN Ic-bl hosts (PTF~10xem and 11qcj, both of their SDSS spectra are indeed classified as starbursts by the MPA-JHU catalog) 
and one SN-GRB (XRF020903). 
In general, the PTF SN Ic/Ic-bl and SN-GRB host galaxies follow the main sequence of star formation, 
with a few moderate outliers as starbursts above the sequence. 

We now compare the absolute levels of SFRs between the SN host samples. 
The side panels show the CDFs of three samples (PTF SNe Ic, Ic-bl, and SN-GRBs) for the $M_*$ and SFR, respectively. 
The open symbols are included in these CDFs, but the basic conclusions that we draw from the hypothesis tests on the CDFs 
are the same even if the open symbols are excluded. 
The overall $M_*$  for the three samples seem to follow by eye the order of 
$M_{*\rm GRB} \simeq M_{*\rm Ic-bl} < M_{*\rm Ic}$, and the overall SFRs follow the same order of 
${\rm SFR_{GRB} \simeq SFR_{Ic-bl} < SFR_{Ic}}$; however, these trends have not yet been statistically tested for actual significance. 
All the hypothesis tests that we have applied (the K--S test, the Wilcoxon rank sum test, and the Anderson Darling test) 
show the same results for the distributions of both $M_*$s and SFRs: 
the difference between the SN-GRB and SN Ic-bl hosts is not statistically significant, 
but it is between the SN Ic-bl and Ic hosts, assuming a significance level, $\alpha = 0.05$. 
The SNe Ic occur in galaxies with a large range of $M_*$s and SFRs, 
whereas the SNe Ic-bl and SN-GRBs only reside in galaxies with low $M_*$s and low absolute levels of SFRs. 
However, the bulk of star formation is contributed by more massive galaxies with higher SFRs in the local universe \citep[e.g.][]{Blanton05}. 
Indeed, the GRB hosts do not form a representative subset of all star forming galaxies, 
and thus it is unlikely that GRBs are unbiased tracers of the overall star formation in the local universe 
\citep[e.g.,][]{Stanek06, Graham13, Vergani15, chen17}, at least out to the redshifts we are probing here ($z<0.3$). 

We next investigate the relative enhancement of star formation between samples by comparing the SFRs of objects with similar $M_*$s. 
Specifically, we take out the effect of increasing SFRs with $M_*$s by fitting an SFR vs. $M_*$ relation to each sample. 
All the open symbols (
less/un-reliable values) are excluded from this analysis. 
Note that all the extreme outliers that may affect the fit are open symbols.
The comparison between the fitted lines in Figure \ref{fig:SFR} shows that the SN Ic-bl relation (blue dashed line) 
lies slightly above the SN Ic relation (black dotted line), and that the SN Ic relation lies slightly above the SN-GRB relation (red solid line), 
over most of the $M_*$ ranges in common. 
The GRB hosts are well known by their vigorous star formation activities, 
and we find that the PTF SN Ic/Ic-bl hosts experience even slightly enhanced star formation relative to the SN-GRB hosts. 
However, as for the comparison between the $M-Z$ relations in section \ref{sec:mz}, 
the differences between the SFR vs. $M_*$ relations of the hosts of different kinds of SNe are not statistically significant. 

\citet{Kelly14} found that host galaxies of SNe Ic-bl are not substantially more strongly star forming for their $M_*$ than 
other core-collapse host galaxies and the SDSS star-forming population. 
We also compare the PTF SN Ic/Ic-bl and SN-GRB host galaxies with galaxies from the SDSS. 
We derive the average SDSS trend in Figure \ref{fig:SFR} (yellow dash-dotted line) 
from a representative subset of all the SDSS star forming galaxies (see section \ref{sec:sdss}). 
With the MPA-JHU SFRs and $M_*$s that we adopt for this subset, we can reproduce the average sSFR vs. $M_*$ relation 
in \citet{Salim07} (UV-based SFRs in that work), 
as a benchmark of the local star forming sequence defined by all the pure star-forming galaxies from the SDSS (with no AGN contribution). 
Over the $M_*$ ranges in common, all three average SFR--$M_*$ relations of the SN host galaxies 
lie slightly above the same relation of the SDSS galaxies, i.e., the SN host galaxies on average experience somewhat more enhanced star formation 
relative to the star forming galaxies from the SDSS. 
However, the SN-GRB SFR--$M_*$ relation (red shaded band) still falls within the 68\% confidence interval of the SDSS SFR--$M_*$ relation (yellow shaded band), 
i.e., the difference in the average relations is not statistically significant. 
Similarly, most of the SN host galaxies fall within the 68\% prediction interval of the SDSS SFR--$M_*$ relation, 
i.e., we cannot rule out the hypothesis that the SN host galaxies are drawn from the same underlying population as the SDSS 
star-forming galaxies in such an SFR vs. $M_*$ diagram. 

Finally, we compare the PTF SN Ic/Ic-bl and SN-GRB host galaxies with galaxies from the LVL survey. 
Unlike the SDSS sample, which contains only the star forming galaxies, 
the LVL sample is inclusive of all the local galaxies within 11~Mpc, including also the non-star-forming ones that are far below the main sequence. 
As expected, the LVL SFR--$M_*$ relation (gray dashed line, with the 68\% prediction interval as a gray shaded band)  
is below the same relations of all the other samples. 
In particular, the LVL SFR--$M_*$ relation falls below the same relation of the SN-GRB hosts: 
the star formation of the SN-GRB hosts is enhanced relative to the local galaxies from the LVL survey with similar masses. 
However, again, this offset is not statistically significant, given the large confidence intervals of the SFR--$M_*$ relations for both 
the SN-GRB and LVL samples. 

In summary, {\it in terms of the absolute levels of SFRs and $M_*$s, the SN-GRB and PTF SN Ic-bl hosts are comparable, 
and they are on average below those of the PTF SN Ic hosts. 
However, in terms of the relative enhancement of star formation activity as gauged by the SFR vs. $M_*$ relation 
(preference to the starbursts), 
all three SN host samples have similar SFR vs. $M_*$ relations that are only slightly above the same relation as defined by 
the star-forming galaxies from the SDSS. }

\subsection{High sSFR vs. Low Metallicity: Which one is driving SN-GRB and SN Ic-bl production?}
\label{sec:degeneracy}

\begin{figure*}[ht!]
\epsscale{1.2}
\plotone{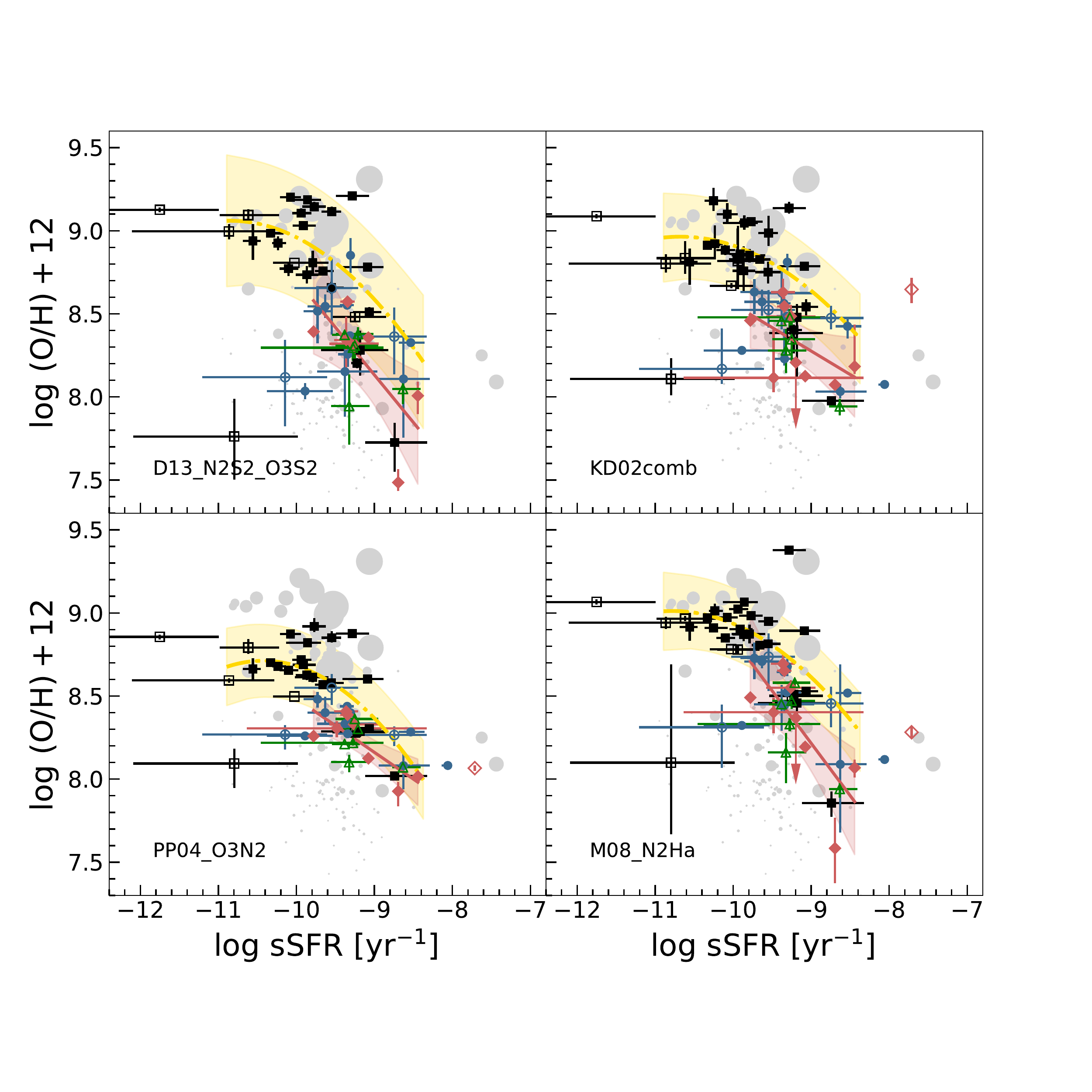}
\vspace{-1.8cm}
\caption{
Specific SFR (sSFR) vs. metallicity of the PTF SN Ic and PTF Ic-bl host galaxies compared to the SN-GRB hosts, and local galaxies from the SDSS and LVL survey. 
Definitions of the symbols, lines and color shaded areas are the same as those in Figures \ref{fig:BPT} and \ref{fig:MZ}, and \ref{fig:SFR}. As in Figure \ref{fig:MZ}, the metallicity of GRB130702A/SN2013dx is an upper limit, indicated by an arrow.
High absolute sSFRs and low metallicities are two features both present in the host samples of the 
SN-GRBs and PTF SNe Ic-bl. However, as gauged by a comparison with the SDSS galaxies, low metallicity is likely to be a fundamental cause for the formation of SN-GRB progenitors, whereas high sSFR is only a consequence of the low metallicity (see text).
}
\label{fig:sSFR}
\end{figure*}

In section \ref{sec:mz}, we showed that the PTF SN Ic/Ic-bl and SN-GRB hosts 
have metallicities that are on average lower than typical star-forming galaxies of the same $M_*$s. 
In section \ref{sec:sfsq}, we showed that the PTF SN Ic/Ic-bl and SN-GRB hosts 
have SFRs that are on average slightly higher than typical star-forming galaxies of the same $M_*$s. 
To explain these two effects, \citet{Mannucci11} argued that the apparent GRB preference 
for low-metallicity hosts is due to a more fundamental mass, metallicity, and SFR relation. 
This proposed relationship, which is an extension of the well-known $M-Z$ relation, claims that
the metallicity of a galaxy of a given stellar mass is anti-correlated with its SFR \citep{Mannucci10}, a relationship that is highly debated in the field, with some works confirming it (e.g., \citealt{sanders18,hirschauer18}) while others are not able to reproduce it (e.g., \citealt{kashino16,sanchez18}.

In any case, \citet{Mannucci11} and \citet{Kocevski11} suggest that 
the low metallicities of GRB hosts may not be intrinsic to their formation, but rather a consequence 
of the preference for starburst galaxies.
Meanwhile, these preferences could have the same physical origin as the preference towards low-mass galaxies: 
low-mass galaxies tend to have low metallicities and high sSFRs in the general galaxy population. 

In this section, we compare these two competing factors: low metallicity and high sSFR, in order to answer the question of 
which one is the key requirement for the formation of SN-GRB and SN Ic-bl progenitors. 
In this comparison, the sSFR is chosen over SFR to characterize the star formation activity in the host galaxies, 
because the former parameter should be more informative for the local environments of SN progenitors. 
Like the global $M_*$ of a galaxy is a summation of the $M_*$s from all local regions, 
the global SFR of a galaxy is a summation of the SFRs from all local star forming regions, 
i.e., the $M_*$ and SFR are expected to scale up with a galaxy size. 
However, a local star-forming region in principle cannot know about the global $M_*$ or SFR. 
In contrast, like metallicity, the global sSFR of a galaxy is a luminosity-weighted average of the 
same values from all local star forming regions. 
In terms of a more physical property that may be related to the local environment for the progenitor formation, 
a high sSFR could, for example, indicate a young stellar population, which could therefore contain very massive stars capable 
of forming SN-GRBs and SNe Ic-bl \citep{chen17}. 
Alternatively, a high sSFR could be interpreted as evidence for a progenitor favored by an altered (e.g., top-heavy) IMF 
or dense clusters with abundant dynamical interactions \citep{Perley16b}.

The frequent observation of very high specific star-formation rates
within the population of host galaxies for a given supernova type
necessarily implies a progenitor with a short delay time, since it is
unlikely to be older than the ongoing starburst.  However, if a short
delay time is the only feature that distinguishes the progenitor of a
rare SN type from the broader population of core-collapse explosions,
this cannot explain the rarity of explosion in massive or more modestly
star-forming hosts.  Massive galaxies form plenty of stars outside of
the starburst phase, and our own Local Group is not lacking in quite
massive stars even though the Milky Way and LMC have unremarkable
specific star-formation rates.  Therefore while the occasional presence
of SNe in very starburst galaxies, i.e., those with large sSFRs, can be seen as an indicator of a
young progenitor age \citep{leloudas15,schulze18}, other factors are necessary to explain a high
frequency of such occurrences.  Metallicity could indirectly increase
this frequency, since metal-poor galaxies are typically dwarfs with
burstier star-formation histories \citep{Lee09} likely to form more stars during burst
phases.  If low-metallicity starbursts produce an excess of massive stars (i.e., top-heavy IMF, \citealt{crowther10,schneider18}),
or of close binaries, this could be even more effective in increasing
the rate.

Figure \ref{fig:sSFR} show a diagram of sSFR vs. metallicity in the four calibrations that we choose. 
Indeed, all data points from the PTF SN Ic/Ic-bl and SN-GRB hosts demonstrate a weak anti-correlation 
between the sSFR and metallicity. Relative to the PTF SN Ic hosts, the SN-GRB and PTF SN Ic-bl hosts 
show a preference to the upper left portion in all the panels that is associated with both low metallicities and high sSFRs. 
We test if the preference to high sSFRs is statistically significant. 
Similar to the hypothesis test results on the differences between the metallicity distributions for the three samples, 
the sSFR distributions of the SN-GRB and PTF SN Ic-bl hosts are comparable ($p>0.05$), 
but the PTF SN Ic-bl hosts have significantly higher sSFRs than the PTF SN Ic hosts ($p<0.05$). 
The same results hold regardless of whether we include the objects with less reliable SFR estimates in the comparison, 
for all the  tests that we have applied (the K--S test, the Wilcoxon rank sum test, and the Anderson Darling test). 


As the next step, we will break the degeneracy between low metallicity and high sSFR in order to disentangle the needed conditions 
for the formation of SN-GRB and SN Ic-bl progenitors.
In order to address this issue, we compare the hosts of SN-GRBs and PTF SNe Ic-bl with the star forming galaxies 
from the SDSS on the same diagram. 
In this diagram, the two competing factors - low metallicity and high sSFR - have opposite effects on 
the sSFR vs. metallicity relation of the SN host galaxies with respect to the same relation of the SDSS galaxies. 
The SDSS galaxies set the baseline for this comparison. 
If the sSFR vs. metallicity relation of the SN host galaxies is below the same relation of the SDSS galaxies, i.e, skewed towards lower metallicities, 
then a low metallicity is likely the dominant effect.  
Otherwise, if the sSFR vs. metallicity relation of the SN host galaxies is above that of the SDSS galaxies, 
it is likely that a high sSFR is the dominant effect. 

In Figure \ref{fig:sSFR}, over the sSFR range in common and across all the four metallicity calibrations, 
the sSFR vs. metallicity relation of the SN-GRB hosts (red solid line) lies well below the same relation of the SDSS galaxies (yellow dash dotted line). 
 This comparison demonstrates that the SN hosts are biased towards low metallicities even if we take the excess sSFRs into account. 
 In addition, if it were merely a sSFR effect such that the SN-GRB host galaxies just produced more massive stars for their galaxy mass than comparisons samples like the SDSS galaxies, as suggested by \citet{Savaglio15}, then the SN Ic hosts should also exhibit high sSFRs, since SNe Ic are also explosions of massive stars, like SN-GRBs. However, this is not what we observe $-$ - rather, PTF SN Ic hosts have lower sSFRs than than those of PTF SNe Ic-bl and of SN-GRBS and more in-line of those of SDSS galaxies. Therefore, low metallicity is likely to be the fundamental cause for the formation of SN-GRB progenitors, whereas a high sSFR is only a consequence of the low metallicity. 
 We note that the hosts with unreliable SFR estimates (open symbols) or metallicities as upper limits (arrows pointing to the left) are excluded from the fits.


The result of the same analysis but with the LVL galaxies as a baseline is ambiguous. 
The anti-correlation between sSFRs and metallicities appears to be much weaker among the LVL galaxies, 
which may be partly due to the fact that the metallicities of the LVL galaxies are in heterogeneous calibrations. 

 {\it In summary, we show that the hosts of SN-GRBs and of PTF SNe Ic-bl both exhibit high absolute sSFRs and low metallicities. In addition, we are able to break the degeneracy between the two competing factors by comparing our samples to the SDSS galaxies and to PTF SN Ic host galaxies. We find that low metallicity is likely to be 
the fundamental cause for the formation of SN-GRB progenitors, whereas high sSFR is only a consequence of low metallicity.}

\subsection{Hosts of Weird Transients and Transients with Uncertain SN Subtype}

As discussed in section \ref{sec:sample}, we consider six weird transients and transients with uncertain SN subtype, 
in addition to the PTF SNe Ic/Ic-bl with clean IDs. 
These transients are included because they are all potential SNe Ic or Ic-bl, 
or SN Ic with peculiar behaviors (PTF~12gzk). We follow the same approach as for the PTF SNe Ic/Ic-bl with clean IDs 
to observe these hosts (Table \ref{tab:obs}) and to derive their host properties, including line fluxes (Table \ref{tab:emission}), 
metallicites (Table \ref{tab:PTFZ}), $M_*$s and SFRs (Table \ref{tab:SED}). 

They are shown in all the plots as green open triangles. They do not form a unified class of SN hosts and thus 
we do not summarize their distributions by CDFs or linear fits as we do for the other SN host samples. 
However, We note here that they generally occupy similar parameter spaces in all the plots as the SN-GRB and PTF SN Ic-bl hosts do: 
the hosts of weird transients and transients with uncertain SN subtype have overall low absolute levels of metallicities, $M_*$, and SFRs. 
In particular, the host of PTF~12gzk is the least luminous \citep{benami12} and least massive among all the SN hosts ($M_* = 10^{6.95}~M_\odot$). 
In terms of the 2D distributions in the $M-Z$ (Figure \ref{fig:MZ}), SFR vs. $M_*$ (Figure \ref{fig:SFR}), and 
sSFR vs. metallicity diagrams (Figure \ref{fig:sSFR}), the hosts of weird transients and transients with uncertain SN subtype are not outliers 
that significantly deviate from the average relations as defined by the SN-GRB or PTF SN Ic-bl hosts. 
Therefore, we do not expect qualitative changes to our analysis results on the PTF SN Ic-bl sample 
by including the addition SN Ic-bl candidates from these transients. 
Meanwhile, low $M_*$ hosts with low SFRs and low metallicities already exist in the sample of PTF SN Ic hosts with clean IDs. 
We also do not expect qualitative changes to our analysis results of the PTF SN Ic sample by including a few more SN Ic candidates 
with similarly low $M_*$s, SFRs, and metallicities. 

In conclusion, these hosts of weird transients and transients with uncertain SN subtype exhibit similar distributions as the SN-GRB or PTF SN Ic-bl hosts in the various diagrams of host galaxy properties. 
They do not change our conclusions about sample comparisons.

\section{Discussion: Implication for Jet-production and SN-GRB progenitor models $-$ A Unified SN-GRB, SN Ic-bl and SN Ic picture} 
\label{sec:discussion}

In the previous section \ref{sec:result} we found that while SN-GRBs prefer somewhat more metal-poor environments than PTF SNe Ic-bl, this preference is not statistically significant. Our finding that SN-GRBs and SN Ic-bl thus inhabit statistically similar environments could be due to the fact that indeed both types of explosions had the same low-metallicity progenitors which were able to produce a GRB - and that was not detected for the cases of SNe Ic-bl $-$ whether those GRBs be off-axis, choked, low-luminosity, missed entirely, or otherwise. Here off-axis GRBs refer to GRBs that produced gamma-ray emission which was not observed on earth, due to the jet axis not being aligned with our line-of-sight, while their more spherical SN component was observed (though SN Ic-bl themselves, as well as CCSNe more generally, are not fully spherical, e.g., \citealt{leonard06, modjaz08_doubleoxy,mazzali01}). Indeed, since cosmological GRBs are highly collimated events with beaming angles of $\sim$10 degrees \citep{Frail01, Ryan2015, racusin09, zhang15}, a large number of GRBs should be viewed off-axis (though see  \citet{Ryan2015} for the claim that even the GRBs are seen slightly off-axis). We discuss the evidence whether the PTF SNe Ic-bl in our sample harbored such off-axis GRBs below, in subsection~\ref{sec:off-axis-grbs} .

Indeed \cite{moesta15} showed that the ingredients for jet production, namely large-scale dynamo and strong MHD turbulence, are set up in rapidly rotating stars, which would more preferably occur in low-metallicity environments via a number of different mechanisms that include binaries (e.g.,\citealt{levan16}). Once the jet is produced and travels through the star, it leaves an observable imprint on the stellar envelope set into motion in form of broad lines in the SN spectrum, indicating significant amount of mass accelerated to high velocities, as shown for the first time in \citealt{Barnes18}, who perform an end-to-end simulation of SN-GRBs via state-of-the-art hydrodynamic and spectral synthesis simulations. Thus broad lines would be an indicator of a central engine. 
 
 Broad lines would presumably be produced even when the jet becomes choked and does not fully reach the surface of the star - however, one may expect lower SN velocities for chocked jet-engines. Indeed, \citet{modjaz16} showed that SNe Ic-bl without observed GRBs (including three of the PTF SN Ic-bl in our host sample) have generally lower velocities and less broadened lines than SNe Ic-bl with GRBs, but higher values and more broadening than SNe Ic. On the other hand, the spectra of SNe Ic-bl without observed GRBs may also show lower velocities than those of SN-GRBs because the GRB was not pointing towards earth - future work that extends the pioneering work of \citet{Barnes18} is needed to disentangle the effects of a choked jet from an off-axis GRB on the SN spectrum.


The natural implication of the work by \citet{Barnes18} is that the engine of SNe Ic-bl both with and without GRBS may be similar, but very different from those of their SNe Ic cousins. Indeed, we show here that PTF SNe Ic prefer environments that have higher metallicities than SNe Ic-bl and SN-GRBs, that is, SNe Ic have environments probably not conducive for jet production. Thus our results are in strong tension with the suggestion that \textbf{all} SNe Ib/c produced jets that were choked \citep{sobacchi17}.

 
While we break the degeneracy between high sSFR and low metallicity as the driver for SN Ic-bl production by favoring low metallicity (Section~\ref{sec:degeneracy}), it may not be the only necessary ingredient for producing SN Ic-bl. \cite{Kelly14} showed that a (small) sample of SN Ic-bl from a melange of untargeted surveys and low-$z$ GRB have other host properties in common, namely much higher SFR- and mass-densities compared to SDSS galaxies. They suggest that dense star forming regions may be preferentially producing massive binary systems that give rise to SNe Ic-bl with and without GRBs. The SN Ic hosts in their sample seem to have more typical densities, similar to SDSS galaxies, supporting the notion that not all subtypes of Stripped SNe live and die in the same environments as GRBs and thus probably did not all produce jets.

Lastly, our finding that the mass-metallicity relations as defined by SN-GRBs are not significantly different from the same relations as defined by the SDSS galaxies (in most of our considered metallicity scales and in most of our tests) contradicts earlier works based on smaller samples that used a metallicity scale which is probably not appropriate for the comparison. Thus, our finding resolves the dilemma of why SN-GRBs would have chosen a completely different galaxy population rarely found in the local universe of star forming galaxies - because they do not.


 



\subsection{Did the PTF SN Ic-bl Harbor Off-axis GRBs?}
\label{sec:off-axis-grbs}
 
 Here we discuss whether the PTF SNe Ic-bl could have harbored off-axis GRBs. A common diagnostic for the presence of off-axis GRBs is the detection of late-time radio emission from the relativistic ejecta that has decelerated over time and is emitting more isotropically. Such an extensive radio search was conducted by \citep{Corsi16} who observed a total of 15 broad-lined SNe Ic from PTF and iPTF, including $\sim$ 2/3 of the SN Ic-bl in our sample (namely the 9 SNe Ic-bl out of 14 SNe Ic-bl in our sample: 10bzf, 10qts, 10xem, 10aavz, 11cmh, 11img, 11lbm, 11qcj, 12as) at late times with the VLA.

For three out of the 15 SNe in Corsis's sample they detected radio emission, which they cannot securely rule out as due to off-axis low-luminosity GRBs expanding into a constant-density medium ISM with $n_{ISM}\sim$ 10. Two of those SNe are in our sample  (PTF11cmh and 11qcj) and even for the non-detections, the radio limits are not deep enough to exclude on-axis GRBs similar to the radio-quiet and low-energy GRB060218 and off-axis GRBs similar to the low-energy, but radio-loud GRB 980425 or high-energy off-axis GRBs such as GRB031203 expanding into a low-density environment.


 
Thus we conclude that our finding that SN-GRBs and PTF SNe Ic-bl prefer the same low-metallicity environment could be simply due to the fact that the PTF SNe Ic-bl also hosted GRBs, but whose jetted gamma-ray emission was not seen by us because of the viewing angle effects --  with two distinct possibilities of why they were not observed in the radio:  1) the GRBs were of low radio luminosity - and indeed those may be the most common kinds of GRBs (e. g., \citealt{soderberg06_06aj}) 2)  the GRBs were of high luminosity but were expanding into a low-density medium.

\section{Caveats and Future Work}
\label{sec:caveats}

\subsection{Sample Bias}

Samples of galaxies are biased by the survey techniques that define them, with further impact from additional selection criteria 
that are applied. In this section, we discuss the impacts of survey and selection biases on our conclusions, for both the SDSS and PTF SN samples. 

The SDSS legacy galaxy redshift sample has an apparent {\it r}-band magnitude limit of 17.77~mag. 
It misses a larger fraction of galaxies with low luminosities at a higher redshift (below the yellow dash-dotted line in Figure \ref{fig:LD}). 
To first order, this incompleteness is a function of $M_*$, and thus it has mild impact on our studies of the $M-Z$ (Figure \ref{fig:MZ}) 
or SFR vs. $M_*$ (Figure \ref{fig:SFR}) relations, which are both functions of $M_*$. 
Considering that metallicities are strongly correlated with $M_*$, the impact of incompleteness on 
our analysis of the sSFR vs. metallicity (Figure \ref{fig:sSFR}) is also minor. We note however that the incompleteness has to be taken into account
in order to quantitatively address the question whether the SN-GRB hosts are unbiased tracers of the overall population of star-forming galaxies 
\citep[e.g.,][]{Stanek06, Graham13, Vergani15}. 

If the sample is only affected by this magnitude limit, we can potentially correct for the incompleteness by applying a volume weight \citep[e.g.,][]{Huang12b}. 
However, the volume correction only accounts for the fact that the fainter objects are observed closer by. 
In order to obtain valid metallicities from strong line methods, we require a BPT class of star-forming, 
and thus the non-starforming galaxies and AGNs are further eliminated. Meanwhile, the S/N cutoff in all major emission lines may result 
in an additional bias against metal poor galaxies with weak [N{\sc ii}]$\lambda6584$ lines. 
The bias with respect to the metallicity, $M_*$ and SFR that are introduced by these selection criteria cannot be easily recovered. 
Therefore, our subset of SDSS galaxies only represents a population of star-forming galaxies that can reproduce the $M-Z$ relation in 
\citet{Kewley08} and the main star-forming sequence in \citet{Salim07}. 
It is not a complete sample, but it ensures a fair comparison with the previous works that usually 
use a similar SDSS population as the baseline of local galaxies. 

Unlike the targeted SN surveys which preferentially monitor the most luminous galaxies in order to maximize the chance of SN detection, 
the PTF as an untargeted survey avoids such bias to massive galaxies with overall high metallicities. 
However, the PTF survey relies on automated detection and verification pipelines to identify all the candidates of transients \citep{Brink13}, 
which may introduce a bias against transients that occur in the brightest regions of luminous galaxies. 
Specifically, the initial candidates are extracted from a subtraction image, which is constructed by subtracting a new image obtained nightly from 
a deep reference image of the static sky. The subtraction images are noisy in the brightest regions of luminous galaxies, 
and thus any transients there may be missed by the detection pipeline. This effect has a higher impact on the completeness of 
the PTF SN Ic sample, relative to the Ic-bl sample, due to the fact that the PTF SNe Ic are preferentially found in 
the central bright regions of luminous galaxies \citep[e.g.,][]{Anderson15}. If more SNe Ic hosts with high luminosities 
were to be added, the difference between the SN Ic and Ic-bl hosts would become even more significant. Therefore, we do not expect any qualitative 
change to our main conclusions due to the incompleteness of the PTF survey. 
Another potential bias may have been introduced when the PTF transients were chosen for spectroscopic classification, since only a minority of PTF transients could be followed up spectroscopically. However, \citep{Perley16b} argued that in the first few years of PTF, from which our sample is drawn, any spectroscopic classification bias did not introduce a large impact on the obtained host galaxy population, since the host galaxies of PTF CCSNe \citep{Arcavi10} traced the general population of star forming galaxies well. Nevertheless, any spectroscopic classification bias will be mitigated by the Zwicky Transient Facility \citep[ZTF;][]{ztf1, ztf2} since they have a dedicated low-R prism, the SED machine \citep{blagorodnova18} that is supposed to obtain spectroscopic IDs for all transients - yielding bias-free samples.

\subsection{Small Number Statistics and SN-bias}

Compared to the effect of survey bias, our conclusion is more vulnerable to the effect of small number statistics. 
Although our analysis is based on the largest homogenous samples of untargeted SNe Ic/Ic-bl from a single survey to date, 
as well as on all the SN-GRBs hosts published recently, we still only have 14 SNe Ic-bl with clean IDs 
and 10 SN-GRBs in total. The power of our statistical tests is highly limited by the small sample sizes, especially 
for the SNe Ic-bl and SN-GRBs. In the 2D analysis, the confidence intervals of the SN-GRB relations are so 
broad at a fixed $M_*$ that the differences between the relations of different samples are always insignificant. 
Our conclusions are mostly drawn from the $p$-values, which only account for type I error rates. 
However, the small sample sizes of the SN Ic-bl and SN-GRBs may lead to type II errors, i.e., 
the tests lack power to tell apart the true small differences between the intrinsic populations. 

Larger sample sizes of untargeted SNe Ic-bl and SN-GRBs are highly desired to enhance the test power. For SN-GRBs, every nearby system should be observed and published. Indeed the host of the recent long-duration GRB171205A may be a massive galaxy (see GCN by \citealt{perley17_grb17host} and \citealt{wang18}), though HST data are required to exclude the possibility that the GRB did not originate in a satellite galaxy of that massive galaxy, as was the case for GRB 130702A \citep{Kelly14}.
For SNe, current and upcoming surveys, such as the Zwicky Transient Facility \citep[ZTF;][]{ztf1, ztf2} and the Large Synoptic Survey Telescope \citep[LSST;][]{Ivezic08}, are expected to yield a large number of transients, which will solve this problem of small number statistics.

For our science goals, we have chosen to compare our SN hosts to hosts of GRBs that have accompanied SNe. However, there are a number of GRBs with no observed SNe (e.g., \citealt{dado18}), with the most famous being GRB 060614 with very deep SN limits (e.g., \citealt{galyam06}), though some of which may not be bona fide long-duration GRBs, but rather short-duration GRBs with extended tails (e.g, \citealt{caito09}). Those short-duration GRBs masquerading as long GRBs are not expected to be connected to core-collapse SNe  or may be altogether another type of GRB\citep{gehrels06}. Future work should include a complete host galaxy study of all GRBs within a certain volume, some of which do have very different host properties than those of long-duration GRBs and are more similar to those of short-GRBs (e.g. \citealt{levesque07} for GRB060605).

\subsection{Integrated vs. IFU studies}

Ideally, we want to probe the metallicities of SN progenitor stars by measuring their metallicities from their immediate environments 
that center at the exact SN sites. However, our long-slit observations limit our ability to obtain such ideal measurements. 

In terms of the centering accuracy, the majority of our host spectra were obtained long after the SNe themselves had faded. 
During observations, the exact SN sites are precisely localized by offsetting from the reference stars and the SN sites are put at the slit centers. 
However, the slit center locations are not easy to find during SN host spectrum extraction, and thus we use the positions of standard stars 
that are always in the slit center for guidance.
As a result, we can only localize the positions of SN sites to an accuracy 
that is comparable to the seeing. Due to the fact that the exact SN sites are not determined, we instead try to center 
the aperture for spectrum extraction at the nearest H$\alpha$ emission peak in the slit. 

In terms of the very immediate environment, due to the seeing effect and the requirement to gain a sufficient S/N for the spectrum, 
the size of the aperture for spectrum extraction corresponds to a larger physical scale than what would be ideal. 
By setting the aperture sizes to be twice the seeing, the SN sites safely fall within the apertures in most cases, 
and thus we name such apertures SN sites. However, rather than the metallicities at the exact SN sites, our metallicity measurements 
are closer to the metallicities at the nearest H{\sc ii} regions, with the caveats in mind that a cluster of H$\alpha$ emission 
may be a result of the superposition of multiple H{\sc ii} regions and that the truly nearest H{\sc ii} region may not be in slit. 
As found by the CALIFA IFS observations, the nearest H{\sc ii} regions provide the best approximations of local metallicities at the 
exact SN sites (unbiased and with the least scatter), compared to either global metallicities 
or local metallicities estimated using gradients \citep{Galbany16}. 

Despite the limitations of long-slit observations, we try our best to characterize the metallicities at the SN sites 
by the approximations of metallicities at the nearest H{\sc ii} regions. These apertures probe a typical physical scale of 2-3 kpc. 
They may be considered local for the SN Ic hosts that are massive, but are indeed closer to global for the low mass SN Ic-bl hosts. 
In contrast to our long-slit observations that are still integrated by nature, IFU observations with high angular resolutions 
provide a more comprehensive way to study the very immediate environments of SN explosions. 

For example, \citet{Kruehler17} present the spatially resolved spectroscopy of environments obtained with MUSE 
throughout the host galaxy of GRB980425/SN1998bw. They suggest that the common strong line calibrations using the 
$O3N2$ or $N2$ diagnostics lead to unrealistic variations in metallicities on sub-kpc scales, which are likely due to the variations 
in ionization parameters. Using a more realistic calibration that relies on H$\alpha$, [N{\sc ii}], and [S{\sc ii}] \citep{Dopita16}, 
those authors derive an immediate metallicity of $\log ({\rm O/H}) + 12 \sim 8.2$ at the SN site, $\log ({\rm O/H}) + 12 \sim 8.3$ 
for a galaxy-integrated spectrum, and $\log ({\rm O/H}) + 12 \sim 8.4$ for the nearby WR region. Among the four calibrations 
that we choose, the D13\_N2S2\_O3S2 calibration should be the most comparable. For the host of GRB980425/SN1998bw we obtain
a metallicity of $\log ({\rm O/H}) + 12 \sim 8.3$ in the D13\_N2S2\_O3S2 calibration based on data of \citet{Christensen08}, 
which is broadly consistent with their results. 

While pioneering IFU studies have been conducted that combine high spatial resolutions, sensitivity, and broad wavelength coverage to measure 
the oxygen abundance of the immediate environs of different types of SNe \citep{kuncarayakti13,Galbany16,kuncarayakti18} and of some SN-GRBs (e.g., \citealt{Christensen08,Kruehler17}) we suggest such a study for a large SN set from untargetted surveys, such as ours. Such a study would provide unarguably better constraints on the properties of SN progenitor stars.  In addition we suggest also wide-field, deep studies to constrain any galaxy companions or prior mergers that may have caused the episode of star formation responsible for the SN-GRB progenitor (e.g., \citet{Izzo17}). In addition, we suggest deeper spectra of all objects in order to measure the iron abundance from the faint emission line of [Fe III] $\lambda$4658, since iron is the element important for stellar evolution, and since using oxygen as a tracer in young galaxies such as GRB galaxies may overestimate the iron abundance \citep{hashimoto18}.


\section{Conclusions} \label{sec:concl}

In this work, we present the largest set of host-galaxy spectra of 28 SNe Ic and 14 SN Ic-bl, all discovered by the same untargeted survey, namely the Palomar Transient Factory (PTF), before 2013, as well as hosts of five SN Ic/Ic-bl candidates with uncertain IDs. These spectra were taken with the Keck, Palomar and Gemini-South telescopes and are supplemented by the SDSS spectra for the hosts of one PTF SN Ic and one PTF SN Ic-peculiar, and by the broadband photometry data from SDSS, Pan-STARRS, and {\it GALEX} for all host galaxies. 
Taking advantage of the {\it pyMCZ} code that was developed by the SNYU group, we calculate the metallicities from the host spectra in various calibrations, including recently introduced ones. 
The global $M_*$s, SFRs, and sSFRs are derived from SED fitting to the UV-to-optical bands following a Bayesian approach. 
Combined with the emission line fluxes and global stellar properties from the literatures for 10 SN-GRB hosts, 
we compare three SN host samples with each other (SN Ic vs. SN Ic-bl vs. SN-GRB), 
and with typical local galaxies from the SDSS and LVL, in order to uncover if any of these SNe occur preferentially in certain types of galaxies 
over others, in terms of their metallicities, $M_*$s, SFRs, and sSFRs. 
Such an SN host environment study provides constraints on the properties of SN progenitors, which are not directly observable 
from the pre-explosion images for these SN types. 

Compared to prior works, our work has the following strengths. 

\begin{enumerate}

  \item Our work is based on a homogenous dataset of the largest samples of SNe Ic and Ic-bl to date that are selected from a single 
  untargeted survey. Therefore, these host galaxies are not biased towards massive galaxies with systematically higher metallicities. 
  
  \item We carefully designed the strategies for observation and spectrum extraction, so that the line fluxes probe the metallicities of 
  immediate environments of the SN explosions to the best extent as is allowed by our long-slit observations. In particular, 
  the far background regions that we choose are free from the extended emission from the host galaxies, and 
  the emission lines are corrected for stellar absorption. 
  These measures all result in more reliable line ratios and thus metallicity measurements. 
  
  \item We calculate (recalculate) the gas-phase metallicities of the PTF SNe Ic/Ic-bl hosts, as well as those of SN-GRB hosts and SDSS galaxies, 
  with the {\it pyMCZ} code, such that the metallicities of different samples can be compared for the same calibrations in a self-consistent manner. 
  Meanwhile, we present the results in multiple calibrations to ensure that any observed trends are independent of the calibrations adopted. 
  The four calibrations we choose (namely KD02, PP04, M08 and D13) include different strong line methods, including both theoretical and hybrid ones, 
  as well as calibrations based on several different line diagnostics ($N2O2$, $R_{23}$, $O3N2$, $N2$, etc.). 
  Some of them are recent calibrations and thus are not considered by the previous works that compare the SN Ic-bl and GRB 
  hosts with the SDSS galaxies. 
  
  \item We carefully gather reliable UV-to-optical SEDs for the PTF SN Ic/Ic-bl hosts, and apply the same methods to derive the 
  $M_*$, SFRs, and sSFRs, as the MPA-JHU group does for the SDSS galaxies. In addition, if the SDSS photometry pipeline 
  suffers significantly from shredding, we use the NS-Atlas to recover the bad pipeline photometry. We also make sure that these global values for the 
  SN Ic/Ic-bl hosts make fair comparisons with those for the SN-GRB hosts, which we compiled from the literatures. 
  These global properties are therefore directly comparable across samples. 
  
  \item We identify one SN-GRB host with potential AGN contamination in a BPT diagram, as had been suggested in the literature, which may render its 
  metallicity and SFR measurements incorrect. This has a particular impact on the statistical analysis, 
  due to both the very small sample size of SN-GRBs and the extreme line ratios due to AGN contamination. 
  
  \item We adopt rigorous analytical methods to draw quantitative results. For example, we always present the uncertainties 
  along with the estimates for the physical parameters that are derived in this work, we apply survival analysis for 
  upper limits, and we conduct hypothesis tests to assess the statistical significance of the observed differences in the underlying populations. 
  
\end{enumerate}

Given these improvements, we reach the following main conclusions:

\begin{enumerate}

  \item As suggested by our comparisons between the distributions of line ratios, as well as of the metallicities in various calibrations (our Table~\ref{tab:OHdist}), 
  the three samples appear to follow the order of $Z_{\rm GRB}  \lessapprox Z_{\rm Ic-bl} < Z_{Ic}$  $-$  however, 
  the differences between the SN-GRB and Ic-bl hosts are not statistically significant (our Table~\ref{tab:Htest}), whereas those between the SN Ic-bl and Ic hosts are. 
    
  \item Given the $M-Z$ relations, the higher metallicities of the SN Ic hosts may be explained by the fact that they are overall more massive. 
  We isolate this effect by comparing the $M-Z$ relations between samples. Indeed, the SN-GRBs follow an average $M-Z$ relation that is 
  below that for the SN Ic-bl hosts, and the average $M-Z$ relations as defined by the SN Ic, Ic-bl, and SN-GRB hosts all fall slightly below 
  the same relations as defined by the SDSS galaxies. However, none of these differences in the average relations is statistically significant at the 95\% confidence level. Our result contradicts earlier works based on smaller samples that used a metallicity scale which is probably not appropriate for the comparison. Thus, our finding resolves the dilemma of why SN-GRBs would have chosen a completely different galaxy population not found in the local universe of star forming galaxies - because they do not.

  \item In terms of the absolute levels of global SFRs and $M_*$s, the SN-GRB and SN Ic-bl hosts are comparable, and they are significantly 
  below those of the SN Ic hosts. However, in terms of the relative enhancement of star formation activity as gauged by the main sequence of 
  star formation, all three SN host samples follow similar SFR vs. $M_*$ relations that are only slightly above the same relation as defined by the 
  star forming galaxies from the SDSS.  
  
  \item We show that the hosts of SN-GRBs and of PTF SNe Ic-bl both exhibit high absolute sSFRs and low metallicities. However, we are able to break the degeneracy between these two factors by comparing our samples to SDSS galaxies. We find that low metallicity is likely to be 
the fundamental cause for the formation of SN-GRB progenitors, whereas high sSFR is only a consequence of low metallicity. Thus we resolve a big debate in the field.

\end{enumerate}

Since we find that SN-GRB and PTF SN Ic-bl prefer statistically similar environments, in particular low metallicity, we suggest that the PTF SNe Ic-bl may have produced jets that were choked inside the star or were able tp break out of the star as GRBs that either were off-axis GRBs or low-luminosity radio-quiet on-axis GRBs. Thus broad lines in the spectra of SNe Ic may be a good indicator for the presence of a jet that either was able to breakout of the star \citep{Barnes18} or not. However, PTF SNe Ic live and die in very different environments than both SN-GRBs and PTF SN Ic-bl, namely in higher-metallicity regions with lower sSFRS, and may have had different progenitors than SN-GRBs and PTF SN Ic-bl $-$ there is no evidence supporting jet formation, in contrast to suggestions by Sobacchi et al. (2017) of all SNe Ib/c producing jets that are choked.


However, we should bear the following caveats in mind. 
The truly nearest HII region to the SN site or the birth cloud of the SN progenitor star may be outside of the slit that we chose 
to cover both the SN site and the galaxy center, so that a long-slit observation may still miss the region that is more representative 
of the progenitor properties. Most importantly, the strength of our statistical tests are limited by the small sample sizes of the SN-GRBs and SNe Ic-bl. 
Future transient surveys that provide even larger samples of SNe Ic-bl and with homogeneous dataset of their host galaxies 
will help to overcome this problem of small number statistics. Future IFU studies that combine high spatial resolution, sensitivity, 
and broad wavelength coverage will better probe the physical conditions of the progenitor stars, and in the context of different regions across the 
whole galaxies. Adopting a similar analytical approach as in the current work, such observations in the future will provide further insights to 
the progenitor models for the formation of SNe, especially the SN-GRBs and SNe Ic-bl.

We suggest that radio searches conduct even deeper follow-up observations of specifically SNe Ic-bl from untargeted galaxy searches in low metallicity environments as they appear to be the best candidates for hosting off-axis GRBs, as well as extending them to later times since GRBs may stay collimated longer than previously thought \citep{Ryan2015}.



\acknowledgements
We are grateful to Andrew MacFadyen and Brian Metzger for useful discussions. We kindly thank S. Ben-Ami, T. G. Brink,, K. I. Clubb, O. D. Fox , M. L. Graham, A. A. Miller, I. Shivvers, J. Silverman and O. Yaron for co-observing at Keck and C. Ott for PI'ing some proposals submitted before September 2015 and for co-observing at Keck on 2014-11-21.
M.M. and the SNYU group are supported by the NSF CAREER award AST-1352405, by the NSF award AST-1413260 and by a Humboldt Faculty Fellowship.Y.-Q. L. was supported in part by NYU GSAS Dean's Dissertation Fellowship. 
Support for I.A. was provided by NASA through the Einstein Fellowship Program, grant PF6-170148.  
A.G.-Y. is supported by the EU via ERC grant No. 725161, the Quantum Universe I-Core program, the ISF, the BSF Transformative program and by a Kimmel award. 


Furthermore, this research has made use of NASA's
Astrophysics Data System Bibliographic Services
(ADS)  and the NASA/IPAC Extragalactic Database (NED) which is operated by the Jet Propulsion Laboratory, California Institute
of Technology, under contract with the
National Aeronautics and Space Administration.

Some of the data presented herein were obtained at the W. M. Keck Observatory, which is operated as a scientific partnership among the California Institute of Technology, the University of California and the National Aeronautics and Space Administration. The Observatory was made possible by the generous financial support of the W. M. Keck Foundation. The authors wish to recognize and acknowledge the very significant cultural role and reverence that the summit of Maunakea has always had within the indigenous Hawaiian community.  We are most fortunate to have the opportunity to conduct observations from this mountain.
Gemini Observatory Telescopes (North and South) proposal IDs: GN-2011A-Q-93 and GS-2011A-C-5, PI: Modjaz. which are based on observations obtained at the Gemini Observatory (processed using the Gemini IRAF package), which is operated by the Association of Universities for Research in Astronomy, Inc., under a cooperative agreement with the NSF on behalf of the Gemini partnership: the National Science Foundation (United States), the National Research Council (Canada), CONICYT (Chile), Ministerio de Ciencia, Tecnolog\'{i}a e Innovaci\'{o}n Productiva (Argentina), and Minist\'{e}rio da Ci\^{e}ncia, Tecnologia e Inova\c{c}\~{a}o (Brazil).

This research has made use of the GHostS database (www.grbhosts.org), which is partly funded by Spitzer/NASA grant RSA Agreement No. 1287913.

This material is based upon work supported by AURA through the National Science Foundation under AURA Cooperative Agreement AST 0132798 as amended. 

Funding for the Sloan Digital Sky Survey IV has been provided by the Alfred P. Sloan Foundation, the U.S. Department of Energy Office of Science, and the Participating Institutions. SDSS-IV acknowledges
support and resources from the Center for High-Performance Computing at
the University of Utah. The SDSS web site is www.sdss.org.

SDSS-IV is managed by the Astrophysical Research Consortium for the 
Participating Institutions of the SDSS Collaboration including the 
Brazilian Participation Group, the Carnegie Institution for Science, 
Carnegie Mellon University, the Chilean Participation Group, the French Participation Group, Harvard-Smithsonian Center for Astrophysics, 
Instituto de Astrof\'isica de Canarias, The Johns Hopkins University, 
Kavli Institute for the Physics and Mathematics of the Universe (IPMU) / 
University of Tokyo, Lawrence Berkeley National Laboratory, 
Leibniz Institut f\"ur Astrophysik Potsdam (AIP),  
Max-Planck-Institut f\"ur Astronomie (MPIA Heidelberg), 
Max-Planck-Institut f\"ur Astrophysik (MPA Garching), 
Max-Planck-Institut f\"ur Extraterrestrische Physik (MPE), 
National Astronomical Observatories of China, New Mexico State University, 
New York University, University of Notre Dame, 
Observat\'ario Nacional / MCTI, The Ohio State University, 
Pennsylvania State University, Shanghai Astronomical Observatory, 
United Kingdom Participation Group,
Universidad Nacional Aut\'onoma de M\'exico, University of Arizona, 
University of Colorado Boulder, University of Oxford, University of Portsmouth, 
University of Utah, University of Virginia, University of Washington, University of Wisconsin, 
Vanderbilt University, and Yale University.
We  gratefully  acknowledge  NASA's  support  for  construction,  operation,  and  science
analysis  of  the  GALEX  mission,  developed  in  cooperation  with  the
Centre  National  d'Etudes  Spatiales  of  France  and  the  Korean  Ministry of Science and Technology.
These results are based in part on observations obtained at the Hale Telescope, Palomar Observatory, as part of a collaborative agreement between Caltech, JPL and Cornell University.
We gratefully acknowledge NASA's support for construction, operation, and science analysis of the GALEX mission, developed in cooperation with the Centre National d'Etudes Spatiales of France and the Korean Ministry of Science and Technology.

 \facility{GALEX, Gemini:Gillett, Gemini:South, Hale, Keck:I (LRIS), Keck:II (DEIMOS), Sloan}


\appendix

\section{Notes on Individual SN-GRBs}
\label{app:grbsn}

Here follows some individual notes for all SN-GRBs,  discussed in Section \ref{sec:sngrbhosts}.

\begin{itemize}

  \item GRB980425/SN1998bw -- We adopt the host line fluxes presented in \citet{Christensen08}, 
because they are based on IFU observations and are detected at the SN site. 
However, no [O{\sc iii}] $\lambda4959$ line flux nor its flux uncertainty is reported therein.  
We assume the [O{\sc iii}] $\lambda4959$ flux to be 1/3 that of [O{\sc iii}] $\lambda5007$, and the flux uncertainties to be at the 10\% level.

  \item XRF020903, GRB030329/SN2003dh, and GRB/XRF060218/SN2006aj -- 
For these three SN-GRB hosts we use the host line fluxes from \citet{han10} who re-analyze archival VLT spectra of GRB host galaxies. Being a systematic study of multiple sources, \citet{han10} correct for stellar absorption 
by modeling the continuum and stellar absorption lines, as we do for the PTF SN hosts. 

  \item GRB031203/SN2003lw -- We use the host line fluxes averaged from four VLT observations in \citet{Margutti07}. 
Because this source is very close to the Galactic plane, the far-IR map from \citet{Schlegel98} is unreliable along our line of sight towards this galaxy.
In order to correct for Galactic extinction, we follow \citet{Margutti07} and assume $E(B-V)_{\rm MW} = 0.72$~mag. 


  \item GRB100316D/SN2010bh -- We use line fluxes from \citet{Izzo17}, which are extracted at the SN site from IFU observations, and have been corrected for stellar absorption in the host galaxy and extinction due to the Milky Way. We note that if we use the line ratios of \citet{levesque11_10bh}, who do not correct for stellar absorption, nor provide uncertainties, then we compute a lower metallicity for this host than when using the values from \citet{Izzo17}.
  

  \item GRB120422A/SN2012bz -- We adopt the line fluxes measured by \citet{Schulze14} 
for the galaxy center rather than for the explosion site, because fewer emission lines of interest are detected at the explosion site, 
including the H$\beta$ line that is only marginally detected at the explosion site. 
According to \citet{Schulze14}, a slightly higher metallicity is derived at the explosion site ($\log ({\rm O/H}) + 12 \gtrsim 8.57 \pm 0.05$) 
than for the galaxy center ($\log ({\rm O/H}) + 12 = 8.43 \pm 0.01$), in the PP04\_N2H$\alpha$ calibration \citep{Pettini04}. 
Therefore, with our choice of galaxy center spectrum over SN site spectrum, we are likely to slightly underestimate the metallicity of the
immediate SN-GRB environment in this case. 

  \item GRB130427A/SN2013cq -- \citet{Xu13} report the final metallicity values in a specific calibration, but not the emission line fluxes on which that value is based. 
We requested the original spectra from the authors and measured the line fluxes via IRAF's {\it splot}. 

  \item GRB130702A/SN2013dx -- We note that this is the first GRB host galaxy that has been recognized to be the satellite dwarf galaxy of a more massive, metal-rich galaxy \citep{Kelly13}. The [N{\sc ii}] flux for its host galaxy 
is an upper limit in \citet{Kelly13} and the [O{\sc iii}] lines are unavailable, so that we can only obtain upper limits on metallicities in the 
KD02comb ($N2$-based in this case) and M08\_N2H$\alpha$ calibrations (see section \ref{sec:meta}). 

  \item GRB161219B/SN2016jca -- \citet{Ashall17} report the final metallicity range of the host galaxy for various calibrations, 
but not the emission line fluxes on which that range is based. We requested the emission line fluxes from the authors. 
The fluxes for the host galaxy of this most recent event are from an afterglow spectrum. 
We adopt the $M_*$ and SFR values for this host from \citet{cano17_obs_guide}. 
Note that the emission line fluxes are also available in \citet{cano17_obs_guide}, which result in line ratios that are 
consistent with the values that we adopt. However, the H$\beta$ line is unavailable in \citet{cano17_obs_guide}, 
which is important for extinction estimation.

\end{itemize}

Due to the limited access to raw spectra, we are not able to run {\it platefit} on this sample of the SN-GRB host galaxies 
to correct for stellar absorption. 
The only SN-GRB hosts for which stellar absorption has been removed are from \citet{han10} 
(XRF020903, GRB030329/SN2003dh, and GRB/XRF060218/SN2006aj) 
and from \citet{Izzo17} (GRB100316D/SN2010bh). 
We note that the stellar absorption correction is mostly ignored in the other SN-GRB works. 
However, the GRB hosts are generally known to be vigorously star forming so that stellar absorption is expected to be minimal.

\section{More details about the metallicity calibrations and scales we adopt in this paper}
\label{app:zcalib}

\begin{itemize}

\item
The KD02comb calibration is theoretical. 
It automatically chooses the optimal calibration from KD02\_N2O2 \citep{Kewley02}, KK04\_N2H$\alpha$ and KK04\_R$_{23}$ \citep{Kobulnicky04}, 
given the input line fluxes, and is implemented by {\it pyMCZ} following Appendix 2.2 and 2.3 from \citet{Kewley08}. 
The $N2O2$ ratio is not influenced by the diffused ionized gas \citep{Zhang16}, and is 
not sensitive to the ionization parameter, $q$, defined as the 
number of hydrogen-ionizing photons passing through a unit area per second, divided by the 
hydrogen density of the gas. 
Moreover, the $N2O2$ ratio is a strong monotonic function of the metallicity above log ([N{\sc ii}]/[O{\sc ii}]) $> -1.2$ \citep{Kewley02}. 
For log ([N{\sc ii}]/[O{\sc ii}]) $> -1.2$, or $\log ({\rm O/H}) + 12 \gtrsim 8.4$, the KD02comb calibration 
chooses KD02\_N2O2, if the desired line fluxes are available. 
For log ([N{\sc ii}]/[O{\sc ii}]) $< -1.2$, or $\log ({\rm O/H}) + 12 < 8.4$, the KD02comb calibration 
uses an average of the KK04\_R$_{23}$ lower branch and M91\_R$_{23}$ lower branch. 
The relation between oxygen abundance and $R_{23}$ is double valued, 
but parameter $q$ can break the degeneracy. 
To calculate the $R_{23}$-based metallicities, {\it pyMCZ} follows the recommendation in \citet{Kewley08} 
to derive a consistent $q$ and metallicity solution via an iterative approach. 
Note that the sample of 500 SDSS galaxies reveals a mild effect of discontinuity in the metallicity distribution 
around $\log ({\rm O/H}) + 12 \sim 8.4$ in the KD02comb calibration. 
A similar gap is more evident in the KD02 calibration, originally introduced in \citet{Kewley02}. 
This discontinuity may be caused by the S/N cutoff in line detections that are applied on the SDSS sample. 
We discuss the impact of this effect in section \ref{sec:ohdist}.

\item
The D13\_N2S2\_O3S2 calibration is theoretical. 
It is the only calibration out of the four that relies on the [S{\sc ii}] $\lambda6717$ and [S{\sc ii}] $\lambda6731$ lines. 
This calibration is calculated by the Python module, {\it pyqz} (version 0.4, the first publicly released version), 
following the MAPPINGS IV simulations \citep{Dopita13}. 
In particular, the photoionization models used in \citet{Kewley02} and in \citet{Kewley04} are updated by including new atomic data within 
a modified photoionization code and by assuming a more realistic $\kappa$ distribution for the energy of the electrons 
in the H{\sc ii} regions, rather than the simple Maxwell-Boltzmann distribution assumed in prior works. 

\item
The PP04\_O3N2 calibration is hybrid. It is a linear calibration with $O3N2$ \citep{Pettini04}. 
This calibration is included mostly for comparison reason, because it is commonly used in the literature 
investigating SN and GRB environments \citep[see][and references therein]{modjaz11}. 
We note that the PP04 calibrations are superseded by the M13 ones \citep{Marino13}, 
including another $O3N2$-based calibration, M13\_O3N2. 
Working with empirical calibrations, \citet{Marino13} find a significantly shallower slope in the relationship 
between the $N2$ and $O3N2$ ratios and the metallicity. 
We confirmed by our PTF SN hosts that the M13\_N2 and M13\_O3N2 calibrations give systematically lower metallicities 
for more metal-rich systems, relative to all the other theoretical or hybrid calibrations that we have presented here. 
As expected, purely empirical calibrations give lower metallicity values. 
However, the M13\_O3N2 calibration is calibrated over 
$-1 < \log$ [ ([O{\sc iii}]$\lambda5007$/H$\beta$) / ([N{\sc ii}]$\lambda6584$/H$\alpha$) ] $< 1.7$. 
This calibration is only valid for half of the SN-GRB hosts. 
To avoid the problem of small number statistics, it is critical to use a calibration that is valid for more SN-GRB hosts, 
and thus we choose the PP04\_O3N2 calibration over M13\_O3N2. 

\item
The M08\_N2H$\alpha$ calibration relies on only two lines with small separation in wavelength, H$\alpha$ and $N2$, and thus is available for all of the 
host galaxies of PTF SNe and SN-GRBs. The employed [N{\sc ii}]$\lambda6584$ line saturates 
at high metallicities, and thus usually the $N2$-based calibrations saturate at high-metallicity galaxies 
- at $\log({\rm O/H}) + 12 \gtrsim 8.8$ \citep{Kewley08}. 
As a hybrid calibration, M08\_N2H$\alpha$ in particular is $T_e$-based at low metallicities but uses KD02 photoionization models 
at high metallicities, and thus is less affected by such saturation. 
However, the sample of 500 SDSS galaxies reveals a clear discontinuity in the metallicity distribution 
around $\log ({\rm O/H}) + 12 \sim 9.2$ in the M08\_N2H$\alpha$ calibration. 
To derive a smooth $M-Z$ relation, we eliminate the SDSS galaxies with  $\log ({\rm O/H}) + 12 > 9.2$ in the M08\_N2H$\alpha$ calibration. 
Note that among all the PTF SN Ic/Ic-bl and SN-GRB hosts, only one SN Ic host has $\log ({\rm O/H}) + 12 > 9.2$ in the M08\_N2H$\alpha$ calibration 
and thus is affected by the discontinuity. 

\end{itemize}

\section{Comparison with Literature Values for all Samples}
\label{app:allgalscomparison}

In order to perform a self-consistent analysis, we have calculated metallicities for the SDSS galaxies and the 
host galaxies of all the PTF SNe Ic/Ic-bl and SN-GRBs in our sample using the same approach and code. 
Here, we compare our metallicity values with the literature values for the same galaxies for all our samples (PTF SNe, SN-GRBs and SDSS).

\begin{itemize} 

\item SDSS galaxies: For the SDSS galaxies, we can reproduce the \citet{Kewley08} $M-Z$ relation with the metallicities  in the PP04\_O3N2 calibration that we have in common. 


\item PTF Host galaxies: For the hosts of PTF SNe Ic/Ic-bl, we compare our values with those in \citet{valenti12} for PTFbov/SN2011bm and in \citet{sanders12} for seven PTF Ic/Ic-bl hosts. In summary, their values are in excellent agreement with our results, with ours being superior since we either have higher S/N data, provide detailed error bars or better IDs for the SNe.

\citet{valenti12} obtain line fluxes from the "vicinity" of PTFbov/SN2011bm and calculate a metallicity in the PP04\_O3N2 scale that is consistent with our value (though they do not provide error bars).
\citet{sanders12} presents seven PTF SN Ic/Ic-bl hosts in their sample, 
including PTF~10bzf (SN Ic-bl), which is listed as SN~2010ah in \citet{sanders12}. 
However, there have been changes in the IDs of some of the PTF SNe due to our improved 
SN identification (see section \ref{sec:sample}): 
PTF~09q is listed as an SN Ic in \citet{sanders12}, but we re-classify it as an SN Ia; thus, it is eliminated from this work; 
PTF~10bip and PTF~10vgv are classified as SNe Ic in \citet{sanders12}, 
but we update the ID of PTF~10bip to be of uncertain SN type, and PTF~10vgv to be an SN Ic-bl, as shown in \citet{modjaz16}. 
In conclusion, there are three SNe Ic-bl, two SNe Ic, and one SN with an uncertain type
in common between our and their sample of the PTF SNe Ibc, according to our improved IDs. 
Among them, the host of PTF~10aavz (SN Ic-bl) has weak emission line detections 
that are not sufficient to estimate the metallicities in \citet{sanders12}, 
whereas we have obtained the metallicities for this host based on our spectroscopic observations. 
Amongst the six different metallicity calibrations that are considered by \citet{sanders12}, there are two for which we overlap 
(KD02comb and PP04\_O3N2). In \citet{sanders12}, the metallicity in the KD02comb calibration is only reported for PTF~11hyg, 
and the metallicity in the PP04\_O3N2 calibration is reported for PTF~10bzf, PTF~10bip, and PTF~11hyg. 
Only three out of the 
five SNe hosts that we have in common can be compared in the same calibrations. 
Their values are in excellent agreement with ours within $1\sigma$, except for PTF~11hyg in PP04\_O3N2. 
The line ratios for the host of PTF~11hyg are measured at the SN site, which is not coincident with the nucleus, 
but suggest an ionization source from AGN, according to the BPT diagram (see section \ref{sec:bpt}), which is unphysical.
Our line fluxes suggest an ionization source from star formation at the SN site and therefore more reliable.
In short, our metallicities agree with those in \citet{sanders12} in almost all cases, and in the one case they do not, we think our data are superior since they indicate the more physically plausible scenario that the SN site was not an AGN.

\item Host galaxies of SN-GRBs:
For the host galaxies of SN-GRBs, we compare our metallicities with those derived in the same papers 
from which we collected the line fluxes. Based on the same input of line fluxes, the metallicity results are generally 
in agreement within the uncertainties for the same calibrations as expected. 
We note that the metallicities from the literature have no uncertainties reported for 
GRB130702A/SN2013dx \citep[][upper limit]{Kelly13} and GRB161219B/SN2016jca \citep{Ashall17}. 
If the exact same calibration used in the literature is not calculated by {\it pyMCZ}, 
we choose the one that is based on the same line ratios. 
For example, line fluxes for three of the following SN-GRB hosts are drawn from \citet{han10}, who re-analyze all archival VLT data of nearby GRB host galaxies (z$<$1), including those of SN-GRBs: 
XRF020903, GRB030329/SN2003dh, and GRB/XRF060218/SN2006aj. 
Those authors report the metallicities in the K99 calibration \citep{Kobulnicky99}, which is an $R_{23}$-based one. 
All these three hosts have low metallicities and low $N2O2$ ratios, so that their {\it pyMCZ}-derived 
metallicities for the KD02comb calibration are also $R_{23}$-based. 
We therefore compare the literature values in KK99 scale from \citet{han10} with the metallicities we compute in the KD02comb calibration,
and find excellent agreement for all these three hosts. 
The only exception with a large discrepancy comes from the comparison for GRB031203/SN2003lw , 
which may be due to AGN contamination (see section \ref{sec:bpt}). 
Note that \citet{Margutti07} and \citet{han10} derived different line ratios for this host from almost the 
same observational data (\citealt{han10} uses additional archival VLT data from 2005 that \citet{Margutti07} does not show) . 
The \citet{han10} values are less reliable in this case 
(private communication with Xuhui Han and Francois Hammer) 
and we adopt the line fluxes from \citet{Margutti07} for GRB031203/SN2003lw. \citet{Margutti07} report a $T_e$-based metallicity of 
$\log({\rm O/H}) + 12 = 8.12\pm0.04$ and \citet{han10} report 
one of 8.11$\pm$0.11 in the K99 calibration. 
Similarly, {\it pyMCZ} yields a low metallicity in both the PP04\_O3N2 and M08\_N2H$\alpha$ calibrations, 
but a high metallicity in the KD02comb calibration: $\log({\rm O/H}) + 12 = 8.647_{-0.082}^{+0.069}$. 
It is outside of the grid in \citet{Dopita13} so that the value in the D13\_N2S2\_O3S2 calibration is not calculated. 
This source has a high $N2O2$ ratio with log ([N{\sc ii}]/[O{\sc ii}]) $> -1.2$. 
Without the additional information of the auroral [O{\sc iii}] $\lambda4363$ line, 
{\it pyMCZ} places it on the upper branch of $R_{23}$ and 
its metallicity value for the KD02comb calibration is based on the $N2O2$ diagnostic. 
The host spectra of GRB031203/SN2003lw have most likely contribution from AGN activity. 
The effect of an AGN contribution is small for the $N2O2$-based calibrations, 
but larger for 
the calibrations using [O{\sc iii}] as a diagnostic \citep[e.g.][]{Kewley04, Kewley08}. 
We therefore keep its $N2O2$-based metallicity in the KD02comb calibration, which is much higher than our 
$O3N2$-based value, or the $R_{23}$- and $T_e$-based values from the literature. 
As mentioned in section \ref{sec:bpt}, we present our analysis results with and without this object in order to assess its impact, 
and plot it with a different symbol in our plots. 

\end{itemize}

\section{Identifying AGNs from the BPT diagram}
\label{app:bpt}

Usually all relevant lines are required to be detected with $S/N > 3$ to place the source on a BPT diagram, but we include 
sources with detected lines even at $S/N < 3$ in Figure \ref{fig:BPT}, in order to not introduce systematic bias (see section \ref{sec:meta}). 
However, there are three hosts with non-detections in certain lines such that they are excluded from the diagram. 
Two of them are SN Ic hosts (PTF~10qqd and PTF~10yow), with non-detections in [O{\sc iii}]$\lambda5007$. 
The third is an SN-GRB host (GRB130702A/SN2013dx), with non detection in [O{\sc iii}]$\lambda5007$ 
and only upper limit in [N{\sc ii}]$\lambda6584$. 
Following the classification scheme proposed by \citet{Brinchmann04}, they all have [N{\sc ii}]$\lambda6584$/H$\alpha < 0.6$ 
so that we can rule out the possibility of them being low S/N AGNs. 
In addition, all these three have H$\alpha$ lines detected at $S/N > 2$, 
so that we can further classify them as low S/N star forming galaxies. 

All the other host galaxies in our PTF SN and SN-GRB samples are shown in the lower left panel of Figure \ref{fig:BPT}, 
and all of  them lie below the solid division, i.e. they are not classified as AGN, but rather as star forming galaxies.
Another similar diagram is constructed in, e.g., \citet{Kewley01}, which depends on the same emission lines 
except that the [N{\sc ii}]$\lambda6584$ line is substituted by the [S{\sc ii}] $\lambda6717, 31$ lines. 

Thus, the strong line methods that we adopt for the metallicity calculations are valid, 
so are the H$\alpha$-based SFRs for SN-GRB hosts in particular. 
All the hosts form a clean sequence on the diagram. As a sanity check, there is no apparent scatter of host galaxies 
off this sequence, which would have otherwise rendered the line flux measurements suspicious.

We examine if these classifications are robust 
against the effect of extinction. In general, the BPT diagram is designed to be less affected by extinction, 
since it depends on two line ratios each involving two lines with small separation in wavelengths. 
We use line fluxes from Table \ref{tab:emission} (for PTF SNe Ic/Ic-bl, weird transients and transients with uncertain SN subtype) 
and \ref{tab:GRBspec} (for SN-GRBs) to make the plot, 
which are corrected for Galactic extinction but not for internal extinction. 
Applying extinction correction has the effect of shifting the data point 
slightly downwards and to the left in a BPT diagram, though always by a small amount. 
With the extra corrections for internal extinction, all the data points are expected to 
move even further away from the divisions, making no changes to their classifications. 
However, if Galactic extinction correction is not applied, all the data points are expected to move slightly towards the devision, 
which again makes no change to their classifications, except for the host of GRB031203/SN2003lw. 

The host of GRB031203/SN2003lw appears as a moderate outlier above the main distribution, 
but sits just below the two divisions in Figure \ref{fig:BPT}, i.e., still in the regime of star forming galaxy. 
It will shift into the regime of AGN powered objects 
if its line fluxes are not corrected for Galactic extinction.
Note that GRB031203/SN2003lw happens to lie very close to the Galactic plane, and thus the Galactic extinction is both large and highly uncertain. 
We follow \citet{Margutti07} to assume $E(B-V)_{\rm MW} = 0.72$~mag, which results in its current 
location in Figure \ref{fig:BPT}. 
With its data point sitting near the divisions and error bars crossing them, plus large and highly uncertain Galactic extinction, 
this host has a classification that is highly uncertain. 
Based on several diagnostics, \citet{levesque10_grbhosts} conclude that it shows definitive evidence of AGN activity and 
cannot be classified as a purely star-forming galaxy. 
On the contrary, \citet{Margutti07} and \citet{Prochaska04} both classify it as star forming, 
but did not account for the considerable error bars. 

Thus, for the host of GRB031203/SN2003lw, the metallicities derived from strong line methods 
in section \ref{sec:meta} should be taken as approximate, with the same caveat applying to their SFRs, which are based on H$\alpha$. 
As a result, we denote them by open diamonds in all plots to indicate that 
their metallicity and SFR estimates are less reliable. 
For the hypothesis tests on metallicity distributions (see section \ref{sec:ohdist}), we present {\it p}-values in both cases, with and without 
it in the SN-GRB sample. 

For the host galaxy sample of PTF SNe Ic and Ic-bl we do not detect any AGN emission. However, we note that for a majority of our PTF sample, our spectra do not probe the very center of the host galaxy where an AGN would reside. Works that do probe the center of the SN host galaxies, such as the CALIFA survey, do find that a high fraction of SNe hosts founded in a variety of targeted galaxies harbor AGN (22\% of the SNe Ibc hosts, \citep{galbany18}).

\bibliographystyle{apj_short_author}

\clearpage

\begin{turnpage}
\begin{deluxetable*}{llccccccccc}
\tabletypesize{\scriptsize}
\tablenum{2}
\tablewidth{0pt}
\tablecaption{Emission line fluxes of PTF SN host galaxies in units of $10^{-15}{\rm~erg~s^{-1}~cm^{-2}}$, corrected for stellar absorption and Galactic extinction.}
\label{tab:emission}
\tablehead{\colhead{PTF} & 
\colhead{Region} & 
\colhead{Redshift} & 
\colhead{[O{\sc ii}] $\lambda\lambda3726,29$} & 
\colhead{H$\beta$ $\lambda4861$} & 
\colhead{[O{\sc iii}] $\lambda4959$} & 
\colhead{[O{\sc iii}] $\lambda5007$} & 
\colhead{H$\alpha$ $\lambda6563$} & 
\colhead{[N{\sc ii}] $\lambda6584$} & 
\colhead{[S{\sc ii}] $\lambda6717$} & 
\colhead{[S{\sc ii}] $\lambda6731$} \\
\colhead{name} &
\colhead{} &
\colhead{} &
\colhead{} &
\colhead{} &
\colhead{} &
\colhead{} &
\colhead{} &
\colhead{} &
\colhead{} &
\colhead{} }
\startdata
	\multicolumn{11}{c}{SN Ic-bl}\\
\hline
  09sk &    nuc & 0.036 &  16.996$\pm$0.187 &   5.000$\pm$0.033 &   3.762$\pm$0.031 &   9.106$\pm$0.057 &  15.954$\pm$0.046 &   1.650$\pm$0.018 &   2.903$\pm$0.021 &   1.990$\pm$0.020\\ 
10aavz & SNsite & 0.062 &   0.090$\pm$0.004 &   0.028$\pm$0.003 &   0.002$\pm$0.003 &   0.031$\pm$0.003 &   0.054$\pm$0.004 &   0.010$\pm$0.003 &   0.014$\pm$0.003 &   0.010$\pm$0.003\\ 
 10bzf & SNsite & 0.050 &   0.116$\pm$0.036 &   0.060$\pm$0.005 &   0.027$\pm$0.005 &   0.087$\pm$0.006 &   0.153$\pm$0.007 &   0.013$\pm$0.005 &   0.036$\pm$0.005 &   0.025$\pm$0.005\\ 
 10ciw & SNsite & 0.115 &   4.110$\pm$0.284 &   1.046$\pm$0.087 &   0.546$\pm$0.077 &   2.077$\pm$1.102 &   3.960$\pm$0.093 &   0.703$\pm$0.076 &   0.908$\pm$0.054 &   0.617$\pm$0.047\\ 
 10qts & SNsite & 0.090 &  19.599$\pm$1.018 &   8.673$\pm$0.456 &   7.891$\pm$0.463 &  22.479$\pm$0.588 &  22.354$\pm$0.702 &   0.425$\pm$0.504 &   1.552$\pm$0.725 &   1.341$\pm$0.986\\ 
 10tqv & SNsite & 0.080 &   0.089$\pm$0.006 &   0.026$\pm$0.002 &   0.013$\pm$0.002 &   0.040$\pm$0.002 &   0.075$\pm$0.003 &   0.004$\pm$0.002 &   0.011$\pm$0.003 &   0.009$\pm$0.002\\ 
 10vgv & SNsite & 0.015 &  24.929$\pm$2.019 &  12.483$\pm$0.705 &   7.180$\pm$0.673 &  22.342$\pm$0.749 &  38.253$\pm$1.306 &   6.204$\pm$0.922 &   3.626$\pm$0.900 &   2.414$\pm$1.159\\ 
 10xem & SNsite & 0.056 &  12.023$\pm$0.093 &  10.542$\pm$0.062 &  18.984$\pm$0.111 &  42.328$\pm$0.252 &  48.036$\pm$0.062 &   1.618$\pm$0.008 &   1.517$\pm$0.008 &   1.263$\pm$0.008\\ 
 11cmh & SNsite & 0.105 &   0.164$\pm$0.029 &   0.024$\pm$0.006 &   0.005$\pm$0.006 &   0.017$\pm$0.006 &   0.068$\pm$0.004 &   0.013$\pm$0.006 &   0.016$\pm$0.002 &   0.012$\pm$0.003\\ 
 11gcj & SNsite & 0.148 &   0.479$\pm$0.011 &   0.130$\pm$0.006 &   0.098$\pm$0.005 &   0.286$\pm$0.006 &   0.340$\pm$0.004 &   0.028$\pm$0.003 &   0.054$\pm$0.006 &   0.041$\pm$0.004\\ 
 11img & SNsite & 0.158 &   0.135$\pm$0.008 &   0.026$\pm$0.006 &   0.028$\pm$0.004 &   0.071$\pm$0.005 &   0.141$\pm$0.009 &   0.012$\pm$0.004 &   0.018$\pm$0.004 &   0.015$\pm$0.005\\ 
 11lbm & SNsite & 0.039 &   0.525$\pm$0.010 &   0.169$\pm$0.002 &   0.080$\pm$0.002 &   0.314$\pm$0.003 &   0.672$\pm$0.006 &   0.039$\pm$0.003 &   0.122$\pm$0.004 &   0.077$\pm$0.003\\ 
 11qcj & SNsite & 0.028 &   5.300$\pm$0.447 &   1.303$\pm$0.010 &   1.120$\pm$0.010 &   3.290$\pm$0.018 &   3.472$\pm$0.021 &   0.355$\pm$0.008 &   0.617$\pm$0.010 &   0.448$\pm$0.009\\ 
  12as & SNsite & 0.033 &   0.622$\pm$0.029 &   0.133$\pm$0.002 &   0.055$\pm$0.001 &   0.180$\pm$0.002 &   0.601$\pm$0.005 &   0.089$\pm$0.003 &   0.120$\pm$0.004 &   0.076$\pm$0.003\\ 
\hline
	\multicolumn{11}{c}{SN Ic}\\
\hline
 09iqd &   HII3 & 0.034 &   0.341$\pm$0.062 &   0.130$\pm$0.005 &   0.047$\pm$0.005 &   0.125$\pm$0.006 &   0.590$\pm$0.018 &   0.189$\pm$0.014 &   0.095$\pm$0.014 &   0.067$\pm$0.013\\ 
 10bhu & SNsite & 0.036 &   0.937$\pm$0.129 &   0.284$\pm$0.023 &   0.088$\pm$0.024 &   0.208$\pm$0.024 &   0.965$\pm$0.020 &   0.228$\pm$0.014 &   0.304$\pm$0.017 &   0.190$\pm$0.016\\ 
 10fmx & SNsite & 0.044 &   0.249$\pm$0.041 &   0.071$\pm$0.013 &   0.026$\pm$0.013 &   0.037$\pm$0.014 &   0.223$\pm$0.013 &   0.068$\pm$0.012 &   0.049$\pm$0.012 &   0.031$\pm$0.011\\ 
 10hfe & SNsite & 0.048 &   3.633$\pm$0.270 &   0.991$\pm$0.018 &   0.282$\pm$0.014 &   0.716$\pm$0.018 &   2.419$\pm$0.024 &   0.697$\pm$0.013 &   0.627$\pm$0.013 &   0.463$\pm$0.012\\ 
 10hie & SNsite & 0.067 &   0.299$\pm$0.036 &   0.103$\pm$0.013 &   0.075$\pm$0.014 &   0.241$\pm$0.019 &   0.268$\pm$0.010 &   0.005$\pm$0.006 &   0.040$\pm$0.008 &   0.043$\pm$0.009\\ 
 10lbo & SNsite & 0.052 &   0.529$\pm$0.036 &   0.183$\pm$0.017 &   0.013$\pm$0.017 &   0.117$\pm$0.018 &   0.501$\pm$0.021 &   0.138$\pm$0.018 &   0.135$\pm$0.016 &   0.075$\pm$0.015\\ 
 10ood & SNsite & 0.060 &   1.353$\pm$0.132 &   0.502$\pm$0.011 &   0.427$\pm$0.011 &   1.190$\pm$0.017 &   1.769$\pm$0.020 &   0.186$\pm$0.010 &   0.217$\pm$0.010 &   0.166$\pm$0.010\\ 
 10osn &    HII & 0.038 &   0.168$\pm$0.201 &   0.104$\pm$0.007 &   0.018$\pm$0.007 &   0.050$\pm$0.009 &   0.584$\pm$0.013 &   0.220$\pm$0.010 &   0.093$\pm$0.009 &   0.067$\pm$0.008\\ 
 10qqd & SNsite & 0.080 &   0.032$\pm$0.031 &   0.071$\pm$0.011 &   ... &   ... &   0.867$\pm$0.021 &   0.187$\pm$0.014 &   0.145$\pm$0.015 &   0.095$\pm$0.018\\ 
 10tqi &   HII2 & 0.038 &   0.940$\pm$0.015 &   0.374$\pm$0.003 &   0.078$\pm$0.002 &   0.285$\pm$0.003 &   1.229$\pm$0.007 &   0.285$\pm$0.005 &   0.280$\pm$0.005 &   0.183$\pm$0.004\\ 
 10wal & SNsite & 0.028 &   0.124$\pm$0.010 &   0.055$\pm$0.002 &   0.005$\pm$0.002 &   0.025$\pm$0.002 &   0.177$\pm$0.004 &   0.046$\pm$0.003 &   0.052$\pm$0.004 &   0.033$\pm$0.005\\ 
 10xik & SNsite & 0.071 &   0.187$\pm$0.011 &   0.063$\pm$0.002 &   0.032$\pm$0.002 &   0.130$\pm$0.002 &   0.164$\pm$0.004 &   0.014$\pm$0.003 &   0.031$\pm$0.003 &   0.015$\pm$0.003\\ 
 10yow &   HII1 & 0.025 &   0.063$\pm$0.029 &   0.241$\pm$0.021 &   ... &   ... &   1.287$\pm$0.022 &   0.385$\pm$0.014 &   0.096$\pm$0.014 &   0.105$\pm$0.015\\ 
 10ysd & SNsite & 0.096 &   0.555$\pm$0.128 &   0.667$\pm$0.017 &   0.032$\pm$0.015 &   0.107$\pm$0.017 &   2.377$\pm$0.021 &   1.025$\pm$0.013 &   0.413$\pm$0.012 &   0.301$\pm$0.011\\ 
 10zcn & SNsite & 0.021 &   0.233$\pm$0.110 &   0.192$\pm$0.008 &   0.014$\pm$0.007 &   0.029$\pm$0.007 &   0.826$\pm$0.014 &   0.270$\pm$0.010 &   0.164$\pm$0.009 &   0.090$\pm$0.011\\ 
 11bov & SNsite & 0.022 &   1.035$\pm$0.050 &   0.270$\pm$0.006 &   0.238$\pm$0.006 &   0.660$\pm$0.009 &   0.659$\pm$0.006 &   0.066$\pm$0.003 &   0.166$\pm$0.004 &   0.083$\pm$0.003\\ 
 11hyg & SNsite & 0.029 &   0.158$\pm$0.038 &   0.200$\pm$0.008 &   0.017$\pm$0.007 &   0.050$\pm$0.007 &   1.136$\pm$0.014 &   0.459$\pm$0.010 &   0.184$\pm$0.010 &   0.123$\pm$0.011\\ 
 11ixk & SNsite & 0.024 &   0.451$\pm$0.194 &   0.161$\pm$0.016 &   0.003$\pm$0.014 &   0.087$\pm$0.015 &   0.475$\pm$0.013 &   0.176$\pm$0.011 &   0.114$\pm$0.011 &   0.107$\pm$0.011\\ 
 11jgj & SNsite & 0.040 &   0.319$\pm$0.115 &   0.173$\pm$0.015 &   0.034$\pm$0.013 &   0.045$\pm$0.014 &   1.005$\pm$0.016 &   0.340$\pm$0.010 &   0.162$\pm$0.011 &   0.139$\pm$0.011\\ 
 11klg &   HII2 & 0.026 &   0.115$\pm$0.007 &   0.109$\pm$0.002 &   0.011$\pm$0.002 &   0.019$\pm$0.002 &   0.393$\pm$0.006 &   0.159$\pm$0.003 &   0.088$\pm$0.004 &   0.057$\pm$0.003\\ 
 11rka & SNsite & 0.074 &  89.090$\pm$1.447 &  57.711$\pm$0.601 &  64.998$\pm$0.653 & 179.874$\pm$1.029 & 180.433$\pm$1.291 &   3.248$\pm$0.550 &  13.202$\pm$0.602 &   8.788$\pm$0.647\\ 
 12cjy & SNsite & 0.044 &   0.452$\pm$0.030 &   0.359$\pm$0.011 &   0.015$\pm$0.009 &   0.033$\pm$0.010 &   1.161$\pm$0.012 &   0.407$\pm$0.008 &   0.238$\pm$0.008 &   0.160$\pm$0.007\\ 
 12dcp & SNsite & 0.031 &   6.228$\pm$0.112 &   1.788$\pm$0.037 &   0.780$\pm$0.021 &   2.233$\pm$0.045 &   6.878$\pm$0.034 &   1.497$\pm$0.014 &   1.090$\pm$0.013 &   0.802$\pm$0.012\\ 
 12dtf & SNsite & 0.062 &   0.884$\pm$0.052 &   0.348$\pm$0.009 &   0.302$\pm$0.009 &   0.962$\pm$0.015 &   1.609$\pm$0.013 &   0.156$\pm$0.007 &   0.192$\pm$0.008 &   0.142$\pm$0.007\\ 
 12fgw & SNsite & 0.055 &   2.597$\pm$0.074 &   1.602$\pm$0.019 &   0.248$\pm$0.015 &   0.742$\pm$0.018 &   8.329$\pm$0.029 &   2.454$\pm$0.013 &   1.309$\pm$0.011 &   0.891$\pm$0.010\\ 
 12hvv &    nuc & 0.029 &   0.705$\pm$0.328 &   0.172$\pm$0.019 &   0.034$\pm$0.019 &   0.100$\pm$0.019 &   0.498$\pm$0.015 &   0.136$\pm$0.012 &   0.173$\pm$0.015 &   0.087$\pm$0.012\\ 
 12jxd & SNsite & 0.025 &   0.854$\pm$0.017 &   0.504$\pm$0.004 &   0.139$\pm$0.003 &   0.240$\pm$0.003 &   2.353$\pm$0.010 &   0.805$\pm$0.006 &   0.527$\pm$0.006 &   0.349$\pm$0.006\\ 
 12ktu & SNsite & 0.031 &   0.292$\pm$0.109 &   0.406$\pm$0.010 &   0.025$\pm$0.008 &   0.054$\pm$0.008 &   1.648$\pm$0.016 &   0.567$\pm$0.011 &   0.229$\pm$0.011 &   0.177$\pm$0.011\\ 
\hline
	\multicolumn{11}{c}{weird/uncertain SN subtype}\\
\hline
  09ps & SNsite & 0.107 &   0.757$\pm$0.047 &   0.244$\pm$0.019 &   0.162$\pm$0.023 &   0.329$\pm$0.034 &   0.930$\pm$0.016 &   0.083$\pm$0.011 &   0.169$\pm$0.010 &   0.120$\pm$0.009\\ 
 10bip &    nuc & 0.051 &   ... &   0.135$\pm$0.008 &   0.171$\pm$0.009 &   0.409$\pm$0.010 &   0.636$\pm$0.008 &   0.078$\pm$0.006 &   0.115$\pm$0.005 &   0.100$\pm$0.005\\ 
 10gvb & SNsite & 0.098 &   0.205$\pm$0.083 &   0.074$\pm$0.004 &   0.082$\pm$0.005 &   0.274$\pm$0.007 &   0.293$\pm$0.006 &   0.011$\pm$0.004 &   0.039$\pm$0.006 &   0.022$\pm$0.005\\ 
 10svt & SNsite & 0.031 &   1.188$\pm$0.037 &   0.546$\pm$0.009 &   0.459$\pm$0.009 &   1.303$\pm$0.015 &   1.613$\pm$0.019 &   0.096$\pm$0.011 &   0.172$\pm$0.015 &   0.116$\pm$0.013\\ 
 12gzk & SNsite & 0.148 &   ... &   0.750$\pm$0.016 &   0.645$\pm$0.015 &   1.928$\pm$0.022 &   2.322$\pm$0.020 &   0.051$\pm$0.006 &   0.125$\pm$0.007 &   0.091$\pm$0.007\\ 
 12hni & SNsite & 0.106 &   ... &   1.012$\pm$0.023 &   1.246$\pm$0.018 &   3.639$\pm$0.031 &   3.170$\pm$0.029 &   0.266$\pm$0.010 &   0.370$\pm$0.011 &   0.312$\pm$0.011\\
\enddata
\end{deluxetable*}
\end{turnpage}

\begin{turnpage}
	\begin{deluxetable}{lccccccccc}
		\tabletypesize{\scriptsize}
		\tablenum{3}
		\tablewidth{0pt}
		\tablecaption{SN-GRB host galaxy emission line fluxes, in units of $10^{-17}{\rm~erg~s^{-1}~cm^{-2}}$, corrected for Galactic extinction.}
		\label{tab:GRBspec}
		\tablehead{\colhead{SN-GRBname} &
			\colhead{${\rm z}$} &
			\colhead{[O{\sc ii}] $\lambda\lambda3726,29$} &
			\colhead{H$\beta$ $\lambda4861$} &
			\colhead{[O{\sc iii}] $\lambda4959$} &
			\colhead{[O{\sc iii}] $\lambda5007$} &
			\colhead{H$\alpha$ $\lambda6563$} &
			\colhead{[N{\sc ii}] $\lambda6584$} &
			\colhead{[S{\sc ii}] $\lambda6717$} &
			\colhead{[S{\sc ii}] $\lambda6731$} \\
			\colhead{} &
			\colhead{} &
			\colhead{} &
			\colhead{} &
			\colhead{} &
			\colhead{} &
			\colhead{} &
			\colhead{} &
			\colhead{} }
		\startdata
		GRB980425/SN1998bw\tablenotemark{a} & 0.00867
		& 260.7$\pm$26.07 & 75.5$\pm$7.55 & 60.9$\pm$6.09 & 182.7$\pm$18.27 & 449.4$\pm$44.94 & 50.3$\pm$5.03 & 91.9$\pm$9.19 & 68.4$\pm$6.84\\
		XRF020903 \tablenotemark{b} & 0.2506
		& 11.48$\pm$0.36 & 8.58$\pm$0.29 & 15.56$\pm$0.31 & 44.11$\pm$0.33 & 26.03$\pm$0.21 & 0.77$\pm$0.1 & 2.14$\pm$0.17 & 1.28$\pm$0.09\\
		GRB030329/SN2003dh\tablenotemark{c} & 0.16867
		& 19.62$\pm$0.32 & 9.8$\pm$0.16 & 11.74$\pm$0.2 & 30.53$\pm$0.26 & 30.94$\pm$0.27 & 0.29$\pm$0.15 & 3.82$\pm$0.09 & 1.89$\pm$0.08\\
		GRB031203/SN2003lw \tablenotemark{d} & 0.10536
		& 529.54$\pm$34.22 & 624.65$\pm$40.06 & 1391.93$\pm$88.99 & 4200.29$\pm$268.59 & 2517.63$\pm$188.04 & 130.84$\pm$9.71 & 86.14$\pm$5.53 & 68.28$\pm$4.37\\
		GRB/XRF060218/SN2006aj  \tablenotemark{e} & 0.03342
		& 224.25$\pm$1.01 & 91.43$\pm$0.65 & 108.4$\pm$0.48 & 291.23$\pm$0.66 & 261.51$\pm$0.54 & 10.73$\pm$0.32 & 15.4$\pm$0.36 & 11.56$\pm$0.41\\
		GRB100316D/SN2010bh \tablenotemark{f} & 0.0592
		& ... & 6.542$\pm$0.142 & 6.688$\pm$0.148 & 19.268$\pm$0.203 & 22.785$\pm$0.208 & 2.15$\pm$0.108 & 3.097$\pm$0.115 & 2.42$\pm$0.109\\
		GRB120422A/SN2012bz \tablenotemark{g} & 0.28253
		& 58.0$\pm$6.7 & 12.8$\pm$0.4 & 8.3$\pm$0.3 & 25.1$\pm$0.5 & 53.6$\pm$0.5 & 8.1$\pm$0.4 & 9.1$\pm$0.2 & 6.7$\pm$0.3\\
		GRB130427A/SN2013cq \tablenotemark{h} & 0.3399
		& 18.55$\pm$1.33 & 5.401$\pm$0.384 & 4.21$\pm$0.448 & 9.766$\pm$0.756 & 18.76$\pm$0.705 & 3.189$\pm$0.583 & 3.477$\pm$1.403 & 7.689$\pm$2.078\\
		GRB130702A/SN2013dx \tablenotemark{i} & 0.145
		& ... & 4.3$\pm$0.52 & ... & ... & 11.10$\pm$0.05 & 0.54 & ... & ...\\
		GRB161219B/SN2016jca \tablenotemark{j}  & 0.1475\tablenotemark{k}
		& 8.31$\pm$0.38 & 2.58$\pm$0.29 & 1.6$\pm$0.28 & 3.99$\pm$0.47 & 5.59$\pm$0.18 & 0.41$\pm$0.12 & ... & ...\\
		\enddata
		\tablenotetext{a}{\citet{Christensen08}, for the SN region. We assume that the flux uncertainties are at the 10\% level and that the line flux for [O{\sc iii}] $\lambda4959$ is 1/3 of that for [O{\sc iii}] $\lambda5007$.} 
		\tablenotetext{b}{\citet{han10}, who correct for stellar absorption.} 
		\tablenotetext{c}{\citet{han10}, who correct for stellar absorption.}  
		\tablenotetext{d}{\citet{Margutti07}, line fluxes averaged from four VLT observations.} 
		\tablenotetext{e}{\citet{han10}, who correct for stellar absorption.}  
		\tablenotetext{f}{
			\citet{Izzo17}, who correct for extinction and stellar absorption. } 
		\tablenotetext{g}{\citet{Schulze14}, at the host center site.} 
		\tablenotetext{h}{spectrum from \citet{Xu13} and we re-measured the line fluxes via {\it splot} in IRAF.}
		\tablenotetext{i}{\citet{Kelly13}. Note that their [N{\sc ii}] flux is a 2$\sigma$ upper limit.} 
		\tablenotetext{j}{\citet{Ashall17}}
		\tablenotetext{k}{\citet{cano17_obs_guide}}
	\end{deluxetable}
\end{turnpage}
			
\clearpage
\begin{deluxetable}{llcccccc}
\tabletypesize{\small}
\tablenum{4}
\tablewidth{0pt}
\tablecaption{pyMCZ Measurements.}
\label{tab:PTFZ}
\tablehead{\colhead{PTFname} & 
\colhead{SN type} & 
\colhead{{$E(B-V)_{\rm host}$}} & 
\colhead{$\log{\rm (O/H)}+12$} & 
\colhead{$\log{\rm (O/H)}+12$} & 
\colhead{$\log{\rm (O/H)}+12$} & 
\colhead{$\log{\rm (O/H)}+12$} \\
\colhead{} &
\colhead{} &
\colhead{(mag)} &
\colhead{D13\_N2S2\_O3S2} &
\colhead{KD02\_COMB} &
\colhead{PP04\_O3N2} &
\colhead{M08\_N2H$\alpha$} }
\startdata
	\multicolumn{8}{c}{SN Ic-bl}\\
\hline
  09sk &  Ic-bl & 0.111$_{-0.007}^{+0.007}$  & 8.368$_{-0.007}^{+0.006}$  & 8.486$_{-0.008}^{+0.009}$  & 8.335$_{-0.002}^{+0.002}$  & 8.522$_{-0.004}^{+0.004}$ \\
10aavz &  Ic-bl & 0.000$_{-0.000}^{+0.000}$  & 8.516$_{-0.193}^{+0.151}$  & 8.631$_{-0.133}^{+0.078}$  & 8.481$_{-0.051}^{+0.044}$  & 8.728$_{-0.127}^{+0.095}$ \\
 10bzf/SN10ah &  Ic-bl & 0.000$_{-0.000}^{+0.000}$  & 8.152$_{-0.270}^{+0.188}$  & 8.625$_{-0.218}^{+0.146}$  & 8.333$_{-0.064}^{+0.052}$  & 8.451$_{-0.161}^{+0.117}$ \\
 10ciw &  Ic-bl & 0.282$_{-0.083}^{+0.092}$  & 8.545$_{-0.070}^{+0.073}$  & 8.572$_{-0.084}^{+0.074}$  & 8.400$_{-0.060}^{+0.096}$  & 8.708$_{-0.041}^{+0.037}$ \\
 10qts &  Ic-bl & 0.000$_{-0.000}^{+0.000}$  & 8.108$_{-0.354}^{+0.294}$  & 8.033$_{-0.041}^{+0.046}$  & 8.081$_{-0.143}^{+0.081}$  & 8.089$_{-0.411}^{+0.601}$ \\
 10tqv &  Ic-bl & 0.007$_{-0.007}^{+0.093}$  & 8.118$_{-0.294}^{+0.226}$  & 8.168$_{-0.090}^{+0.245}$  & 8.268$_{-0.089}^{+0.056}$  & 8.312$_{-0.244}^{+0.138}$ \\
 10vgv &  Ic-bl & 0.070$_{-0.064}^{+0.067}$  & 8.852$_{-0.107}^{+0.104}$  & 8.811$_{-0.055}^{+0.051}$  & 8.399$_{-0.024}^{+0.021}$  & 8.677$_{-0.058}^{+0.048}$ \\
 10xem &  Ic-bl & 0.471$_{-0.006}^{+0.006}$  & ...  & 8.075$_{-0.007}^{+0.007}$  & 8.082$_{-0.001}^{+0.001}$  & 8.118$_{-0.002}^{+0.002}$ \\
 11cmh &  Ic-bl & 0.000$_{-0.000}^{+0.298}$  & 8.654$_{-0.278}^{+0.178}$  & 8.524$_{-0.107}^{+0.118}$  & 8.550$_{-0.103}^{+0.084}$  & 8.737$_{-0.206}^{+0.139}$ \\
 11gcj &  Ic-bl & 0.000$_{-0.000}^{+0.000}$  & 8.257$_{-0.077}^{+0.072}$  & 8.229$_{-0.041}^{+0.179}$  & 8.273$_{-0.017}^{+0.016}$  & 8.444$_{-0.039}^{+0.034}$ \\
 11img &  Ic-bl & 0.657$_{-0.232}^{+0.267}$  & 8.363$_{-0.226}^{+0.176}$  & 8.475$_{-0.069}^{+0.073}$  & 8.265$_{-0.066}^{+0.051}$  & 8.455$_{-0.143}^{+0.102}$ \\
 11lbm &  Ic-bl & 0.333$_{-0.014}^{+0.015}$  & 8.035$_{-0.050}^{+0.049}$  & 8.280$_{-0.020}^{+0.019}$  & 8.260$_{-0.011}^{+0.010}$  & 8.323$_{-0.028}^{+0.026}$ \\
 11qcj &  Ic-bl & 0.000$_{-0.000}^{+0.000}$  & 8.327$_{-0.014}^{+0.014}$  & 8.425$_{-0.074}^{+0.043}$  & 8.284$_{-0.004}^{+0.003}$  & 8.518$_{-0.008}^{+0.008}$ \\
  12as &  Ic-bl & 0.462$_{-0.017}^{+0.017}$  & 8.552$_{-0.020}^{+0.018}$  & 8.562$_{-0.013}^{+0.014}$  & 8.438$_{-0.005}^{+0.005}$  & 8.645$_{-0.012}^{+0.012}$ \\
\hline
	\multicolumn{8}{c}{SN Ic}\\
\hline
 09iqd &     Ic & 0.468$_{-0.049}^{+0.045}$  & 8.996$_{-0.047}^{+0.044}$  & 8.802$_{-0.054}^{+0.057}$  & 8.594$_{-0.013}^{+0.012}$  & 8.941$_{-0.038}^{+0.035}$ \\
 10bhu &     Ic & 0.173$_{-0.083}^{+0.090}$  & 8.657$_{-0.037}^{+0.036}$  & 8.751$_{-0.068}^{+0.062}$  & 8.579$_{-0.020}^{+0.021}$  & 8.814$_{-0.026}^{+0.024}$ \\
 10fmx &     Ic & 0.104$_{-0.104}^{+0.206}$  & 8.938$_{-0.114}^{+0.102}$  & 8.812$_{-0.137}^{+0.081}$  & 8.662$_{-0.060}^{+0.065}$  & 8.916$_{-0.083}^{+0.084}$ \\
 10hfe &     Ic & 0.000$_{-0.000}^{+0.000}$  & 8.782$_{-0.011}^{+0.011}$  & 8.787$_{-0.020}^{+0.021}$  & 8.602$_{-0.005}^{+0.005}$  & 8.893$_{-0.009}^{+0.009}$ \\
 10hie &     Ic & 0.000$_{-0.000}^{+0.045}$  & 7.762$_{-0.260}^{+0.227}$  & 8.108$_{-0.098}^{+0.125}$  & 8.094$_{-0.147}^{+0.088}$  & 8.099$_{-0.431}^{+0.591}$ \\
 10lbo &     Ic & 0.000$_{-0.000}^{+0.058}$  & 8.808$_{-0.075}^{+0.072}$  & 8.853$_{-0.045}^{+0.036}$  & 8.613$_{-0.031}^{+0.031}$  & 8.873$_{-0.055}^{+0.056}$ \\
 10ood &     Ic & 0.210$_{-0.025}^{+0.027}$  & 8.510$_{-0.031}^{+0.031}$  & 8.541$_{-0.051}^{+0.048}$  & 8.304$_{-0.009}^{+0.008}$  & 8.528$_{-0.020}^{+0.019}$ \\
 10osn &     Ic & 0.681$_{-0.071}^{+0.074}$  & 9.106$_{-0.027}^{+0.029}$  & 8.828$_{-0.162}^{+0.201}$  & 8.719$_{-0.025}^{+0.028}$  & 9.022$_{-0.027}^{+0.029}$ \\
 10qqd &     Ic & 1.469$_{-0.146}^{+0.168}$  & ...  & 8.818$_{-0.169}^{+0.195}$  & ...  & 8.779$_{-0.031}^{+0.027}$ \\
 10tqi &     Ic & 0.140$_{-0.010}^{+0.010}$  & 8.758$_{-0.010}^{+0.010}$  & 8.827$_{-0.007}^{+0.008}$  & 8.569$_{-0.003}^{+0.003}$  & 8.806$_{-0.007}^{+0.007}$ \\
 10wal &     Ic & 0.120$_{-0.044}^{+0.043}$  & 8.772$_{-0.044}^{+0.040}$  & 8.884$_{-0.031}^{+0.032}$  & 8.656$_{-0.015}^{+0.016}$  & 8.849$_{-0.027}^{+0.029}$ \\
 10xik &     Ic & 0.000$_{-0.000}^{+0.000}$  & 8.282$_{-0.154}^{+0.117}$  & 8.479$_{-0.359}^{+0.079}$  & 8.288$_{-0.034}^{+0.027}$  & 8.458$_{-0.082}^{+0.065}$ \\
 10yow &     Ic & 0.631$_{-0.087}^{+0.097}$  & ...  & 9.181$_{-0.061}^{+0.078}$  & ...  & 8.909$_{-0.017}^{+0.018}$ \\
 10ysd &     Ic & 0.222$_{-0.027}^{+0.027}$  & 9.209$_{-0.013}^{+0.014}$  & 9.136$_{-0.033}^{+0.039}$  & 8.875$_{-0.021}^{+0.024}$  & 9.378$_{-0.012}^{+0.010}$ \\
 10zcn &     Ic & 0.411$_{-0.042}^{+0.048}$  & 9.115$_{-0.028}^{+0.031}$  & 8.986$_{-0.077}^{+0.104}$  & 8.851$_{-0.030}^{+0.038}$  & 8.950$_{-0.019}^{+0.019}$ \\
 11bov/SN11bm &     Ic & 0.000$_{-0.000}^{+0.000}$  & 8.204$_{-0.031}^{+0.029}$  & 8.402$_{-0.140}^{+0.033}$  & 8.286$_{-0.008}^{+0.008}$  & 8.511$_{-0.016}^{+0.016}$ \\
 11hyg/SN11ee &     Ic & 0.693$_{-0.040}^{+0.042}$  & 9.185$_{-0.018}^{+0.017}$  & 9.046$_{-0.039}^{+0.047}$  & 8.820$_{-0.018}^{+0.020}$  & 9.066$_{-0.016}^{+0.017}$ \\
 11ixk &     Ic & 0.030$_{-0.030}^{+0.115}$  & 8.926$_{-0.043}^{+0.041}$  & 8.922$_{-0.087}^{+0.111}$  & 8.679$_{-0.027}^{+0.031}$  & 9.013$_{-0.038}^{+0.042}$ \\
 11jgj &     Ic & 0.717$_{-0.086}^{+0.088}$  & 9.094$_{-0.030}^{+0.036}$  & 8.835$_{-0.095}^{+0.102}$  & 8.792$_{-0.039}^{+0.051}$  & 8.966$_{-0.016}^{+0.015}$ \\
 11klg &     Ic & 0.235$_{-0.024}^{+0.022}$  & 9.126$_{-0.015}^{+0.015}$  & 9.087$_{-0.013}^{+0.013}$  & 8.856$_{-0.015}^{+0.014}$  & 9.066$_{-0.015}^{+0.017}$ \\
 11rka &     Ic & 0.090$_{-0.013}^{+0.013}$  & 7.726$_{-0.175}^{+0.117}$  & 7.977$_{-0.012}^{+0.011}$  & 8.018$_{-0.026}^{+0.022}$  & 7.856$_{-0.084}^{+0.069}$ \\
 12cjy &     Ic & 0.124$_{-0.031}^{+0.033}$  & 9.145$_{-0.024}^{+0.030}$  & 9.055$_{-0.017}^{+0.017}$  & 8.919$_{-0.035}^{+0.050}$  & 8.984$_{-0.011}^{+0.011}$ \\
 12dcp &     Ic & 0.300$_{-0.020}^{+0.021}$  & 8.807$_{-0.006}^{+0.006}$  & 8.669$_{-0.015}^{+0.015}$  & 8.497$_{-0.004}^{+0.004}$  & 8.782$_{-0.004}^{+0.004}$ \\
 12dtf &     Ic & 0.485$_{-0.025}^{+0.028}$  & 8.481$_{-0.026}^{+0.024}$  & 8.384$_{-0.063}^{+0.057}$  & 8.281$_{-0.008}^{+0.007}$  & 8.501$_{-0.016}^{+0.014}$ \\
 12fgw &     Ic & 0.604$_{-0.012}^{+0.012}$  & 9.031$_{-0.003}^{+0.003}$  & 8.861$_{-0.009}^{+0.009}$  & 8.688$_{-0.004}^{+0.004}$  & 8.903$_{-0.002}^{+0.003}$ \\
 12hvv &     Ic & 0.017$_{-0.017}^{+0.122}$  & 8.735$_{-0.053}^{+0.054}$  & 8.759$_{-0.131}^{+0.148}$  & 8.627$_{-0.032}^{+0.035}$  & 8.870$_{-0.040}^{+0.039}$ \\
 12jxd &     Ic & 0.495$_{-0.009}^{+0.009}$  & 8.985$_{-0.005}^{+0.004}$  & 8.912$_{-0.007}^{+0.006}$  & 8.701$_{-0.002}^{+0.002}$  & 8.971$_{-0.004}^{+0.004}$ \\
 12ktu &     Ic & 0.354$_{-0.026}^{+0.027}$  & 9.201$_{-0.016}^{+0.017}$  & 9.099$_{-0.052}^{+0.067}$  & 8.874$_{-0.019}^{+0.022}$  & 8.974$_{-0.010}^{+0.011}$ \\
\hline
	\multicolumn{8}{c}{weird/uncertain SN subtype}\\
\hline
  09ps &     Ic/Ic-bl & 0.292$_{-0.078}^{+0.083}$  & 8.309$_{-0.088}^{+0.073}$  & 8.347$_{-0.131}^{+0.116}$  & 8.362$_{-0.025}^{+0.026}$  & 8.472$_{-0.050}^{+0.042}$ \\
 10bip &     Ic/Ic-bl & 0.504$_{-0.061}^{+0.065}$  & 8.378$_{-0.050}^{+0.042}$  & ...  & 8.302$_{-0.013}^{+0.013}$  & 8.580$_{-0.027}^{+0.027}$ \\
 10gvb &  SLSN/Ic-bl & 0.329$_{-0.058}^{+0.061}$  & 7.945$_{-0.231}^{+0.191}$  & 8.280$_{-0.135}^{+0.129}$  & 8.102$_{-0.062}^{+0.045}$  & 8.161$_{-0.185}^{+0.118}$ \\
 10svt &     Ib/c & 0.032$_{-0.020}^{+0.021}$  & 8.296$_{-0.079}^{+0.070}$  & 8.479$_{-0.061}^{+0.049}$  & 8.218$_{-0.016}^{+0.015}$  & 8.331$_{-0.042}^{+0.037}$ \\
 12gzk & Ic-peculiar & 0.081$_{-0.022}^{+0.022}$  & 8.048$_{-0.091}^{+0.072}$  & 7.943$_{-0.053}^{+0.046}$  & 8.071$_{-0.017}^{+0.015}$  & 7.941$_{-0.052}^{+0.046}$ \\
 12hni &     SLSN/Ic-bl & 0.091$_{-0.024}^{+0.025}$  & 8.373$_{-0.024}^{+0.025}$  & 8.457$_{-0.007}^{+0.004}$  & 8.211$_{-0.006}^{+0.006}$  & 8.451$_{-0.013}^{+0.013}$ \\
\enddata
\end{deluxetable}
\clearpage

\begin{deluxetable}{lccccc}
\tabletypesize{\small}
\tablenum{5}
\tablewidth{0pt}
\tablecaption{PYMCZ computations for SN-GRB host galaxies: reddening and metallicities}
\label{tab:GRBZ}
\tablehead{\colhead{SN-GRBname} &
\colhead{{$E(B-V)_{\rm host}$}} &
\colhead{$\log{\rm (O/H)}+12$} &
\colhead{$\log{\rm (O/H)}+12$} &
\colhead{$\log{\rm (O/H)}+12$} &
\colhead{$\log{\rm (O/H)}+12$} \\
\colhead{} &
\colhead{(mag)} &
\colhead{D13\_N2S2\_O3S2} &
\colhead{KD02\_COMB} &
\colhead{PP04\_O3N2} &
\colhead{M08\_N2H$\alpha$} }
\startdata
	GRB980425/SN1998bw & 0.743$_{-0.15}^{+0.143}$ & 8.32$_{-0.072}^{+0.066}$ & 8.485$_{-0.049}^{+0.068}$ & 8.329$_{-0.025}^{+0.024}$ & 8.55$_{-0.049}^{+0.046}$ \\
XRF020903 & 0.059$_{-0.035}^{+0.035}$ & 8.007$_{-0.111}^{+0.083}$ & 8.183$_{-0.045}^{+0.252}$ & 8.016$_{-0.019}^{+0.018}$ & 8.068$_{-0.058}^{+0.049}$ \\
GRB030329/SN2003dh & 0.1$_{-0.019}^{+0.018}$ & 7.485$_{-0.051}^{+0.081}$ & 8.073$_{-0.018}^{+0.019}$ & 7.927$_{-0.091}^{+0.057}$ & 7.584$_{-0.21}^{+0.183}$ \\
GRB031203/SN2003lw & 0.347$_{-0.098}^{+0.097}$ & ... & 8.647$_{-0.082}^{+0.069}$ & 8.066$_{-0.016}^{+0.017}$ & 8.282$_{-0.037}^{+0.038}$ \\
GRB/XRF060218/SN2006aj & 0.0$_{0.0}^{+0.007}$ & 8.357$_{-0.02}^{+0.021}$ & 8.125$_{-0.006}^{+0.007}$ & 8.125$_{-0.004}^{+0.004}$ & 8.194$_{-0.011}^{+0.011}$ \\
GRB100316D/SN2010bh & 0.199$_{-0.023}^{+0.022}$ & 8.393$_{-0.032}^{+0.027}$ & 8.46$_{-0.001}^{+0.001}$ & 8.259$_{-0.007}^{+0.007}$ & 8.491$_{-0.017}^{+0.016}$ \\
GRB120422A/SN2012bz & 0.385$_{-0.033}^{+0.035}$ & 8.573$_{-0.025}^{+0.026}$ & 8.545$_{-0.118}^{+0.057}$ & 8.387$_{-0.009}^{+0.008}$ & 8.652$_{-0.018}^{+0.018}$ \\
GRB130427A/SN2013cq & 0.195$_{-0.076}^{+0.087}$ & 8.343$_{-0.164}^{+0.139}$ & 8.63$_{-0.101}^{+0.081}$ & 8.407$_{-0.03}^{+0.029}$ & 8.694$_{-0.071}^{+0.062}$ \\
GRB130702A/SN2013dx & 0.0$_{0.0}^{+0.021}$ & ... & 8.20\tablenotemark{a} & ... & 8.37\tablenotemark{a}\\
GRB161219B/SN2016jca & 0.0$_{0.0}^{+0.0}$ & ... & 8.115$_{-0.086}^{+0.289}$ & 8.305$_{-0.053}^{+0.043}$ & 8.402$_{-0.126}^{+0.09}$ \\ 
\enddata
\tablenotetext{a}{upper limit}
\end{deluxetable}
	

\begin{deluxetable}{lccccccccc}
\tabletypesize{\small}
\tablenum{6}
\tablewidth{480pt}
\tablecaption{SED fitting of PTF Ic and Ic-bl hosts: input photometry and output parameters }
\label{tab:SED}
\tablehead{\colhead{PTF} &
\colhead{{\it u}-mag} &
\colhead{{\it g}-mag} &
\colhead{{\it r}-mag} &
\colhead{{\it i}-mag} &
\colhead{{\it z}-mag} &
\colhead{FUV-mag} &
\colhead{NUV-mag} &
\colhead{$\log M_*$} &
\colhead{$\log {\rm SFR}$} \\
\colhead{Name} &
\colhead{(mag)} &
\colhead{(mag)} &
\colhead{(mag)} &
\colhead{(mag)} &
\colhead{(mag)} &
\colhead{(mag)} &
\colhead{(mag)} &
\colhead{($M_\odot$)} &
\colhead{($M_\odot~{\rm yr}^{-1}$)} }
\startdata
	\multicolumn{10}{c}{SN Ic-bl}\\
\hline
	09sk & 18.741$\pm$0.033 & 17.811$\pm$0.008 & 17.522$\pm$0.007 & 17.301$\pm$0.01 & 17.134$\pm$0.021 & 19.596$\pm$0.151 & 19.127$\pm$0.08 & 8.93$_{-0.06}^{+0.1}$ & -0.38$_{-0.11}^{+0.24}$ \\
10aavz & 20.827$\pm$0.157 & 19.575$\pm$0.026 & 19.056$\pm$0.022 & 18.892$\pm$0.029 & 18.588$\pm$0.071 & 21.68$\pm$0.186 & 21.089$\pm$0.12 & 8.96$_{-0.09}^{+0.09}$ & -0.75$_{-0.16}^{+0.15}$ \\
10bzf/SN10ah & 20.111$\pm$0.11 & 19.382$\pm$0.024 & 19.019$\pm$0.025 & 18.701$\pm$0.032 & 18.574$\pm$0.091 & 21.631$\pm$0.317 & 21.637$\pm$0.197 & 8.75$_{-0.07}^{+0.1}$ & -0.61$_{-0.35}^{+0.42}$ \\
10ciw & 20.669$\pm$0.122 & 19.651$\pm$0.024 & 19.187$\pm$0.024 & 18.884$\pm$0.026 & 18.808$\pm$0.071 & 21.387$\pm$0.412 & 21.301$\pm$0.317 & 9.39$_{-0.08}^{+0.1}$ & -0.22$_{-0.22}^{+0.21}$ \\
10qts & 22.562$\pm$0.372 & 21.884$\pm$0.079 & 21.509$\pm$0.086 & 22.162$\pm$0.208 & 21.208$\pm$0.394 & 21.846$\pm$0.493 & 21.215$\pm$0.433 & 7.53$_{-0.18}^{+0.19}$ & -1.08$_{-0.2}^{+0.23}$ \\
10tqv & 23.212$\pm$0.799 & 21.686$\pm$0.103 & 21.085$\pm$0.084 & 20.926$\pm$0.121 & 20.498$\pm$0.305 & ... & ... & 8.46$_{-0.16}^{+0.17}$ & -1.66$_{-1.1}^{+0.48}$ \\
10vgv & 16.293$\pm$0.015 & 15.278$\pm$0.003 & 14.755$\pm$0.003 & 14.498$\pm$0.003 & 14.34$\pm$0.008 & 17.845$\pm$0.025 & 16.81$\pm$0.014 & 9.46$_{-0.01}^{+0.35}$ & 0.18$_{-0.0}^{+0.33}$ \\
10xem & 18.547$\pm$0.025 & 17.761$\pm$0.007 & 18.359$\pm$0.012 & 17.861$\pm$0.011 & 18.346$\pm$0.055 & 19.298$\pm$0.042 & 19.056$\pm$0.028 & 8.48$_{-0.01}^{+0.13}$ & 0.42$_{-0.01}^{+0.05}$ \\
11cmh & 21.054$\pm$0.271 & 20.022$\pm$0.044 & 19.572$\pm$0.047 & 19.356$\pm$0.07 & 19.575$\pm$0.333 & ... & ... & 9.08$_{-0.11}^{+0.13}$ & -0.45$_{-0.47}^{+0.32}$ \\
11gcj & 21.197$\pm$0.148 & 20.386$\pm$0.036 & 19.979$\pm$0.034 & 19.842$\pm$0.048 & 19.906$\pm$0.188 & 21.047$\pm$0.034 & 20.913$\pm$0.016 & 9.07$_{-0.14}^{+0.24}$ & -0.28$_{-0.05}^{+0.18}$ \\
11img & 23.858$\pm$1.67 & 21.817$\pm$0.148 & 22.275$\pm$0.294 & 21.74$\pm$0.28 & 21.852$\pm$0.964 & ... & ... & 8.13$_{-0.25}^{+0.28}$ & -0.63$_{-0.33}^{+0.29}$ \\
11lbm & 19.306$\pm$0.082 & 18.121$\pm$0.011 & 17.751$\pm$0.014 & 17.599$\pm$0.018 & 17.505$\pm$0.061 & 20.44$\pm$0.156 & 20.455$\pm$0.182 & 8.88$_{-0.07}^{+0.1}$ & -1.02$_{-0.44}^{+0.38}$ \\
11qcj & 17.256$\pm$0.014 & 16.616$\pm$0.004 & 16.531$\pm$0.005 & 16.457$\pm$0.007 & 16.522$\pm$0.079 & 17.953$\pm$0.022 & 17.71$\pm$0.009 & 8.66$_{-0.05}^{+0.05}$ & 0.13$_{-0.12}^{+0.15}$ \\
12as & 17.617$\pm$0.02 & 16.665$\pm$0.004 & 16.312$\pm$0.004 & 16.11$\pm$0.005 & 15.964$\pm$0.014 & 18.481$\pm$0.011 & 18.072$\pm$0.007 & 9.35$_{-0.07}^{+0.08}$ & -0.01$_{-0.07}^{+0.1}$ \\
\hline
	\multicolumn{10}{c}{SN Ic}\\
\hline
09iqd & ... & 15.043$\pm$0.01 & 14.45$\pm$0.092 & 14.037$\pm$0.01 & 13.872$\pm$0.01 & ... & ... & 10.56$_{-0.1}^{+0.1}$ & -0.34$_{-1.36}^{+0.61}$ \\
10bhu & 18.086$\pm$0.03 & 16.971$\pm$0.006 & 16.554$\pm$0.005 & 16.339$\pm$0.007 & 16.123$\pm$0.016 & 19.278$\pm$0.106 & 18.874$\pm$0.063 & 9.43$_{-0.06}^{+0.08}$ & -0.11$_{-0.15}^{+0.16}$ \\
10fmx & 17.016$\pm$0.019 & 15.287$\pm$0.003 & 14.488$\pm$0.002 & 14.108$\pm$0.002 & 13.794$\pm$0.005 & 18.598$\pm$0.097 & 17.688$\pm$0.048 & 11.08$_{-0.07}^{+0.08}$ & 0.54$_{-0.11}^{+0.08}$ \\
10hfe & 16.134$\pm$0.024 & 15.436$\pm$0.003 & 15.22$\pm$0.004 & 15.01$\pm$0.005 & 14.924$\pm$0.012 & 16.633$\pm$0.008 & 16.228$\pm$0.003 & 10.26$_{-0.22}^{+0.18}$ & 1.18$_{-0.32}^{+0.15}$ \\
10hie & 19.968$\pm$0.135 & 18.448$\pm$0.016 & 18.092$\pm$0.018 & 17.873$\pm$0.021 & 18.015$\pm$0.091 & ... & ... & 9.22$_{-0.07}^{+0.11}$ & -1.59$_{-1.36}^{+0.79}$ \\
10lbo & 19.534$\pm$0.075 & 18.417$\pm$0.015 & 17.914$\pm$0.013 & 17.635$\pm$0.017 & 17.469$\pm$0.051 & 20.817$\pm$0.198 & 20.443$\pm$0.084 & 9.32$_{-0.08}^{+0.09}$ & -0.46$_{-0.23}^{+0.2}$ \\
10ood & 18.991$\pm$0.075 & 17.875$\pm$0.011 & 17.691$\pm$0.012 & 17.457$\pm$0.013 & 17.498$\pm$0.043 & 19.896$\pm$0.054 & 19.474$\pm$0.027 & 9.13$_{-0.04}^{+0.05}$ & 0.09$_{-0.2}^{+0.14}$ \\
10osn & 15.971$\pm$0.023 & 14.712$\pm$0.003 & 13.997$\pm$0.002 & 13.708$\pm$0.002 & 13.452$\pm$0.005 & 17.946$\pm$0.058 & 17.165$\pm$0.027 & 10.87$_{-0.08}^{+0.09}$ & 0.95$_{-0.13}^{+0.13}$ \\
10qqd & 18.68$\pm$0.061 & 17.481$\pm$0.009 & 16.924$\pm$0.008 & 16.598$\pm$0.008 & 16.422$\pm$0.025 & ... & ... & 10.17$_{-0.08}^{+0.09}$ & 0.24$_{-0.26}^{+0.29}$ \\
10tqi & 16.495$\pm$0.013 & 15.418$\pm$0.002 & 15.016$\pm$0.008 & 14.745$\pm$0.003 & 14.585$\pm$0.007 & 17.265$\pm$0.04 & 16.854$\pm$0.024 & 10.22$_{-0.1}^{+0.08}$ & 0.58$_{-0.09}^{+0.09}$ \\
10wal & 16.436$\pm$0.023 & 15.15$\pm$0.003 & 14.525$\pm$0.004 & 14.224$\pm$0.004 & 13.977$\pm$0.008 & 17.704$\pm$0.061 & 17.2$\pm$0.036 & 10.39$_{-0.09}^{+0.08}$ & 0.31$_{-0.13}^{+0.11}$ \\
10xik & ... & 18.851$\pm$0.1 & 18.554$\pm$0.084 & 18.396$\pm$0.052 & 18.377$\pm$0.04 & 20.934$\pm$0.365 & 20.415$\pm$0.265 & 8.93$_{-0.07}^{+0.1}$ & -0.23$_{-0.5}^{+0.32}$ \\
10yow & 14.688$\pm$0.011 & 13.175$\pm$0.001 & 12.381$\pm$0.001 & 11.961$\pm$0.001 & 11.685$\pm$0.002 & 17.108$\pm$0.054 & 16.159$\pm$0.026 & 11.4$_{-0.11}^{+0.1}$ & 1.19$_{-0.13}^{+0.15}$ \\
10ysd & 18.945$\pm$0.037 & 17.701$\pm$0.007 & 17.107$\pm$0.006 & 16.706$\pm$0.006 & 16.464$\pm$0.013 & 21.479$\pm$0.341 & 20.332$\pm$0.135 & 10.4$_{-0.06}^{+0.07}$ & 1.12$_{-0.21}^{+0.23}$ \\
10zcn & 15.403$\pm$0.015 & 14.255$\pm$0.002 & 13.743$\pm$0.005 & 13.419$\pm$0.003 & 13.199$\pm$0.005 & 17.372$\pm$0.055 & 16.602$\pm$0.026 & 10.33$_{-0.05}^{+0.07}$ & 0.79$_{-0.13}^{+0.13}$ \\
11bov/SN11bm & 15.931$\pm$0.009 & 15.078$\pm$0.002 & 14.79$\pm$0.002 & 14.644$\pm$0.002 & 14.559$\pm$0.008 & 16.628$\pm$0.013 & 16.353$\pm$0.003 & 9.44$_{-0.04}^{+0.06}$ & 0.23$_{-0.06}^{+0.1}$ \\
11hyg/SN11ee & 14.163$\pm$0.005 & 12.994$\pm$0.001 & 12.604$\pm$0.001 & 12.285$\pm$0.001 & 12.023$\pm$0.002 & 15.817$\pm$0.009 & 15.335$\pm$0.005 & 10.95$_{-0.07}^{+0.15}$ & 1.11$_{-0.24}^{+0.15}$ \\
11ixk & 15.983$\pm$0.01 & 14.581$\pm$0.002 & 13.924$\pm$0.001 & 13.567$\pm$0.002 & 13.276$\pm$0.004 & 17.165$\pm$0.017 & 16.614$\pm$0.007 & 10.66$_{-0.05}^{+0.13}$ & 0.41$_{-0.06}^{+0.13}$ \\
11jgj & 17.291$\pm$0.026 & 15.624$\pm$0.003 & 14.77$\pm$0.003 & 14.326$\pm$0.003 & 13.936$\pm$0.005 & ... & ... & 10.93$_{-0.09}^{+0.1}$ & 0.32$_{-0.42}^{+0.45}$ \\
11klg & 16.123$\pm$0.042 & 14.319$\pm$0.001 & 13.564$\pm$0.001 & 13.098$\pm$0.001 & 12.884$\pm$0.008 & ... & ... & 10.93$_{-0.1}^{+0.09}$ & -0.88$_{-1.33}^{+0.81}$ \\
11rka & 21.649$\pm$0.199 & 21.201$\pm$0.058 & 21.126$\pm$0.078 & 20.831$\pm$0.092 & 20.722$\pm$0.268 & 21.755$\pm$0.496 & 22.247$\pm$0.402 & 7.86$_{-0.18}^{+0.26}$ & -0.8$_{-0.36}^{+0.28}$ \\
12cjy & 17.341$\pm$0.024 & 16.132$\pm$0.003 & 15.546$\pm$0.004 & 15.222$\pm$0.003 & 14.974$\pm$0.01 & 19.011$\pm$0.108 & 18.287$\pm$0.046 & 10.31$_{-0.07}^{+0.11}$ & 0.56$_{-0.16}^{+0.16}$ \\
12dcp & 15.708$\pm$0.035 & 14.458$\pm$0.002 & 13.912$\pm$0.002 & 13.622$\pm$0.002 & 13.411$\pm$0.006 & ... & ... & 10.57$_{-0.08}^{+0.09}$ & 0.56$_{-0.27}^{+0.3}$ \\
12dtf & 21.047$\pm$0.096 & 19.942$\pm$0.018 & 19.72$\pm$0.02 & 19.298$\pm$0.021 & 19.332$\pm$0.066 & ... & ... & 8.58$_{-0.07}^{+0.12}$ & -0.64$_{-0.28}^{+0.41}$ \\
12fgw & 17.602$\pm$0.016 & 16.398$\pm$0.004 & 15.838$\pm$0.003 & 15.433$\pm$0.003 & 15.211$\pm$0.008 & 19.097$\pm$0.116 & 18.562$\pm$0.06 & 10.42$_{-0.08}^{+0.1}$ & 0.54$_{-0.15}^{+0.15}$ \\
12hvv & 18.454$\pm$0.046 & 17.333$\pm$0.007 & 16.809$\pm$0.007 & 16.545$\pm$0.008 & 16.416$\pm$0.024 & 19.688$\pm$0.055 & 19.288$\pm$0.037 & 9.28$_{-0.08}^{+0.09}$ & -0.57$_{-0.12}^{+0.15}$ \\
12jxd & 15.241$\pm$0.011 & 13.995$\pm$0.001 & 13.324$\pm$0.001 & 12.947$\pm$0.001 & 12.633$\pm$0.002 & 16.358$\pm$0.024 & 15.926$\pm$0.014 & 10.92$_{-0.07}^{+0.03}$ & 0.61$_{-0.05}^{+0.09}$ \\
12ktu & ... & 13.121$\pm$0.012 & 12.528$\pm$0.015 & 12.242$\pm$0.003 & 12.01$\pm$0.004 & 16.223$\pm$0.037 & 15.647$\pm$0.019 & 11.15$_{-0.08}^{+0.1}$ & 1.09$_{-0.14}^{+0.13}$ \\
\hline
	\multicolumn{10}{c}{weird/uncertain SN subtype}\\
\hline
09ps & 20.749$\pm$0.112 & 19.773$\pm$0.019 & 19.482$\pm$0.02 & 19.194$\pm$0.021 & 19.203$\pm$0.085 & ... & ... & 9.04$_{-0.07}^{+0.09}$ & -0.2$_{-0.23}^{+0.33}$ \\
10bip & 20.024$\pm$0.074 & 18.835$\pm$0.012 & 18.591$\pm$0.013 & 18.372$\pm$0.015 & 18.293$\pm$0.05 & 20.998$\pm$0.21 & 20.52$\pm$0.109 & 8.72$_{-0.05}^{+0.07}$ & -0.47$_{-0.26}^{+0.21}$ \\
10gvb & 21.131$\pm$0.202 & 20.125$\pm$0.041 & 19.883$\pm$0.05 & 19.519$\pm$0.051 & 19.884$\pm$0.308 & 21.477$\pm$0.44 & 21.289$\pm$0.308 & 8.81$_{-0.1}^{+0.15}$ & -0.5$_{-0.19}^{+0.28}$ \\
10svt & ... & 18.302$\pm$0.063 & 18.002$\pm$0.05 & 17.958$\pm$0.051 & 17.809$\pm$0.085 & ... & ... & 8.44$_{-0.1}^{+0.13}$ & -0.83$_{-1.13}^{+0.37}$ \\
12gzk & 18.958$\pm$0.194 & 18.782$\pm$0.015 & 18.804$\pm$0.018 & 18.86$\pm$0.03 & 18.61$\pm$0.075 & 19.341$\pm$0.022 & 19.409$\pm$0.005 & 6.95$_{-0.08}^{+0.08}$ & -1.65$_{-0.09}^{+0.14}$ \\
12hni & 19.685$\pm$0.102 & 18.961$\pm$0.018 & 18.624$\pm$0.02 & 18.391$\pm$0.023 & 18.386$\pm$0.09 & 20.203$\pm$0.31 & 19.939$\pm$0.178 & 9.39$_{-0.09}^{+0.16}$ & 0.01$_{-0.14}^{+0.32}$ \\
\enddata
\end{deluxetable}


\end{document}